\documentclass[10pt,journal,compsoc]{IEEEtran}

\usepackage{cite}
\usepackage{amsmath,amssymb,amsfonts}
\usepackage{algorithmic}
\usepackage{graphicx}
\usepackage{textcomp}
\usepackage{xcolor}
\usepackage{amsmath}
\usepackage{caption}
\usepackage{subcaption}
\usepackage{hyperref}
\usepackage{booktabs}
\usepackage{amssymb}
\usepackage{pifont}
\usepackage{multirow}
\usepackage{pdfpages}
\usepackage{tikzpagenodes} 

  \usepackage{cite}
\ifCLASSINFOpdf
\else
\fi

\begin{document}
%
\title{End-to-End Label Uncertainty Modeling in Speech Emotion Recognition using Bayesian Neural Networks and Label Distribution Learning}
%
%
%
%

\author{Navin~Raj Prabhu,~\IEEEmembership{Student Member,~IEEE,}
        Nale~Lehmann-Willenbrock,~\IEEEmembership{Non-Member,~IEEE,}
        and~Timo~Gerkmann,~\IEEEmembership{Senior Member,~IEEE,}
\IEEEcompsocitemizethanks{\IEEEcompsocthanksitem Navin Raj Prabhu and Timo Gerkman are with the Signal Processing Lab, Universit\"at Hamburg, Germany, 20146. \protect
E-mail: navin.raj.prabhu@uni-hamburg.de, timo.gerkmann@uni-hamburg.de
\IEEEcompsocthanksitem Nale Lehmann-Willenbrock is with the Department of Industrial and Organizational Psychology, Universit\"at Hamburg, Germany, 20146.\protect\\
E-mail: nale.lehmann-willenbrock@uni-hamburg.de
}
\thanks{This work was supported by the Landesforschungsförderung Hamburg (LFF-FV79), as part of the research unit  ''Mechanisms of Change in Dynamic Social Interactions''.}

}

\markboth{Journal of \LaTeX\ Class Files,~Vol.~xx, No.~xx, xxx 2022}%
{Raj Prabhu \MakeLowercase{\textit{et al.}}: End-to-End Label Uncertainty Modeling in Speech Emotion Recognition using Bayesian Neural Networks and Label Distribution Learning}
\IEEEtitleabstractindextext{%
\begin{abstract}
    To train machine learning algorithms to predict emotional expressions in terms of arousal and valence, annotated datasets are needed. However, as different people perceive others' emotional expressions differently, their annotations are subjective. To account for this, annotations are typically collected from multiple annotators and averaged to obtain ground-truth labels. However, when exclusively trained on this averaged ground-truth, the model is agnostic to the inherent subjectivity in emotional expressions. In this work, we therefore propose an end-to-end Bayesian neural network capable of being trained on a distribution of annotations to also capture the subjectivity-based label uncertainty. Instead of a Gaussian, we model the annotation distribution using Student's $t$-distribution, which also accounts for the number of annotations available. We derive the corresponding Kullback-Leibler divergence loss and use it to train an estimator for the annotation distribution, from which the mean and uncertainty can be inferred. We validate the proposed method using two in-the-wild datasets. We show that the proposed $t$-distribution based approach achieves state-of-the-art uncertainty modeling results in speech emotion recognition, and also consistent results in cross-corpora evaluations. Furthermore, analyses reveal that the advantage of a $t$-distribution over a Gaussian grows with increasing inter-annotator correlation and a decreasing number of annotations available.
    
\end{abstract}

\begin{IEEEkeywords}
Emotional expressions, annotations, Bayesian neural networks, label distribution learning, end-to-end, speech emotion recognition, uncertainty, subjectivity, $t$-distributions, Kullback-Leibler divergence loss
\end{IEEEkeywords}}

\maketitle

\IEEEdisplaynontitleabstractindextext

%
\IEEEpeerreviewmaketitle

\ifCLASSOPTIONcompsoc
\IEEEraisesectionheading{\section{Introduction}\label{sec:introduction}}
\else
\section{Introduction}
\label{sec:introduction}
\fi

\begin{tikzpicture}[remember picture,overlay]
  \node [draw=black, fill=white, text width=\textwidth, inner sep=2pt, yshift=-1.cm] at (current page text area.south){\footnotesize{Accepted paper. \copyright 2023 IEEE. Personal use of this material is permitted. Permission from IEEE must be obtained for all other uses, in any current or future media, including reprinting/republishing this material for advertising or promotional purposes, creating new collective works, for resale or redistribution to servers or lists, or reuse of any copyrighted component of this work in other works.}};
\end{tikzpicture}

Emotions are typically studied as emotional expressions that others subjectively perceive and respond to \cite{van2009emotions, Schuller2018-xi}. A standard theoretical backdrop for analyzing emotions is the two-dimensional pleasure and arousal framework \cite{russell1980circumplex}, which describes emotional expressions along two continuous, bipolar, and orthogonal dimensions: pleasure-displeasure (\emph{valence}) and activation-deactivation (\emph{arousal}). One way emotions become expressed in social interactions, and therefore accessible for social signal processing (SSP), concerns speech signals. Speech emotion recognition (SER) research spans roughly two decades \cite{Schuller2018-xi}, with ever improving state-of-the-art techniques. As a consequence, research on SER has shown increasing prominence in highly-critical and socially relevant domains, such as health, security, and employee well-being \cite{Schuller2018-xi, dukes2021rise, sridhar2021generative}.

A crucial challenge when studying emotional expressions in terms of arousal and valence is that their annotations are per se \emph{subjective} because different people perceive others' emotional expressions differently \cite{Schuller2018-xi, sridhar2021generative}. To address this, these annotations are typically collected by multiple annotators, and consensus on ground-truth is reached using techniques such as average scores \cite{busso2008iemocap}, majority voting \cite{busso2008iemocap}, or evaluator-weighted mean (EWE) \cite{grimm2005evaluation}. These techniques in principle can lead to loss of valuable information on the inherently subjective nature of emotional expressions, and also tend to mask less prominent emotional traits \cite{sridhar2021generative}. In the context of \textit{reliability} in real-world applications, SER systems not only need to model ground-truth labels but also account for the subjectivity inherent in these labels \cite{Schuller2018-xi, gunes2013categorical}. Moreover, by also capturing subjectivity, SER systems can be efficiently deployed in human-in-the-loop solutions, and aid in the development of algorithms for active learning, co-training, and curriculum learning \cite{sridhar2021generative}.



In this work, we tackle the problem of recognising emotional expressions using speech signals, in terms of \textit{time}- and \textit{value}-continuous arousal and valence. To this, we adopt an \emph{end-to-end} learning framework. Common SER approaches rely on hand-crafted features to model emotion labels \cite{han2020exploring, sridhar2020modeling}. Recently, end-to-end architectures have been shown to deliver state-of-the-art emotion predictions \cite{Tzirakis2018-speech, huang2018end, tzirakis2021-mm}, by \emph{learning} features rather than relying on hand-crafted features. For modeling \textit{subjectivity} in emotions, scholars have suggested that end-to-end learning also promotes learning subjectivity dependent representations \cite{alisamir2021evolution}.




Uncertainty in machine learning (ML) is generally investigated in terms of two broader categories. First, \textit{model uncertainty}, or epistemic uncertainty, accounts for the uncertainty in model parameters, and the resulting uncertainty \textit{can} be reduced given enough data-samples \cite{kendall2017uncertainties, zheng2021uncertainty, maniTACuncert}. Second, \textit{label uncertainty}, or aleatoric uncertainty, captures noise inherent in the data-samples, such as  sensor noise or label noise \cite{kendall2017uncertainties, zheng2021uncertainty}. Label uncertainty \textit{cannot} be reduced even if more data-samples are collected. Label uncertainty has been further categorized into \textit{homoscedastic} uncertainty, which remains constant across data-samples, and \textit{heteroscedastic} uncertainty, whose uncertainty depends on the respective data-sample. This work specifically aims to model the heteroscedastic label uncertainty, henceforth simply mentioned as \textit{label uncertainty}, that corresponds to the \textit{inherent subjectivity in emotion annotations}.

We propose to use \textit{Bayes by Backpropagation} (BBB), a Bayesian neural network (BNN) technique, in order to capture label uncertainty. In ML, stochastic and probabilistic models have mainly been used for uncertainty modeling, through ensemble learning\cite{liu2019accurate}, encoder-decoder architectures\cite{kohl2018probabilistic}, neural processes\cite{t21_interspeech, garnelo2018conditional}, and BNNs\cite{gal2016dropout, blundell2015weight, fang2022integrating}. Among these, the Bayesian frameworks show improved performance over non-Bayesian baselines in previous works \cite{kendall2017uncertainties, gao2017deepldl}, making BNNs such as Monte-Carlo dropout\cite{gal2016dropout} and BBB\cite{blundell2015weight} promising candidates for modeling label uncertainty in SER. BBB learns a distribution over weights to produce \emph{stochastic outputs}, which makes it capable of being trained on a distribution of annotations.

With BBB capable of being trained on a distribution of annotations, we capture label uncertainty using the \textit{label distribution learning} (LDL) technique \cite{gao2017deepldl}, leveraging Kullback-Leibler (KL) divergence-based loss functions. Subjective annotations of emotion create a label distribution to represent the subjectivity in emotions \cite{sridhar2021generative}. For simplicity, histograms \cite{sridhar2021generative, chou2022Typed}, and Gaussians \cite{foteinopoulou2021estimating} have been employed to represent the label distributions. However, Gaussians and histograms with \textit{limited} and \textit{sparse} observations are sensitive to outliers thereby losing their robustness in this scenario\cite{bishop2006pattern, kotz2004multivariate, villa2018objective}. Note here that publicly available SER datasets commonly comprise only limited annotations (e.g., 3 to 6) \cite{MspConv, MspPod, recolaDB, kossaifi2019sewa, raj2020defining}, and there is consensus in the literature that gaining more annotations is expensive and resource inefficient \cite{gatica-perez_automatic_2009, sridhar2021generative}. At the same time, a significant degree of subjectivity in annotations is also well noted \cite{alisamir2021evolution, sridhar2020modeling}, thereby leading to sparse annotations with outliers. To tackle this, in this work, we model emotion annotation distributions as a Student's $t$-distribution, or simply {$t$-distribution}. Kotz et al.\cite{kotz2004multivariate}, and, Bishop\cite{bishop2006pattern}, note that in scenarios of limited and sparse observations with outliers, the $t$-distribution becomes more robust over a Gaussian, by producing robust mean and standard deviation estimates of the distribution.

We derive a KL divergence loss for label uncertainty that quantifies distribution similarity between stochastic emotion predictions, modeled as a Gaussian distribution, and \emph{ground-truth emotion annotations}, modeled as a $t$\emph{-distribution}. Subsequently, we present analyses to reveal the benefits of using $t$-distribution over a Gaussian. We validate the proposed model in two in-the-wild datasets, AVEC'16 \cite{avec16} and MSP-Conversation \cite{MspConv}. We show that the proposed model can aptly capture label uncertainty with state-of-the-art results for both datasets, along with a robust loss curve. To emphasize the benefits of the $t$-distribution, we present experiments studying the impact of the number of emotion annotations available. Finally, we perform an ablation study to understand specific benefits of the respective modules in the architecture.






This work is based on two prior conference contributions \cite{Prabhu2021EndToEndLU, prabhu2022label}, which to the best of our knowledge are the first in the literature to use BBB and LDL in SER. These works were also the first to tackle the problem of limited emotion annotations from an ML perspective. Previously, we only validated the method in one dataset, and with limited experiments\cite{Prabhu2021EndToEndLU, prabhu2022label}. In this extension, we additionally validate the method in a larger and more complex dataset, the MSP-Conversation\cite{MspConv}, along with cross-corpora evaluations. This extension is also the first in the literature to present SER results in this novel dataset\cite{MspConv}. Existing analyses and experiments from \cite{Prabhu2021EndToEndLU, prabhu2022label} were also extended to MSP-Conversation. Moreover, we performed additional experiments that include an experiment to understand the impact of the number of annotations available, and an ablation study. Code for the proposed model and loss function is available online \footnote{\url{https://github.com/sp-uhh/label-uncertainty-ser}}.

\section{Background and Related Work}

\subsection{Ground-truth labels}

To handle subjectivity in emotional expressions, annotations $\{y_{1}, y_{2}, .., y_{a}\}$ for emotions are typically collected from multiple annotators ($a$) \cite{recolaDB, raj2020defining}. The \emph{ground-truth label} is then obtained as the mean $m$ across all annotations from $a$ annotators \cite{abdelwahab2019active},
\begin{equation} \label{eq:mean-annot}
    {m} = \dfrac{1}{a} \sum_{i=1}^a y_{i}.
\end{equation}
Alternatively, the EWE, which weights annotations with inter-annotator correlations, has been proposed as the \emph{gold-standard} $\widetilde{m}$ \cite{grimm2005evaluation}. Both $m$ and $\widetilde{m}$ based approximation of ground-truth leads to loss of information on subjectivity \cite{sridhar2021generative}.

Given a raw audio sequence of $T$ frames $\mathcal{X} = [x_1, x_2, ..., x_T]$, traditional SER approaches aim to estimate either the $m_t$ or $\widetilde{m}_t$ for each time frame $t \in [1,T]$, referred to as $\widehat{m}_t$. The concordance correlation coefficient (CCC) \cite{lawrence1989concordance} has been widely used as a loss function for this task \cite{Schuller2018-xi}. For Pearson correlation $r$, the CCC between $m$ and $\widehat{m}$, for $T$ frames is:
\begin{equation}\label{loss:CCC}
    \mathcal{L}_{\text{CCC}}(m) = {\frac {2r \sigma_{m}\sigma_{\widehat{m}}}{\sigma _{m}^{2}+\sigma_{\widehat{m}}^{2}+(\mu _{m}-\mu _{\widehat{m}})^{2}}},
\end{equation}
where $\mu_{m} = \dfrac{1}{T} \sum_{t=1}^{T}m_t$, $\sigma_{m}^{2} = \dfrac{1}{T} \sum_{t=1}^T (m_t - \mu_m)^2$, and $\mu _{\widehat{m}}$, $\sigma_{\widehat{m}}^{2}$ are obtained similarly for $\widehat{m}$. The CCC metric measures the agreement between two variables, in our case the ground-truth $m$ and its estimate $\widehat{m}$. It ranges from $-$1 to $+$1, with perfect agreement at $+$1. In contrast to Pearson's correlation $r$, CCC takes both the linear correlation and the bias in to account, which makes it preferable over Pearson's correlation as the loss and evaluation metric in SER.

\subsection{Label uncertainty in SER}

As an alternative to exclusively modeling $m_t$ or $\widetilde{m}_t$, previous research has attempted to model ground truth that also explains inter-annotator disagreement, for example by means of soft labels \cite{sridhar2021generative} and entropy of disagreement \cite{steidl2005all}. Sridhar et al.\cite{sridhar2021generative} proposed an auto-encoder technique that jointly models soft- and hard-labels of emotion annotations and subsequently estimates label uncertainty as the entropy on soft-labels. Fayek et al.\cite{fayek2016modeling} and Tarantino et al.\cite{Tarantino2019SelfAttentionFS} proposed to learn soft labels instead of ${m}_t$ with improved performance. Steidl et al.\cite{steidl2005all} quantified label uncertainty using the entropy measure and trained a model to minimize the difference in entropy between model outputs and annotator disagreement. 


Label uncertainty has also been approached as a prediction task by estimating the moments of a distribution \cite{han2017hard, han2020exploring}. Han et al. \cite{han2017hard, han2020exploring} used a multi-task learning (MTL) framework to model the unbiased standard deviation $s$ of $a$ annotators as an auxiliary task,
\begin{equation}\label{eq:pu}
    s = \sqrt{\dfrac{1}{a - 1} \sum_{i=1}^a (y_i - m)^2}.
\end{equation}
Similarly, Dang et al.\cite{dang2018dynamic} captured the temporal dependencies in the annotation signals, using multi-rater Gaussian mixture regression and Kalman filters. Sridhar et al. \cite{sridhar2020modeling} introduced a Monte-Carlo (MC) dropout model to obtain uncertainty estimates from the distribution of stochastic outputs. However, their model was not explicitly trained on any label uncertainty estimate and hence could only capture the model uncertainty, but not the label uncertainty. A similar MC dropout was used by Rizos et al. \cite{schullerBBBUncertainty}, who proposed a meta-learning framework that uses uncertainty estimates to potentially detect highly-uncertain samples and perform soft data selection for the training process.


Research efforts have also been made to estimate emotion annotations as a \textit{distribution}, using LDL \cite{foteinopoulou2021estimating, Prabhu2021EndToEndLU, chou2022Typed, prabhu2022label}. Foteinopoulou et al. \cite{foteinopoulou2021estimating} trained an MTL network using a KL divergence loss that models emotion annotations as a \textit{uni-variate Gaussian} with mean $m$ and unknown variance. Chou et al. \cite{chou2022Typed} used LDL to convert subjective annotations into \textit{histogram}-based distributional labels for training. In our preliminary work \cite{Prabhu2021EndToEndLU}, we modeled emotion annotations as a \textit{Gaussian} using BBB-based uncertainty modeling. Notwithstanding the improved performance of these approaches, a drawback concerns the limited annotations on which previous \emph{histogram} or a \emph{Gaussian} assumptions were based \cite{foteinopoulou2021estimating, Prabhu2021EndToEndLU, chou2022Typed}. These assumptions are susceptible to unreliable $m$ and $s$ for lower values of $a$ and sparsely distributed annotations \cite{bishop2006pattern, kotz2004multivariate}. In our subsequent work \cite{prabhu2022label} and in this extension, we tackle this problem by modeling emotion annotation distribution as a $t$-distribution and show advantages over a Gaussian assumption.

\subsection{On distributions}

A Gaussian distribution $\mathcal{Y}\sim {\mathcal {N}}(\mu ,\sigma ^{2})$ is a continuous probability distribution for a real-valued random variable $y$, with the general form of its probability density function \cite{bishop2006pattern}:
\begin{equation}\label{pdf:gaussian}
p(y\mid \mu, \sigma)={\frac {1}{\sigma {\sqrt {2\pi }}}}e^{-{\frac {1}{2}}\left({\frac {y-\mu }{\sigma }}\right)^{2}}.
\end{equation}
The parameters $\mu$ and $\sigma$ are the mean and standard deviation of the distribution, respectively. Due to its simplicity, Gaussians are often used to model random variables whose distributional family are unknown \cite{blundell2015weight, Prabhu2021EndToEndLU}. However, Gaussians are sensitive to outliers, especially in cases of \textit{limited} and \textit{sparse} observations of the random variable\cite{bishop2006pattern}. In this case, the $t$-distribution is noted to become more robust and realistic over a Gaussian \cite{kotz2004multivariate, bishop2006pattern}.

Student's $t$-distribution, $\mathcal{Y}_t \sim \mathcal{N}(\nu, \mu, \sigma)$, arises when estimating the moments of a normally distributed population in \emph{situations where the sample size is small} \cite{kotz2004multivariate, walpole2017probability}, with the probability density function given by \cite{villa2014objective, villa2018objective},
\begin{equation}\label{pdf:tstudent}
    p(y\mid \nu ,{ {\mu }},{ {\sigma }})=\frac{1}{\mathrm{B} (\frac{1}{2}, \frac{\nu}{2})} \frac{1} {\sqrt{\nu {\sigma }^2}} \left(1+\frac{(y- {\mu })^2}{\nu {\sigma }^2}\right)^{-{\frac {\nu +1}{2}}},
\end{equation}
where $\nu$ denotes the degrees of freedom and B(., .) is the Beta function, for Gamma function $\Gamma$, formulated as,
\begin{equation}
    B(i, j) = \dfrac{\Gamma(i)\,\Gamma(j)}{\Gamma(i+j)}.
\end{equation}
The density function \eqref{pdf:tstudent} is symmetric, and its overall shape resembles the bell shape of a normally distributed variable, except that it has heavier tails, meaning that it better captures values that fall far from its mean (i.e., outliers) \cite{bishop2006pattern, kotz2004multivariate}. The degree of freedom $\nu$, also known as the normality parameter, controls the normality of the distribution and is correlated with the standard deviation of the distribution ${\sigma}$ \cite{kotz2004multivariate, bishop2006pattern}. In \eqref{pdf:tstudent}, the standard deviation $\sigma$ takes the scaled form, where $\sigma$ is scaled using the normality parameter $\nu$: 
\begin{equation}\label{eq:scale-sigma}
     { {\sigma } \, \sqrt{\frac {\nu }{\nu -2}} {\text{  for }}}\nu >2,  
\end{equation}
As $\nu$ increases, the $t$-distribution approaches the normal distribution \cite{villa2018objective}. The normality parameter $\nu$, in our case, allows the $t$-distribution to also account for the number of annotations available. 

The robustness of the $t$-distribution, in cases of \textit{limited} and \textit{sparse} observations of the random variable, is associated with its ability to better capture the outliers by also accounting for the number of observations of the random variable\cite{bishop2006pattern}.  This is the key motivation behind using the $t$-distribution to model the emotion annotations, to produce robust mean and standard deviation estimates by also accounting for the number of annotations available.


\section{Proposed Label Uncertainty Model}

\begin{figure}[t!]
    \centering
    \includegraphics[width=0.48\textwidth]{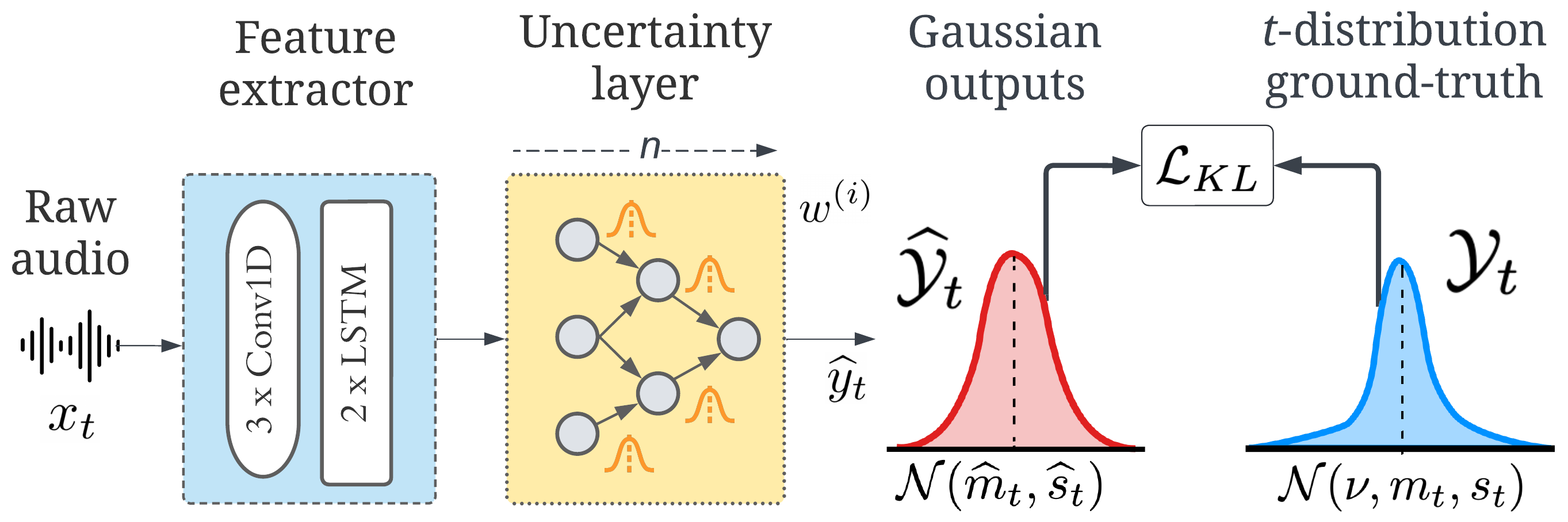}    \caption{Overview of proposed architecture and loss $\mathcal{L}_{\text{KL}}$. $n$: number of forward passes. $w^{(i)}$ and $\widehat{y}_t$: stochastically sampled weight and realization of $\widehat{\mathcal{Y}}_t$, at $i^{th}$ forward pass.}
    \label{Fig:speechEmoBnn}
\end{figure}


In order to better represent subjectivity in emotional expressions, we estimate the \emph{emotion annotation distribution} $\mathcal{Y}_t$ for each time-frame $t$, given raw audio $x_t$. While the true distributional family of subjectively perceived emotions $\mathcal{Y}_t$ is unknown, for simplicity, we can assume that it follows a Gaussian distribution:
\begin{equation}\label{eq:gauss-assumption}
\mathcal{Y}_t \sim \mathcal{N}(m_t, {s_t}^{2}).
\end{equation}
However, with only a limited number of annotations, and in cases where the annotations are sparsely distributed with outliers, we argue that a Gaussian assumption is rather crude \cite{kotz2004multivariate, bishop2006pattern}. Instead, we propose to model the emotion annotations as a $t$-distribution, with degrees of freedom $\nu$:
\begin{equation}\label{eq:tdist-assumption}
\mathcal{Y}_t \sim \mathcal{N}(\nu, m_t, {s_t}^{2}).
\end{equation}
Thus, the goal is to obtain an estimate $\widehat{\mathcal{Y}}_t$ of $\mathcal{Y}_t$ and infer both $\widehat{m}_t$ and $\widehat{s}_t$ from realizations of $\widehat{\mathcal{Y}}_t$.

\subsection{End-to-end DNN architecture}
We propose an end-to-end architecture that uses a feature extractor to learn temporal-paralinguistic features from $x_t$, and an uncertainty layer to estimate $\mathcal{Y}_t$ (see Fig. \ref{Fig:speechEmoBnn}). The feature extractor, inspired by \cite{Tzirakis2018-speech}, consists of three Conv1D layers followed by two stacked long short term memory (LSTM) layers. The uncertainty layer is devised using the BBB technique \cite{blundell2015weight}, comprising three BBB-based MLP.


\subsection{Model uncertainty loss}
Unlike a standard neuron which optimizes a deterministic weight $w$, the BBB-based neuron learns a probability distribution on the weight $w$ by calculating the variational posterior $P(w|\mathcal{D})$ given the training data $\mathcal{D}$\cite{blundell2015weight}. Intuitively, this regularizes $w$ to also capture the inherent uncertainty in $\mathcal{D}$. In contrast to learning a deterministic weight $w$ to exclusively estimate $m_t$, the BBB neuron learns a Gaussian weight distribution $\mathcal{N}(\mu_w, \sigma_w)$, thereby allowing the model to not only estimate $m_t$ but also $s_t$. Estimation of $s_t$ is achieved by calculating the standard deviation of the stochastic estimates obtained from stochastically sampled weights $w^{(i)}$. 
To obtain a non-negative estimate of the standard deviation of the weight distribution $\sigma_w$, we re-parameterize the standard deviation as $\sigma_w = \log(1 + \exp(\rho_w))$ based on an initial estimate $\rho_w$. This way $\theta = (\mu_w, \rho_w)$ can be optimized using simple backpropagation and still ensure a non-negative $\sigma_w$.

For an optimized $\theta$, the predictive distribution $\widehat{\mathcal{Y}}_t$ for $x_t$, is given by $P (\widehat{y}_t|x_t) = \mathbb{E}_{P(w|D)}[P(\widehat{y}_t|x_t, w)]$, where $\widehat{y}_t$ are realizations of $\widehat{\mathcal{Y}}_t$. Unfortunately, the expectation under the posterior of weights is intractable. To tackle this, \cite{blundell2015weight} proposed to learn $\theta$ of a weight distribution $q(w|\theta)$, the variational posterior, that minimizes the Kullback-Leibler (KL) divergence with the true Bayesian posterior, resulting in the negative evidence lower bound (ELBO),
\begin{equation} \label{loss:BBB}
    f(w,\theta)_{\text{BBB}} = \text{KL} \big[q(w|\theta) \| P(w)\big] - \mathbb{E}_{q(w|\theta)} \big[\log P(D|w)\big]. 
\end{equation}

In BBB, stochastic outputs are achieved using multiple forward passes $n$ with stochastically sampled weights $w$, thereby modeling $\widehat{\mathcal{Y}}_t$ using the $n$ stochastic estimates. To account for the stochastic outputs, \eqref{loss:BBB} is approximated as,
\begin{equation}\label{loss:BBB-stoch}
    \mathcal{L}_{\text{BBB}} \approx \sum_{i=1}^{\textit{$n$}} \log q(w^{(i)}|\theta) - \log P(w^{(i)}) - \log P(D|w^{(i)}).
\end{equation}
where $w^{(i)}$ denotes the $i^{th}$ weight drawn from $q(w|\theta)$. The BBB window-size $b$ controls how often new weights are sampled for time-continuous SER. The degree of uncertainty is assumed to be constant within this time period. During testing, the uncertainty estimate $\widehat{s}_t$ is the standard deviation of $\widehat{\mathcal{Y}}_t$, and, $\widehat{m}_t$ is the realization $\widehat{y}_t$ obtained using the mean of the optimized weights $\mu_w$. Obtaining $\widehat{m}_t$ using $\mu_w$ helps overcome the randomization effect of sampling from $q(w|\theta)$, which showed better performances in our case.

Note that variables $n$, $a$, and $\nu$ are closely related to one another. The three variables all denote the number of samples used to model distribution, either $\widehat{\mathcal{Y}}_t$ or ${\mathcal{Y}}_t$. Variable $n$ represents the number of forward passes, thereby the number of stochastic estimates used to model the \textit{estimate} distribution $\widehat{\mathcal{Y}}_t$. Variable $a$ represents the number of annotations used to model the ground-truth distribution ${\mathcal{Y}}_t$. In the probability density function of a $t$-distribution \eqref{pdf:tstudent}, $\nu$ denotes the degree of freedom. In this work, the $\nu$ of a $t$-distribution is set to $a$ enabling the \textit{ground-truth} distribution ${\mathcal{Y}}_t$ to also account for the number of annotations available.

\subsection{Label uncertainty loss}
While \eqref{loss:BBB-stoch} exclusively captures \emph{model uncertainty}, the aim of this work is to also capture \emph{label uncertainty}. For this, using LDL, inspired by \cite{zheng2021uncertainty}, we introduce a \textit{KL divergence-based loss} to fit our model to the annotation distribution $\mathcal{Y}_t$, with either a Gaussian assumption (in Sec.~\ref{sec:gaussKL-deriv}) or a $t$-distribution assumption (in Sec.~\ref{sec:kl-loss-derivation}). 

\subsubsection{Gaussian $\mathcal{Y}_t$ KL divergence}\label{sec:gaussKL-deriv}

For a Gaussian assumption on $\mathcal{Y}_t$ \eqref{eq:gauss-assumption}, the label uncertainty loss, the KL divergence between two Gaussians $\mathcal{Y}_t \sim \mathcal{N}({m_t},{s_t}^{2})$ and $\widehat{\mathcal{Y}}_t \sim \mathcal{N}(\widehat{m_t},\widehat{s_t}^{2})$ is formulated as \cite{bishop2006pattern},
\begin{equation}\label{loss:gauss-KL}
    \mathcal{L}_{KL} (\mathcal{Y}_t\|\widehat{\mathcal{Y}}_t) = \log\left(\frac{\widehat{{s_t}}}{{{s_t}}}\right) + \,\, \frac{{{s_t}}^2 \,+\, ({m_t}-\widehat{m_t})^2}{2\widehat{{s_t}}^2} - \frac {1}{2}.
\end{equation}
The KL divergence is asymmetric, making the order of distributions crucial. In \eqref{loss:gauss-KL}, we choose $\widehat{\mathcal{Y}}_t$ to follow ${\mathcal{Y}}_t$, for a mean-seeking approximation, rather than a mode-seeking one, to capture the full distribution \cite[p.~76]{Goodfellow-et-al-2016}. See Supplementary Sec.~3 for further details on the choice between mean- and mode-seeking approximation using $\mathcal{L}_\textrm{KL}$.


\subsubsection{$t$-distribution $\mathcal{Y}_t$ KL divergence}\label{sec:kl-loss-derivation}

For $\mathcal{Y}_t$ as a $t$-distribution \eqref{eq:tdist-assumption}, we derive the KL divergence between $\mathcal{Y}_t \sim \mathcal{N}(\nu, m_t, {s_t}^{2})$ and the Gaussian outputs $\widehat{\mathcal{Y}}_t \sim \mathcal{N}(\widehat{m_t}, \widehat{{s_t}}^{2})$. Assuming a Gaussian on $\widehat{\mathcal{Y}}$ is fair, as the number of stochastic outputs to model $\widehat{\mathcal{Y}}$ can be controlled using $n$ in \eqref{loss:BBB-stoch}. In this work, we intend to fix $n\geq30$, as a $t$-distribution converges to a stable Gaussian with $30$ samples \cite{villa2014objective, villa2018objective}. As a positive side effect, we result in deriving the KL divergence between a Gaussian and a $t$-distribution, in contrast to between two $t$-distributions, with the latter involving mathematical complexities in calculating intractable expectations for a loss function.

\begin{figure*}[t!]
     \captionsetup[subfigure]{justification=centering}
     \centering
     \begin{subfigure}[b]{0.245\textwidth}
            \includegraphics[width=\textwidth]{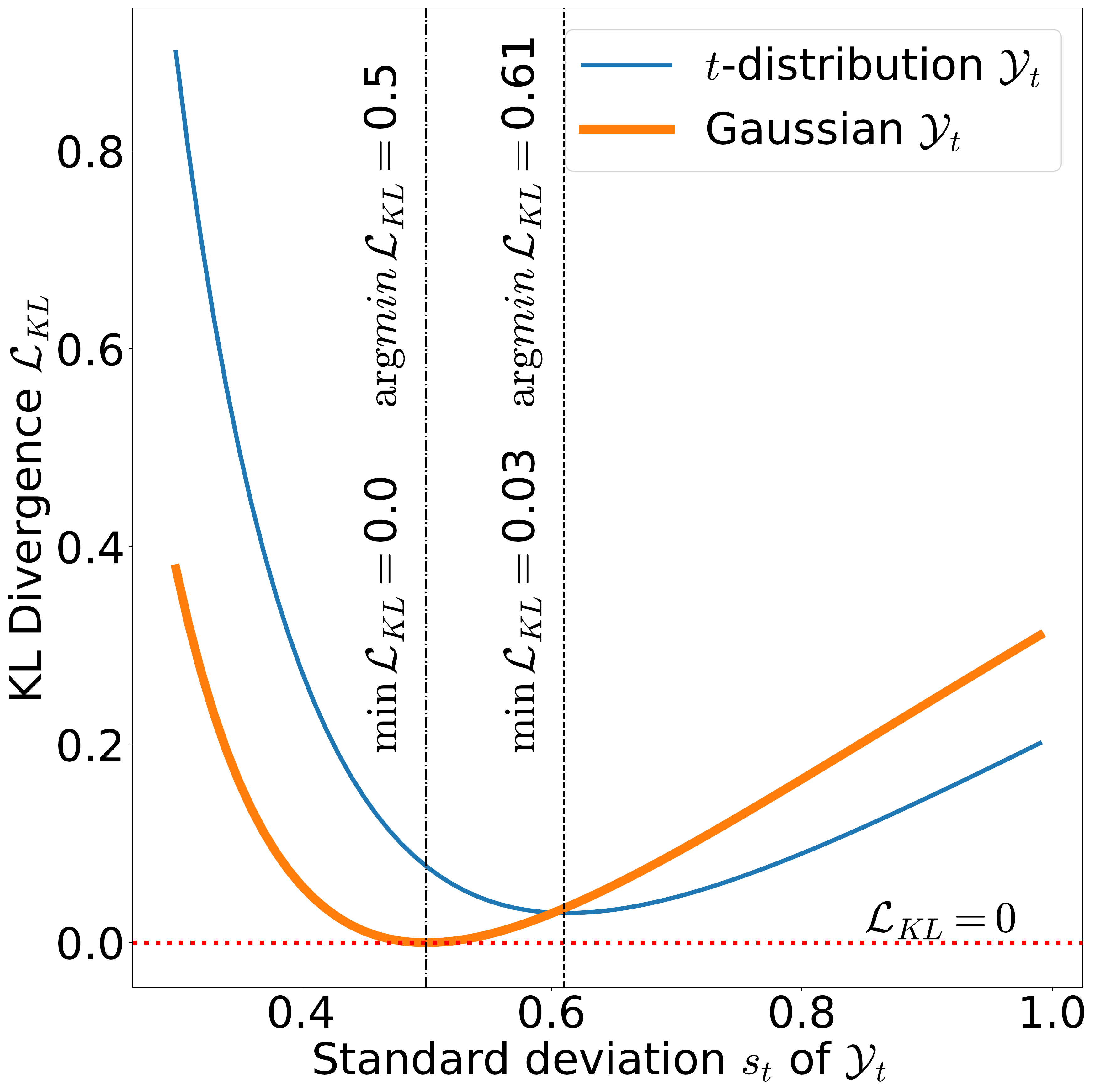}%
         \caption{$\widehat{\mathcal{Y}}_t \sim \mathcal{N}(0, 0.5)$, and $\nu=6$}
      \label{fig:kl-analysis-sigma5_nu6}%
     \end{subfigure}
     \hfill
     \begin{subfigure}[b]{0.245\textwidth}
            \includegraphics[width=\textwidth]{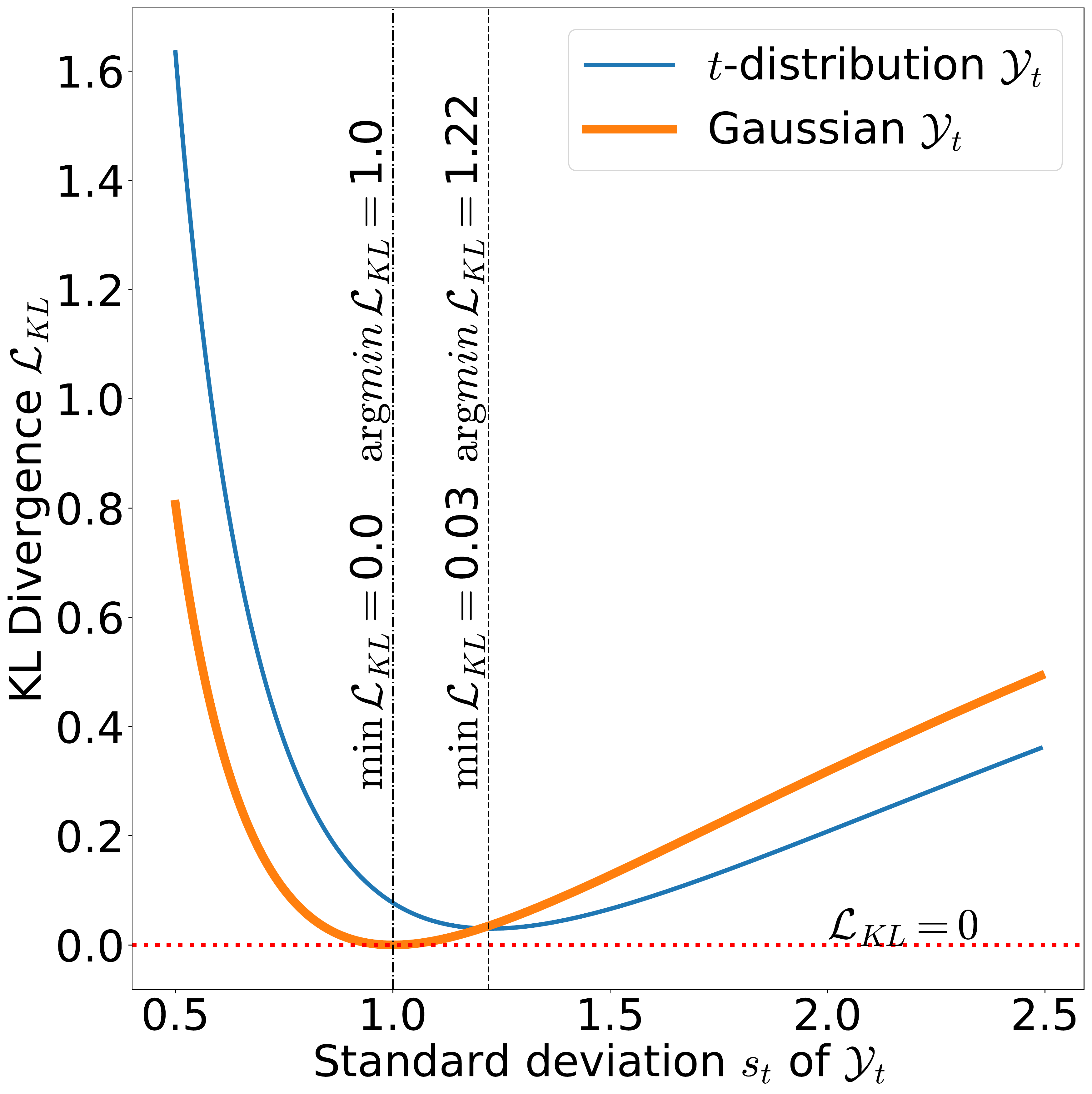}%
         \caption{$\widehat{\mathcal{Y}}_t \sim \mathcal{N}(0, 1)$, and $\nu=6$}
      \label{fig:kl-analysis-sigma1_nu6}%
     \end{subfigure}
     \hfill
     \begin{subfigure}[b]{0.245\textwidth}
            \includegraphics[width=\textwidth]{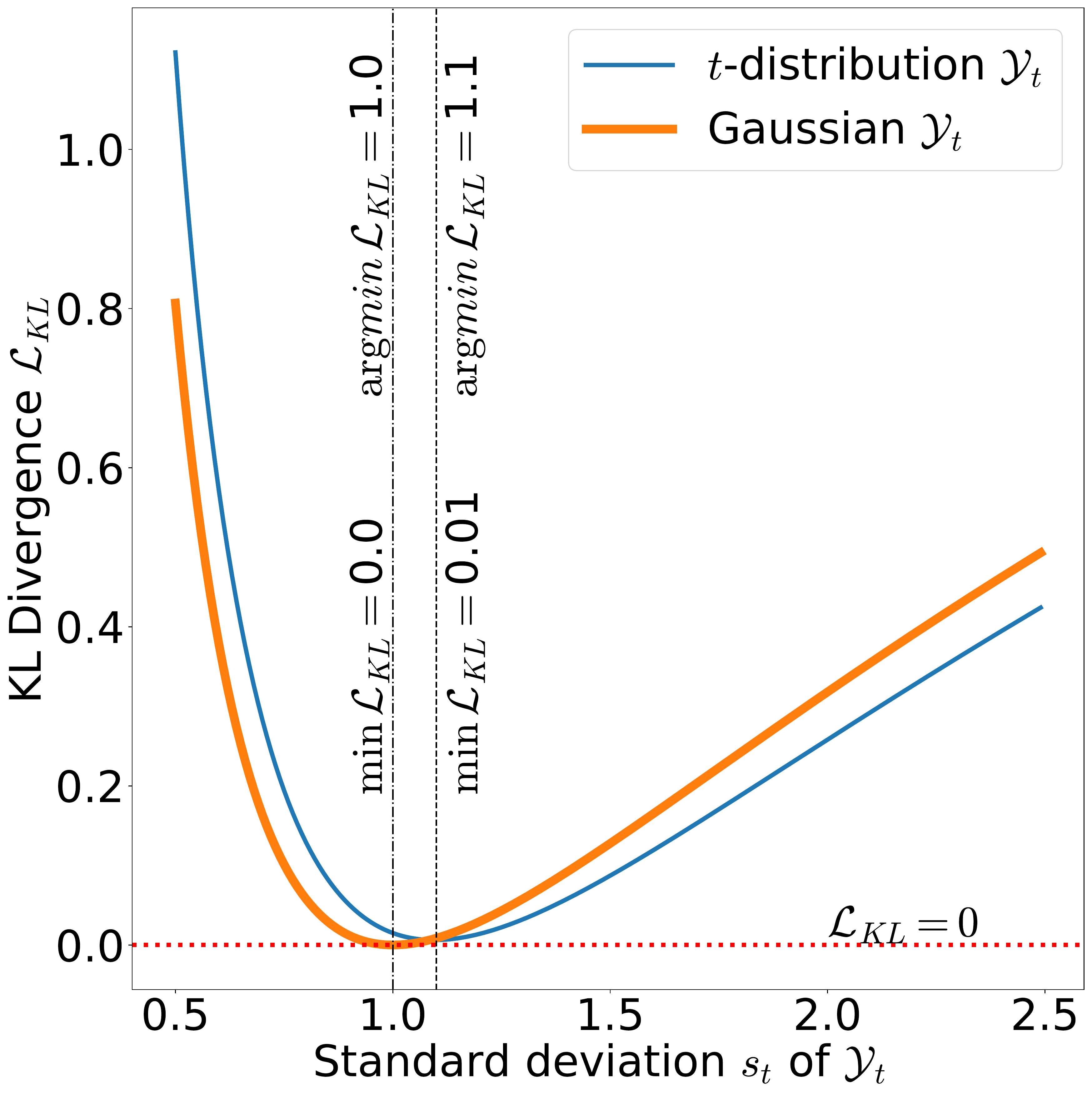}%
         \caption{$\widehat{\mathcal{Y}}_t \sim \mathcal{N}(0, 1)$, and $\nu=12$}
      \label{fig:kl-analysis-sigma1_nu12}%
     \end{subfigure}
     \hfill
     \begin{subfigure}[b]{0.245\textwidth}
            \includegraphics[width=\textwidth]{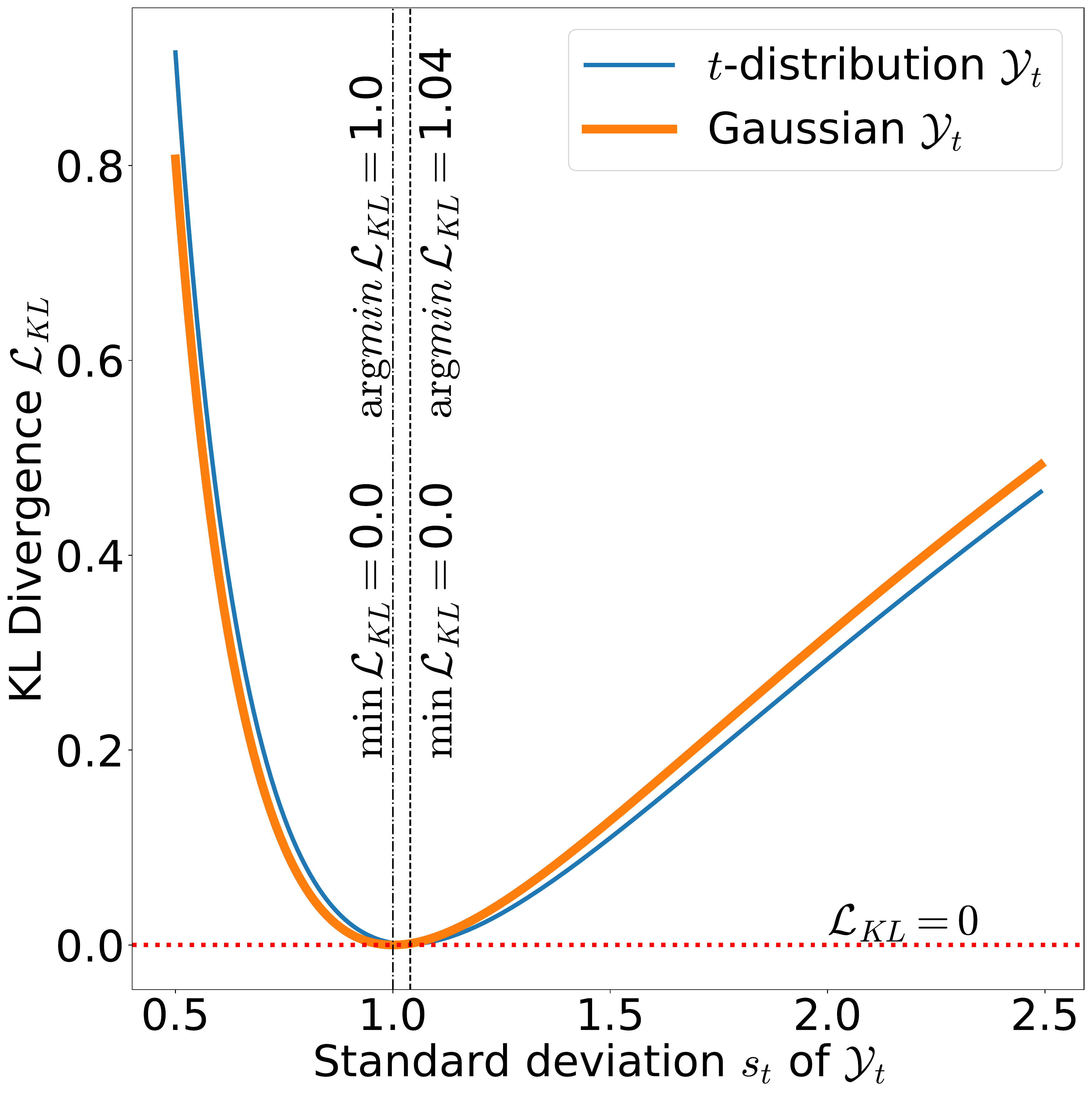}%
         \caption{$\widehat{\mathcal{Y}}_t \sim \mathcal{N}(0, 1)$, and $\nu=30$}
         \label{fig:kl-analysis-sigma1_nu30}
     \end{subfigure}
        \caption{Analysis of the $t$-distribution based KL divergence $\mathcal{L}_{KL}$ \eqref{loss:tdist-KL}, in comparison with Gaussian $\mathcal{L}_{KL}$  \eqref{loss:gauss-KL}.}
        \label{fig:kl-analysis}
\end{figure*}

For a Gaussian $\widehat{\mathcal{Y}}$ (see \eqref{pdf:gaussian}), and a $t$-distributed $\mathcal{Y}$ (see \eqref{pdf:tstudent}), the $\mathcal{L}_{KL}$ is formulated as \cite{kullback1951information, murphy2012machine},
\begin{equation}\label{loss:KL}
    \mathcal{L}_{KL} ({\mathcal{Y}}_t\| \widehat{\mathcal{Y}}_t) = H(\mathcal{Y}_t, \widehat{\mathcal{Y}}_t)- H({\mathcal{Y}}_t),
\end{equation}
where $\text{H}(\cdot, \cdot)$ is the cross-entropy between two distributions, and $\text{H}(\cdot)$ is the entropy of a distribution. The cross-entropy term $\text{H}(\cdot, \cdot)$ in \eqref{loss:KL}, using \eqref{pdf:gaussian}, can be further formulated as,
\begin{eqnarray}
    &    \lefteqn{\text{H} ({\mathcal{Y}}_t, \widehat{\mathcal{Y}}_t) = - \int {\mathcal{Y}}_t(y)\,\, \log\,\widehat{\mathcal{Y}}_t(y)\, dy } \notag 
        \\ 
    &    =& \dfrac{1}{2} \log(2\pi\widehat{{s_t}}^2) + \int \mathcal{Y}_t(y)\,\, \Big( \frac{(y-\widehat{m_t})^2}{2\widehat{{s_t}}^2} \Big) dy  \notag 
        \\
    &    =& \dfrac{1}{2} \log(2\pi\widehat{{s_t}}^2) \,  +  \,  \frac{1}{2\widehat{{s_t}}^2} \Big[ \int \mathcal{Y}_t(y) y^2\, dy \notag 
    \\ 
    && \qquad- 2\widehat{m_t} \int \mathcal{Y}_t(y)\, y\, dy  + \, \widehat{m_t}^2 \int \mathcal{Y}_t(y)\, dy \Big] . \label{eq:init-cross-entropy}
\end{eqnarray}

Noting that $\int \mathcal{Y}_t(y) y^2\, dy = {m_t}^2 + {{s_t}}^2$, $\int \mathcal{Y}_t(y)\, y\, dy = {m_t}$, and $\int \mathcal{Y}_t(y)\, dy = 1$, where ${m_t}$ and ${{s_t}}$ are parameters of the $t$-distribution $\mathcal{Y}_t$, $p(y\mid \nu, {m_t}, {{s_t}})$, the equation \eqref{eq:init-cross-entropy} becomes,
\begin{eqnarray}
    &\lefteqn{} =& \dfrac{1}{2} \log(2\pi\widehat{{s_t}}^2) \,\, + \,\, \frac{1}{2\widehat{{s_t}}^2} \Big[ {{s_t}}^2 \,+\, {m_t}^2 \,-\, 2\widehat{m_t}m_t \,+\, \widehat{m_t}^2 \Big] \notag
    \\
    & \lefteqn{} =& \dfrac{1}{2} \log(2\pi\widehat{{s_t}}^2) \,\, + \,\, \frac{{{s_t}}^2 \,+\, ({m_t}-\widehat{m_t})^2}{2\widehat{{s_t}}^2} \label{eq:final-crossentropy}
\end{eqnarray}
        
        

Finally, using \eqref{eq:final-crossentropy} in \eqref{loss:KL}, our proposed KL divergence is
\begin{equation}\label{loss:tdist-KL}
\boxed{
    \mathcal{L}_{KL} = \frac{1}{2} \log(2\pi\widehat{{s_t}}^2) \,\, + \,\, \frac{{{s_t}}^2 \,+\, ({m_t}-\widehat{m_t})^2}{2\widehat{{s_t}}^2}  - H({\mathcal{Y}}_t) .
    }
\end{equation}

We implement \eqref{loss:tdist-KL} as a custom loss function by extending the \texttt{studentT} pytorch sub-package \cite{paszke2019pytorch}. 

\subsubsection{Comparing Gaussian and $t$-distribution loss} \label{Section:kl-analysis}

While the two loss-functions \eqref{loss:gauss-KL} and \eqref{loss:tdist-KL} have their second term in common, two differences can be noted. Firstly, as \eqref{loss:gauss-KL} calculates the divergence between two similar distributions, $\mathcal{Y}_t$ and $\widehat{\mathcal{Y}}_t$, it includes the logarithm of the ratio between the two Gaussian's standard deviation in its formulation. However, in \eqref{loss:tdist-KL}, the deviations of $\mathcal{Y}_t$ and $\widehat{\mathcal{Y}}_t$ are \textit{separately} quantified using terms $ \frac{1}{2} \log(2\pi\widehat{{s_t}}^2)$ and $H({\mathcal{Y}}_t)$, respectively. Secondly, \eqref{loss:tdist-KL} is dependent on the number of annotations available through scaling ${s}_t$ with the normality factor $\nu$ \eqref{eq:scale-sigma}. 



To further understand the advantages of the $t$-distribution $\mathcal{L}_{KL}$ \eqref{loss:tdist-KL} over the Gaussian $\mathcal{L}_{KL}$ \eqref{loss:gauss-KL}, we plot the $\mathcal{L}_{KL}$ values as a function of varying ${s_t}$, for \eqref{loss:tdist-KL} and \eqref{loss:gauss-KL}. We perform this analysis under four different scenarios, for different values of $\widehat{{s_t}}$ and ${\nu}$, i) Figure \ref{fig:kl-analysis-sigma5_nu6} for scenario $\widehat{{s_t}}=0.5$ and $\nu=6$, ii) Figure \ref{fig:kl-analysis-sigma1_nu6} for scenario $\widehat{{s_t}}=1.0$ and $\nu=6$, iii) Figure \ref{fig:kl-analysis-sigma1_nu12} for scenario $\widehat{{s_t}}=1.0$ and $\nu=12$, and, iv) Figure \ref{fig:kl-analysis-sigma1_nu30} for scenario $\widehat{{s_t}}=1.0$ and $\nu=30$.

From Figure \ref{fig:kl-analysis}, firstly, we see that $\mathcal{L}_{KL}$ behaves differently when the ground-truth $\mathcal{Y}_t$ is modeled as a $t$-distribution \eqref{loss:tdist-KL}, in comparison to the Gaussian assumption \eqref{loss:gauss-KL}. Specifically, from Figure \ref{fig:kl-analysis-sigma5_nu6}, for $\widehat{{s_t}}=0.5$ and $\nu=6$, we see that the minimum $\mathcal{L}_{KL}$ \eqref{loss:tdist-KL} is achieved only at ${{s_t}}=0.61$, in contrast to the Gaussian \eqref{loss:gauss-KL} $\widehat{{s_t}}={{s_t}}=0.5$. While the Gaussian attempts exactly fitting the model to the ground-truth ${{s_t}}=0.5$, the $t$-distribution tries to fit on a more relaxed ${{s_t}}=0.61$ by also considering the reduced degree of freedom $\nu=6$. This behaviour is similar to the confidence intervals calculation using a $t$-distribution \cite[Sec.~9.5]{rees2001book}, where relaxation on ${s_t}$ is noted with respect to $\nu$. Moreover, \cite{bishop2006pattern} associate this relaxed ${{s_t}}$ towards the increased robustness of the $t$-distribution to sparse distributions with outliers.

Secondly, we note that the observed relaxation on ${{s_t}}$ is dependent on two factors, 1) the standard deviation of the stochastic outputs $\widehat{{s_t}}$, and 2) the degree of freedom of the ground-truth $\nu$. From figures \ref{fig:kl-analysis-sigma5_nu6} and \ref{fig:kl-analysis-sigma1_nu6}, we see that, while $\nu$ is constant, the relaxation on ${s_t}$ \emph{increases} along with an increase in $\widehat{{s_t}}$. At $\widehat{{s_t}}=0.5$ a relaxation of $0.11$ is made by the $t$-distribution \eqref{loss:tdist-KL} from $0.5$ to $0.61$, while a larger relaxation of $0.22$ is made for $\widehat{{s_t}}=1.0$. Similarly, from figures \ref{fig:kl-analysis-sigma1_nu12} and \ref{fig:kl-analysis-sigma1_nu30}, we see that, while $\widehat{{s_t}}$ is constant, as $\nu$ increases the relaxation on ${s_t}$ \emph{decreases}. That is, the $t$-distribution \eqref{loss:tdist-KL} starts behaving similar to a Gaussian, in line with literature that states that as the degree of freedom $\nu$ of $t$-distribution increases, the distribution converges into a Gaussian \cite{villa2014objective, villa2018objective, kotz2004multivariate}. This is also in line with our initial motivation behind using the $t$-distribution, which we expected to account for the number of annotations available while fitting on annotation distribution $\mathcal{Y}$.

From an ML and SER perspective, from Figure \ref{fig:kl-analysis}, we note several benefits of $t$-distribution based loss term towards label uncertainty modeling. Firstly, training on a $t$-distribution $\mathcal{L}_{KL}$ \eqref{loss:tdist-KL} leads to training on a relaxed $s_t$, and can lead to better capturing of the \textit{whole} ground-truth label distribution. In other words, this can lead to the $t$-distribution better accounting for sparse annotations with outliers, where a relatively high likelihood is associated along the tails of the distribution, as noted by Bishop \cite{bishop2006pattern}. Secondly, we note that in all cases, the $t$-distribution $\mathcal{L}_{KL}$ \eqref{loss:tdist-KL} values are always higher than Gaussian $\mathcal{L}_{KL}$ for lower values of ${s_t}$ and $\widehat{{s_t}}$. This might lead to larger penalization of the model through the $\mathcal{L}_{KL}$ loss, and may thereby promote better and quicker convergence during training, in comparison to the Gaussian $\mathcal{L}_{KL}$ \eqref{loss:gauss-KL}. Finally, the $t$-distribution $\mathcal{L}_{KL}$ \eqref{loss:tdist-KL} can also adapt to different datasets by also accounting for the number of annotations available during training.




\subsection{Training loss}
The proposed end-to-end uncertainty loss is formulated as,
\begin{equation}\label{eq:end-to-end_loss}
   \mathcal{L} = (1 - \mathcal{L}_{\text{CCC}}(m)) + \mathcal{L}_{\text{BBB}} + \alpha \mathcal{L}_{\text{KL}}.
\end{equation}
 Intuitively, $\mathcal{L}_{\text{CCC}}(m)$ optimizes for mean predictions $m$, $\mathcal{L}_{\text{BBB}}$ optimizes for BBB weight distributions, and $\mathcal{L}_{\text{KL}}$ optimizes for the label distribution $\mathcal{Y}_t$.
 For $\alpha=0$, the model only captures model uncertainty ({MU}). For $\alpha=1$, the model also captures \emph{label uncertainty} ({MU$+$LU} or {$t$-LU}).
 $\mathcal{L}_{\text{CCC}}(m)$ is used as part of $\mathcal{L}$ to achieve faster convergence and jointly optimize for mean predictions. 
 Including $\mathcal{L}_{\text{CCC}}(m)$
 might lead to better optimization of the feature extractor \cite{Tzirakis2018-speech, tzirakis2021-semspeech}.
 
 In Equation \eqref{eq:end-to-end_loss}, $\alpha$ is the tuning parameter that decides how much we want to regularize our model to also account for the label uncertainty. While the proposed models only use two values for the $\alpha$ (0 and 1), as an additional study, we also experimented with varying regularization on the label uncertainty loss term $\mathcal{L}_\textrm{KL}$ (see Supplementary Sec.~5).
 


\section{Experimental Setup}

\subsection{Dataset}\label{sec:dataset}
To validate our proposed methodology, we use two publicly available in-the-wild datasets, with time- and value-continuous annotations for arousal and valence. Firstly, the AVEC'16 \cite{avec16} version of the RECOLA dataset \cite{recolaDB}, which has \textit{2.15hrs} of annotated dyadic interactions. Secondly, the MSP-Conversation dataset, which has \textit{15.15hrs} of annotated interactions with groups of 2-7 interlocutors.


\begin{figure}[t!]
     \centering
     \begin{subfigure}[b]{0.24\textwidth}
        \includegraphics[width=\textwidth]{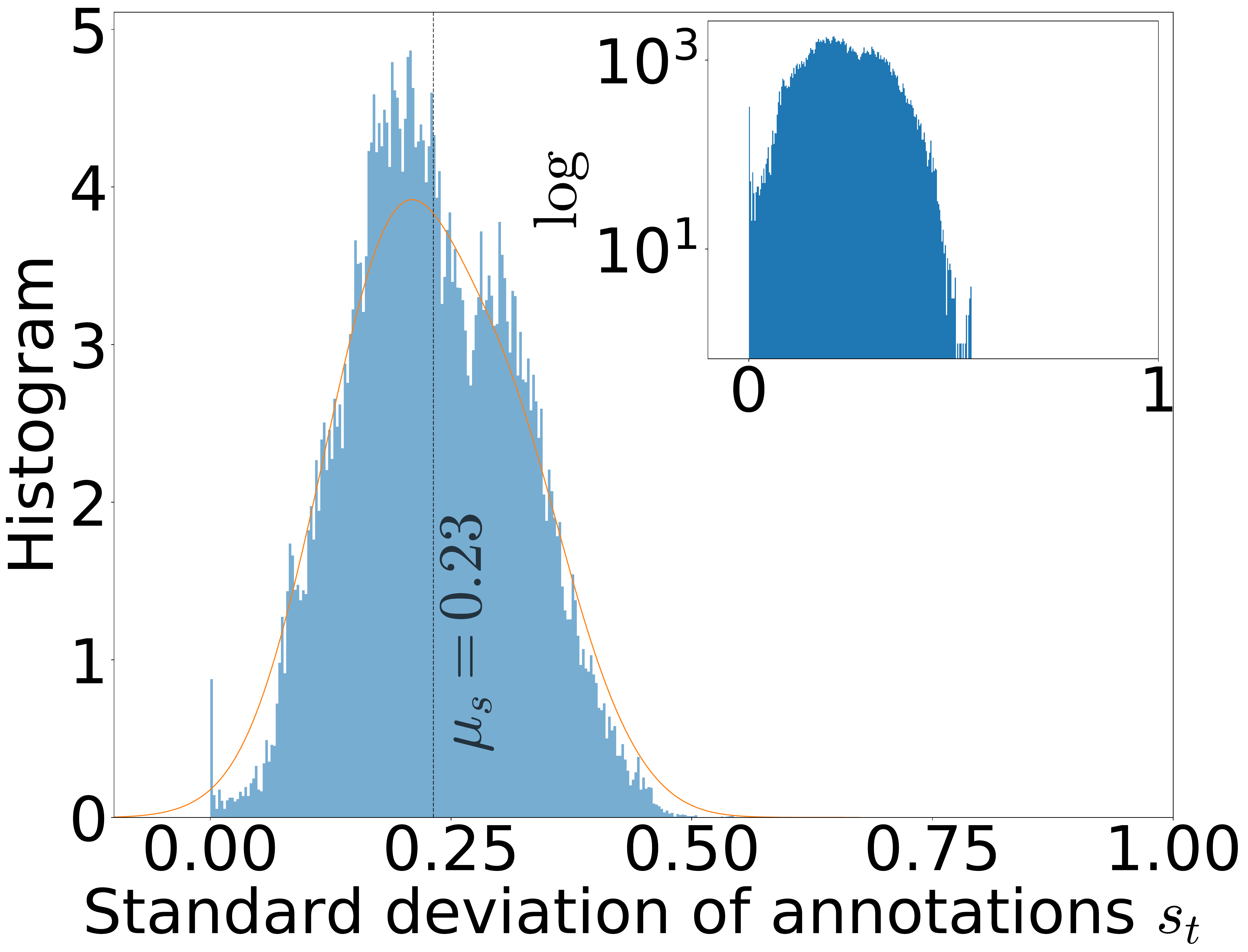}
        \caption{For arousal.}
        \captionsetup{justification=centering}
        \label{Fig:s_dist_arousal_recola}
     \end{subfigure}
     \begin{subfigure}[b]{0.24\textwidth}
        \includegraphics[width=\textwidth]{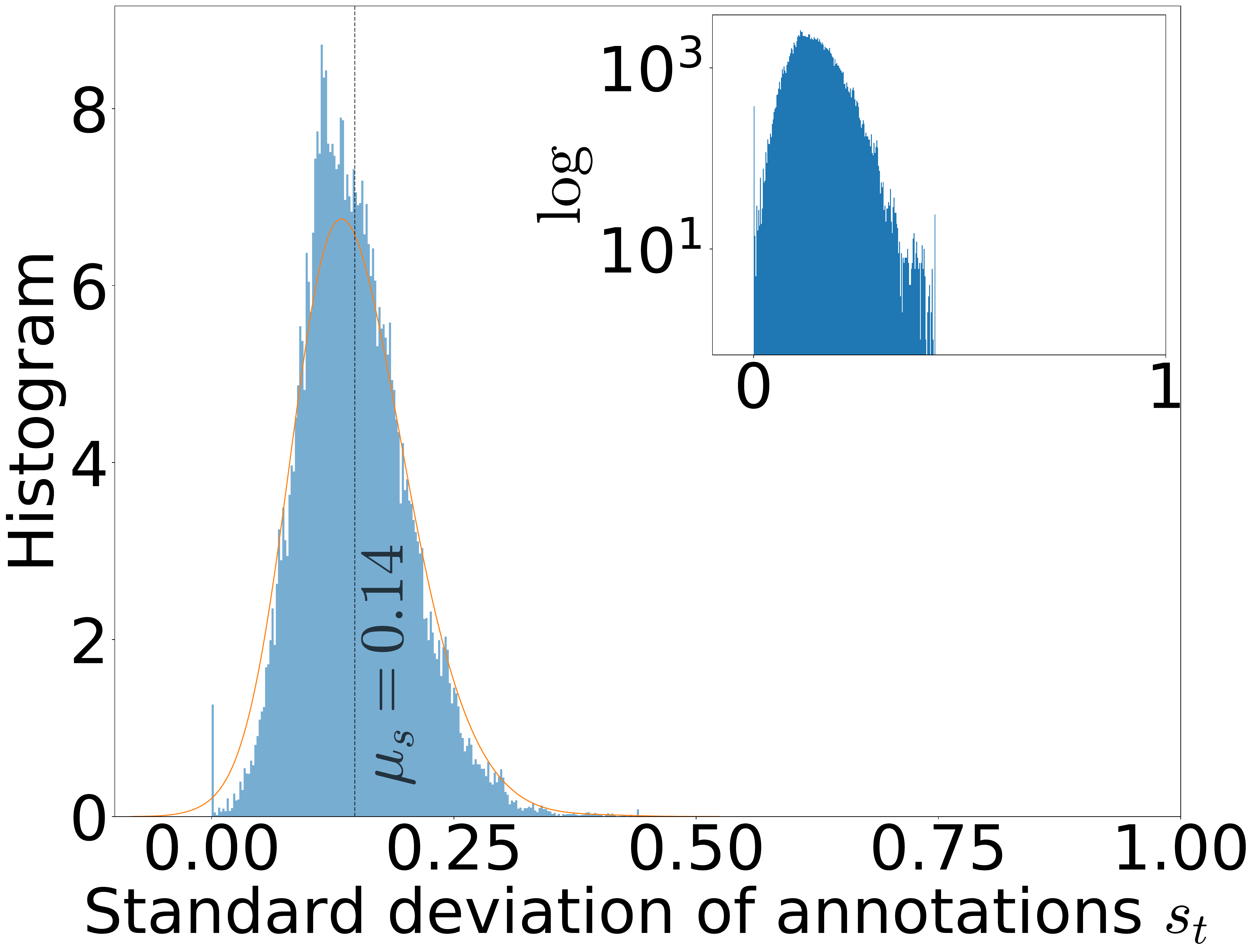}
        \caption{For valence.}
        \captionsetup{justification=centering}
        \label{Fig:s_dist_valence_recola}
     \end{subfigure}
\caption{Histogram of standard deviations $s_t$ in AVEC'16.}
\label{Fig:s_dist_avec}
\end{figure}

\begin{figure}[t!]
     \centering
     \begin{subfigure}[b]{0.24\textwidth}
        \includegraphics[width=\textwidth]{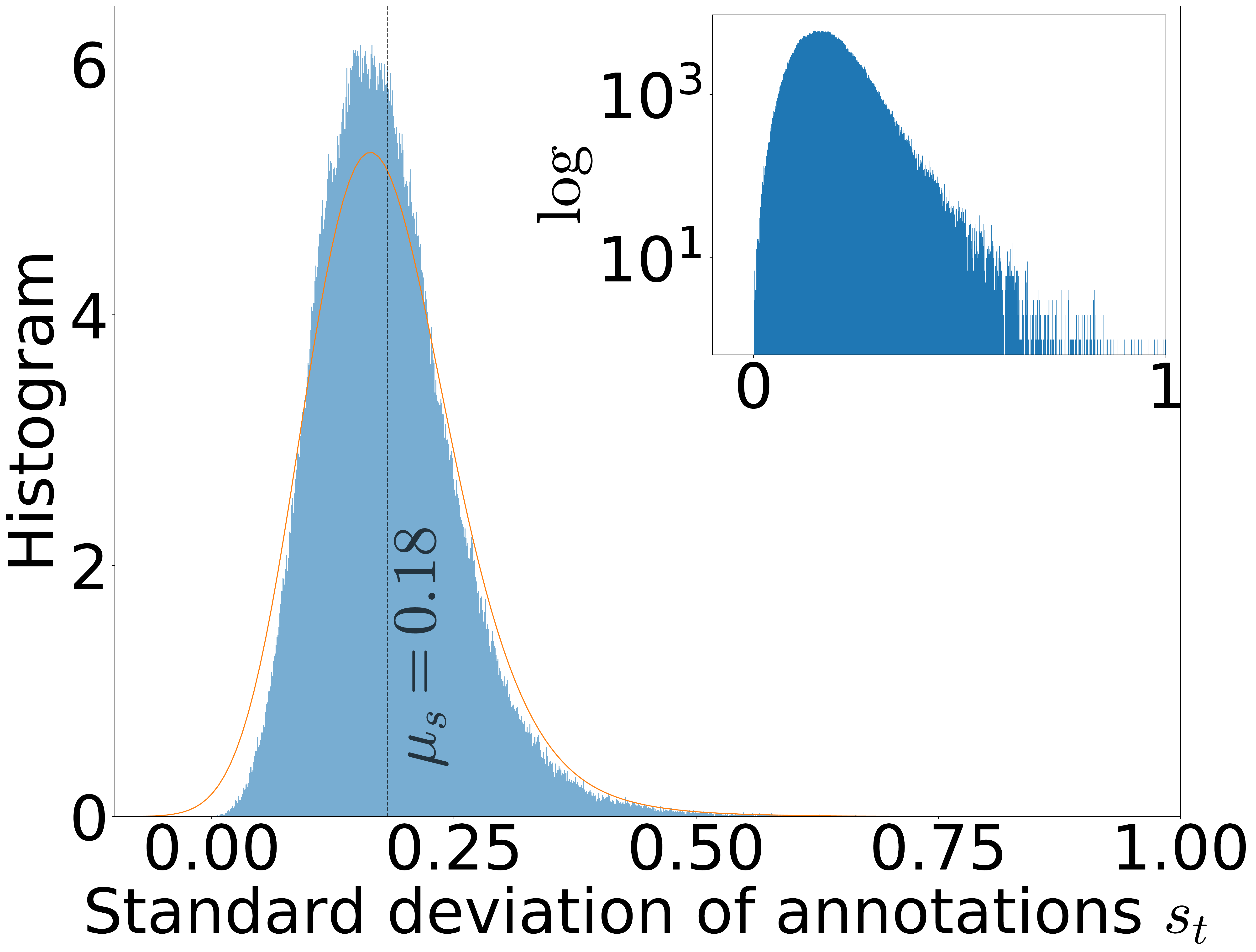}
        \caption{For arousal.}
        \captionsetup{justification=centering}
        \label{Fig:s_dist_arousal_mspconv}
     \end{subfigure}
     \begin{subfigure}[b]{0.24\textwidth}
        \includegraphics[width=\textwidth]{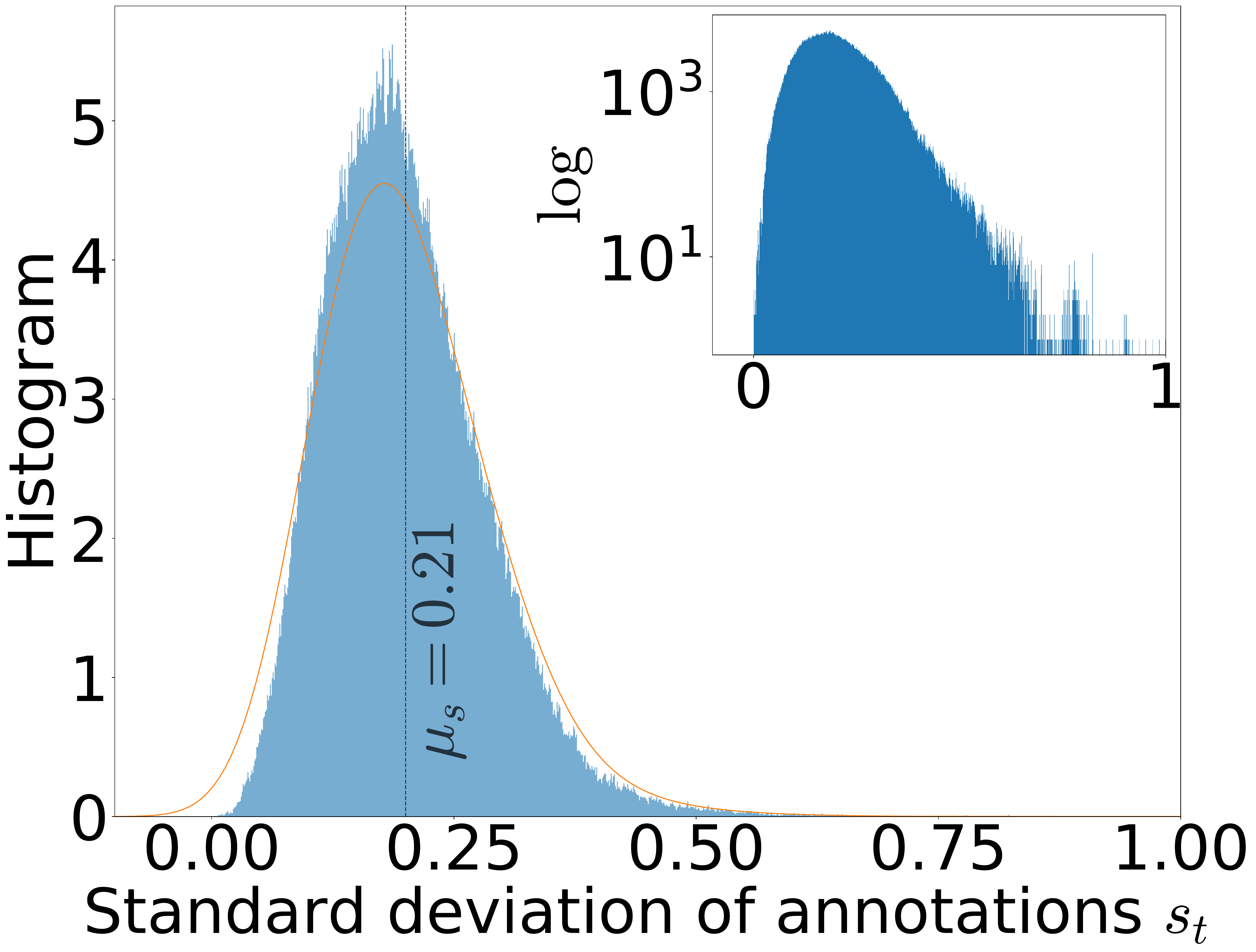}
        \caption{For valence.}
        \captionsetup{justification=centering}
        \label{Fig:s_dist_valence_mspconv}
     \end{subfigure}
\caption{Histogram of standard deviations $s_t$ in MSPConv.}
\label{Fig:s_dist_mspconv}
\end{figure}

\subsubsection{AVEC'16 dataset}



The dataset consists of arousal and valence annotations by $a=6$ annotators at $40$~ms frame-rate, or $25$ frames per second (fps). The arousal and valence annotations in the dataset are distributed on average with $\mu_{{m}} = 0.01$ and $\mu_{{m}} = 0.11$, and $\mu_{s} = 0.23$ and $\mu_{s} = 0.14$, respectively, where $\mu_{s} = \dfrac{1}{T} \sum_{t=1}^{T}s_t$. Further, in Figure \ref{Fig:s_dist_avec} the distribution of $s_t$ is illustrated. It can be noted from Fig. \ref{Fig:s_dist_avec} that $s_t$ distributions are skewed towards high standard deviations $s_t$, thereby indicating the high-level of subjectivity present in the dataset. The high skewness is even more evident in the $\log$-histogram plotted along in Fig. \ref{Fig:s_dist_avec}. The dataset is divided into speaker disjoint partitions for training, development, and testing, with nine $300$~s recordings each. Results with respect to the AVEC'16 are presented only in terms of the development partition, as the annotations for the test partition are not publicly available. Similarly, the hyperparameters are fine-tuned on the train partition for this particular dataset. Note that the \emph{posterior distribution} $P(w|D)$ and the time-shift for post-processing are the only parameters tuned using the partitions. See Supplementary Sec.~1 for the complete list of hyperparameters used.



\begin{table*}
\caption{Comparison on mean $m$, standard deviation $s$, and label distribution estimations $\mathcal{Y}$, in terms of $\mathcal{L}_{\text{ccc}}(m)$, $\mathcal{L}_{\text{ccc}}(s)$, and $\mathcal{L}_\text{KL}$, respectively. Larger CCC indicates improved performance as indicated by $\uparrow$. Lower KL indicates improved performance as indicated by $\downarrow$. ** indicates that the respective approach achieves statistically significant better results than \emph{all} other approaches in comparison. * indicates that it achieves statistically significant better results over \emph{only some} of the approaches in comparison. Results in brackets (.) are for the respective development partition of the dataset.}

\begin{subtable}{\textwidth}
    \centering
    \begin{tabular}{l ccc | ccc}
                \toprule
                & & \textbf{Arousal} & & & \textbf{Valence} & \\
                \addlinespace[0.5em]
                    & $\mathcal{L}_{\text{ccc}}(m) \uparrow$    & $\mathcal{L}_{\text{ccc}}(s) \uparrow$    & $\mathcal{L}_\text{KL} \downarrow$
                    
                    & $\mathcal{L}_{\text{ccc}}(m) \uparrow$    & $\mathcal{L}_{\text{ccc}}(s) \uparrow$    & $\mathcal{L}_\text{KL} \downarrow$
                    
                    \\
                \midrule
                E2E Baseline w/o Temp 
                & {0.581}  & -     & - & 0.129 & -     & -                    \\
                E2E Baseline \cite{Tzirakis2018-speech}     
                & {0.770}  & -     & - & 0.361 & -     & -                    \\
                STL  \cite{han2020exploring}    
                & 0.727        & -   & - & 0.389 & -     & -                 \\ 
                MTL PU  \cite{han2020exploring} 
                & 0.740  & 0.310     & 0.776       & \textbf{0.420**}  & 0.032  & 0.960          \\ 
                MU  \cite{Prabhu2021EndToEndLU}      
                & 0.762        & 0.077    & 0.675  &  0.332 & 0.040   & 0.631    \\
                MU$+$LU  \cite{Prabhu2021EndToEndLU} 
                & 0.751       & {0.361}  & {0.250} &        0.301   & {0.048}   & 0.405  \\
                \textbf{$t$-LU (proposed)} 
                & \textbf{0.782**} & \textbf{0.381**} & \textbf{0.228**} & {0.400*}       & \textbf{0.050*}   & \textbf{0.386**} \\
                \bottomrule
    \end{tabular}
    \caption{Quantitative results on \textbf{AVEC'16} dataset.}
    \label{tab:avec_qaunt_results}
\end{subtable}
\newline
\vspace*{0.2cm}
\newline
\begin{subtable}{\textwidth}
    \centering
    \begin{tabular}{lccc | ccc}
                \toprule
                & & \textbf{Arousal} & & & \textbf{Valence} & \\
                \addlinespace[0.5em]
                    & $\mathcal{L}_{\text{ccc}}(m) \uparrow$    & $\mathcal{L}_{\text{ccc}}(s) \uparrow$    & $\mathcal{L}_\text{KL} \downarrow$
                    
                    & $\mathcal{L}_{\text{ccc}}(m) \uparrow$    & $\mathcal{L}_{\text{ccc}}(s) \uparrow$    & $\mathcal{L}_\text{KL} \downarrow$
                    
                    \\
                \midrule
                E2E Baseline w/o Temp 
                & {0.177} (0.206)  & -     & - & 0.080 (0.115) & -     & -                    \\
                E2E Baseline \cite{Tzirakis2018-speech}    
                & 0.373 (0.407)  & -     & - & 0.192 (0.183) & -     & -                    \\
                STL  \cite{han2020exploring}    
                & 0.292 (0.360)        & -   & - & 0.190 (0.189) & -     & -                 \\ 
                MTL PU  \cite{han2020exploring} 
                & 0.296 (0.363) & 0.107 (0.105)   & 0.527 (0.440) & 0.181 (0.185) & 0.030 (0.030)   & 0.560 (0.450)        \\ 
                MU  \cite{Prabhu2021EndToEndLU}      
                & 0.367 (0.406) & 0.052 (0.067)  & 0.380 (0.410)  & 0.208 (0.220) & 0.022 (0.028)  & 0.451 (0.439)    \\
                MU$+$LU  \cite{Prabhu2021EndToEndLU} 
                & 0.357 (0.397) & 0.111 (0.123)  & 0.370 (0.322) & 0.191 (0.219) & 0.029 (0.032)    & 0.410 (0.396) \\
                \textbf{$t$-LU (proposed)} 
                & \textbf{0.389**} (0.421**) & \textbf{0.118*} (0.134*) & \textbf{0.357**} (0.317**) & \textbf{0.213*} (0.224*) & \textbf{0.032*} (0.035*)    & \textbf{0.373**} (0.382**) \\
                \bottomrule
    \end{tabular}
    \caption{Quantitative results on \textbf{MSPConv} dataset.}
    \label{tab:mspconv_qaunt_results}
\end{subtable}
\label{tab:quant_results}
\end{table*}


\subsubsection{MSP-Conversation dataset}


The MSP-Conversation, or simply \textit{MSPConv}, is approximately $7$ times larger than AVEC'16, comprising of in-the-wild podcasts. The wide range of podcast recordings leads to high variability in terms of population size, group size, and more importantly its emotional content \cite{MspPod, MspConv}, making the MSPConv a more complex dataset to model.


The dataset consists of time- and value-continuous annotations for arousal and valence, performed by at least $a=6$ annotators at $\approx 16$~ms frame-rate, or 60 fps, however not uniform in all cases \cite{MspConv}. For uniform sampling rate, we perform median filtering with a window-size of $500$ms, as suggested in \cite{MspConv}. To keep the sampling rate consistent between the two datasets, for cross-corpora evaluations, we use a step-size of $1/25$s in median filtering. A local normalization, i.e., for each annotated sequence and for each annotator, was performed using zero-mean unit-deviation normalization \cite{recolaDB}, similar to AVEC'16. As illustrated in Figure \ref{Fig:s_dist_mspconv}, in the MSPConv dataset \cite{MspConv}, arousal and valence annotations are distributed on average with $\mu_{{m}} = -0.01$ and $\mu_{{m}} = 0.00$, and $\mu_{s} = 0.18$ and $\mu_{s} = 0.21$, respectively. Further revealing the complexity of MSPConv, when comparing figures \ref{Fig:s_dist_avec} and \ref{Fig:s_dist_mspconv}, we see that the level of subjectivity in MSPConv is higher than the AVEC'16 dataset, where in MSPConv the $s_t$ distribution tail is more skewed towards high subjectivity. The high skewness is more evident in the $\log$-histogram plotted along in Fig. \ref{Fig:s_dist_mspconv}.








In preliminary experiments, we noted that the arousal and valence annotations were prone to periodic distortion noises, especially from particular annotators--- \emph{001}, \emph{007}, and \emph{009}. This could have originated from any technical error or from a human error by the annotator. Directly training on these noisy annotations degraded the performance of all models in comparison. Ignoring the noisy annotations might lead to a loss of information, and might also result in a reduced number of available annotations to derive ground-truth. To reduce these periodic distortions and still retain the inherent annotation information, we use a low-pass filter \cite{butterworth1930theory} with a cut-off frequency of $0.25$Hz. The cut-off frequency was tuned using a Fourier analysis \cite{cooley1965algorithm} followed by a qualitative analysis of the filtered annotations. Filtering was performed only on annotations \textit{with periodic distortions}, i.e., from the three annotators-- \emph{001}, \emph{007}, and \emph{009}.

 


\subsection{Baselines and Proposed model versions}

\textbf{E2E Baselines}: This baseline is a reimplementation of \cite{Tzirakis2018-speech}, with the same end-to-end framework as our proposed model but a multi-layer perceptron instead of the uncertainty layer. The model does not capture any form of uncertainty, and is exclusively trained on the $\mathcal{L}_\text{CCC}(m)$ loss \eqref{loss:CCC}. 

Time-continuous ground-truth annotations contain temporal dependencies \cite{gupta2016modeling}, where an annotation at time $t$ can be expected to have a high correlation with annotations at time $t+1$ and $t-1$. Our proposed architecture accounts for this temporal dependency using two stacked LSTM layers. Moreover, temporal modeling is achieved by batching annotations into sequences of 12s each (300 frames of 40ms each). With this setup, the LSTM operation is performed over the sequence rather than over a single frame, thereby directly learning temporal dependencies. To assess the impact of this temporal modeling, we use an additional baseline: \emph{E2E Baseline w/o Temp} where the \textit{LSTM operation is performed on the feature dimension}, in contrast to the E2E Baseline where the operation is performed on the temporal dimension. This way the number of parameters is kept the same for the two allowing for a fair comparison.



\textbf{MTL Baselines}: From \cite{han2017hard, han2020exploring}, as the baselines, we use the perception uncertainty (\emph{MTL PU}) and single-task models (\emph{STL}). The MTL PU is a label uncertainty model that also models $s_t$ as an auxiliary task. The STL does not capture uncertainty and is exclusively trained on $\mathcal{L}_\text{CCC}(m)$ \eqref{loss:CCC}. For a fair comparison, we reimplemented these baselines. Crucially, the reimplementation also enables us to compare the models in terms of their standard deviation $s$ estimates, which were not presented in Han et al.'s work \cite{han2020exploring}. 




\textbf{Proposed BBB-LDL versions}: We use three versions of the proposed label uncertainty model. Firstly, the \textit{Model Uncertainty} (\emph{MU}) version, which shares the same DNN architecture as the other BBB version but is trained on \eqref{eq:end-to-end_loss} with $\alpha=0$. Secondly, the \textit{Label Uncertainty} (\emph{MU$+$LU}) version also captures the label uncertainty and is trained on \eqref{eq:end-to-end_loss} with $\alpha=1$. The MU$+$LU version however makes a Gaussian assumption on $\mathcal{Y}_t$, thereby $\mathcal{L}_\text{KL}$ follows \eqref{loss:gauss-KL}. Finally, the $t$\textit{-distribution Label Uncertainty} ($t$\emph{-LU}) version, which is trained on the same loss function \eqref{eq:end-to-end_loss} but models $\mathcal{Y}_t$ as a $t$-distribution, and $\mathcal{L}_\text{KL}$ follows \eqref{loss:tdist-KL}.

Finally, for all the models two post-processing techniques are applied, namely, median filtering \cite{Tzirakis2018-speech} and time-shifting \cite{mariooryad2014correcting} (with shifts between 0.04s and 10s). See supplementary Sec.~2 for further detailed information.

\begin{figure*}
     \centering
     \begin{subfigure}[]{0.49\textwidth}
        \includegraphics[width=\textwidth]{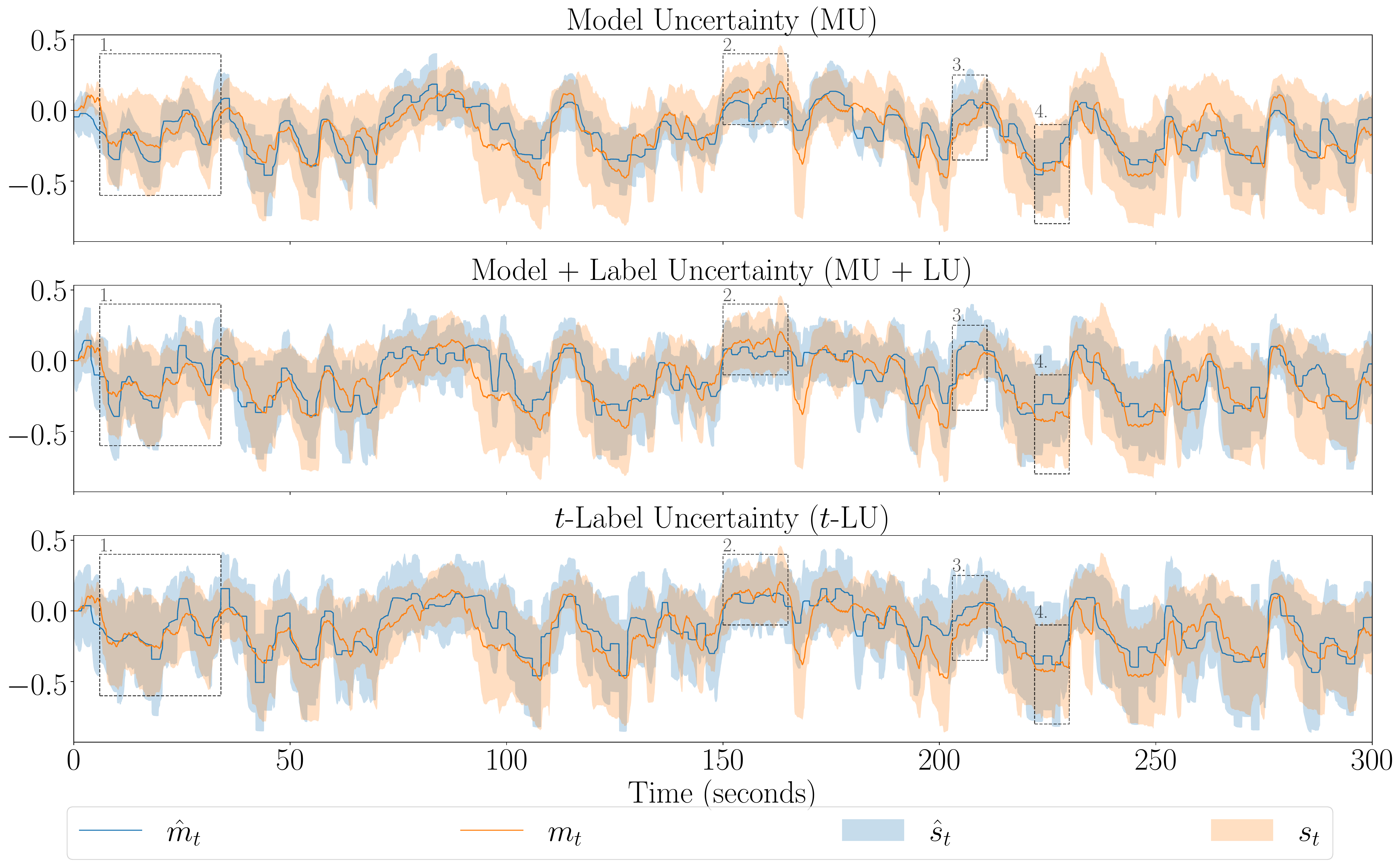}
        \caption{For arousal, in \textbf{AVEC'16} dataset.}
        \captionsetup{justification=centering}
        \label{Fig:results-pred-arousal}
     \end{subfigure}
     \hfill
     \begin{subfigure}[]{0.49\textwidth}
        \includegraphics[width=\textwidth]{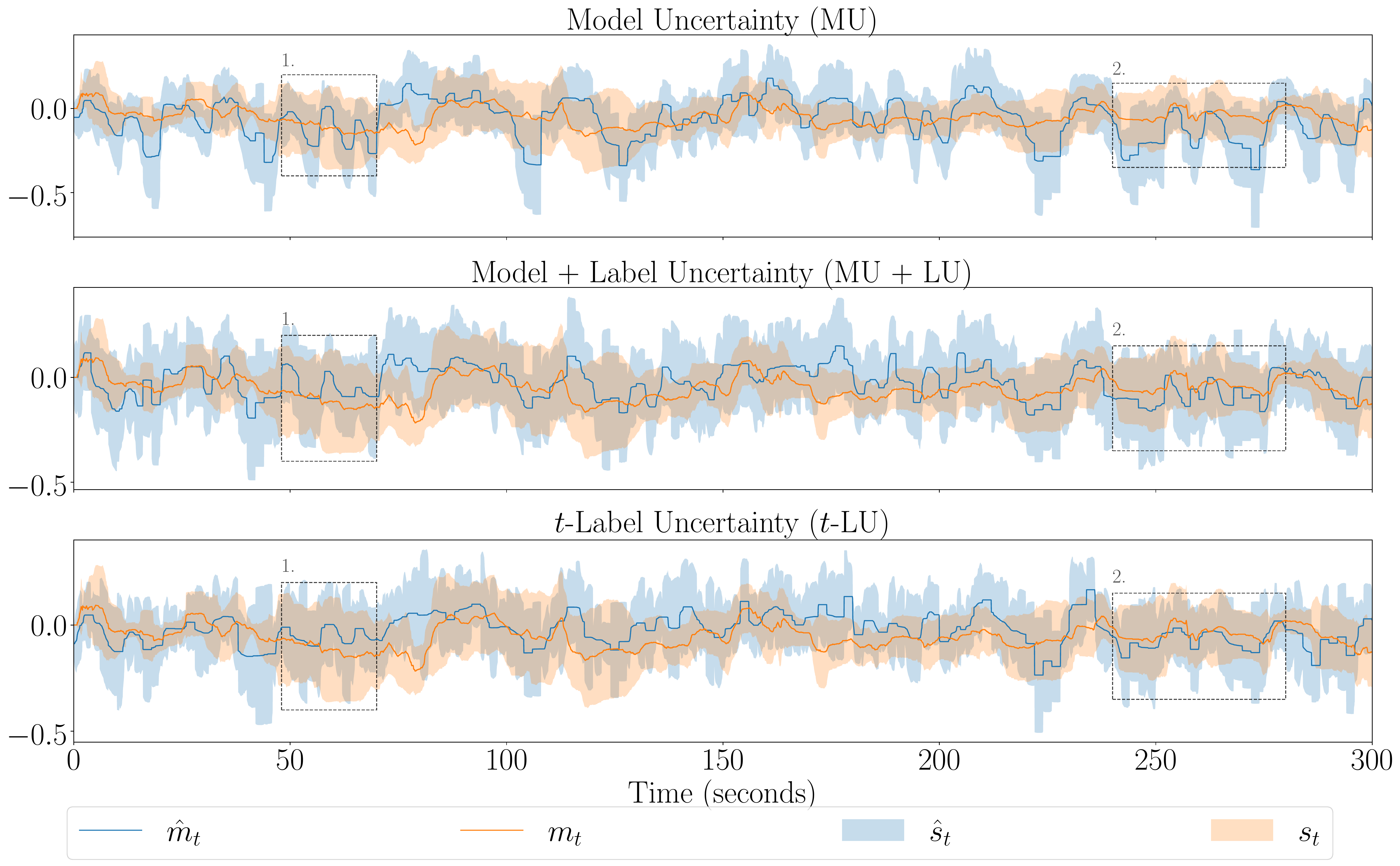}
        \caption{For valence, in \textbf{AVEC'16} dataset.}
        \captionsetup{justification=centering}
        \label{Fig:results-pred-valence}
     \end{subfigure}
 \vfill
 \vspace*{0.2cm}
 \vfill
     \begin{subfigure}[]{0.49\textwidth}
        \includegraphics[width=\textwidth]{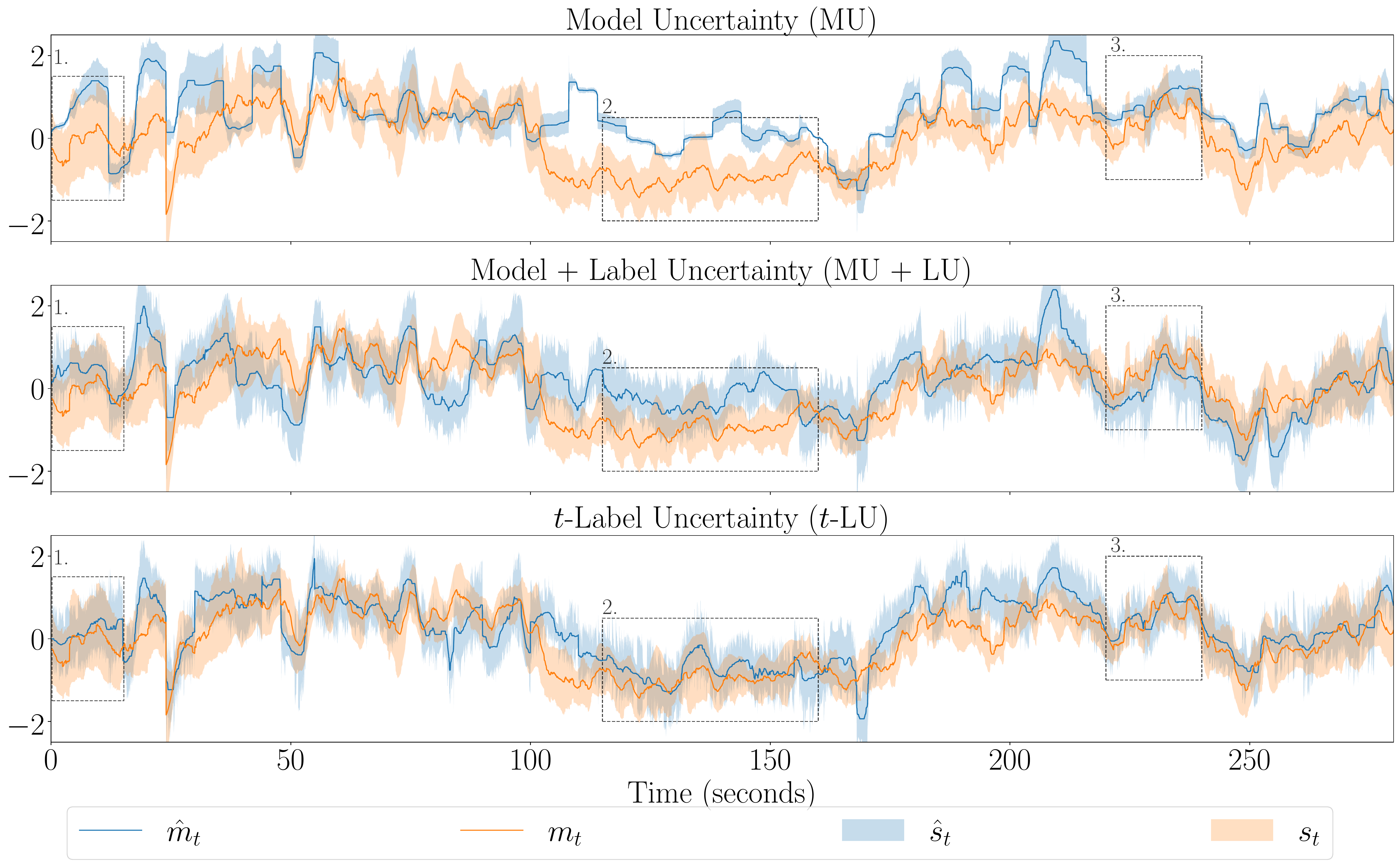}
        \caption{For arousal, in \textbf{MSPConv} dataset.}
        \captionsetup{justification=centering}
        \label{Fig:msp-results-pred-arousal}
     \end{subfigure}
     \hfill
     \begin{subfigure}[]{0.49\textwidth}
        \includegraphics[width=\textwidth]{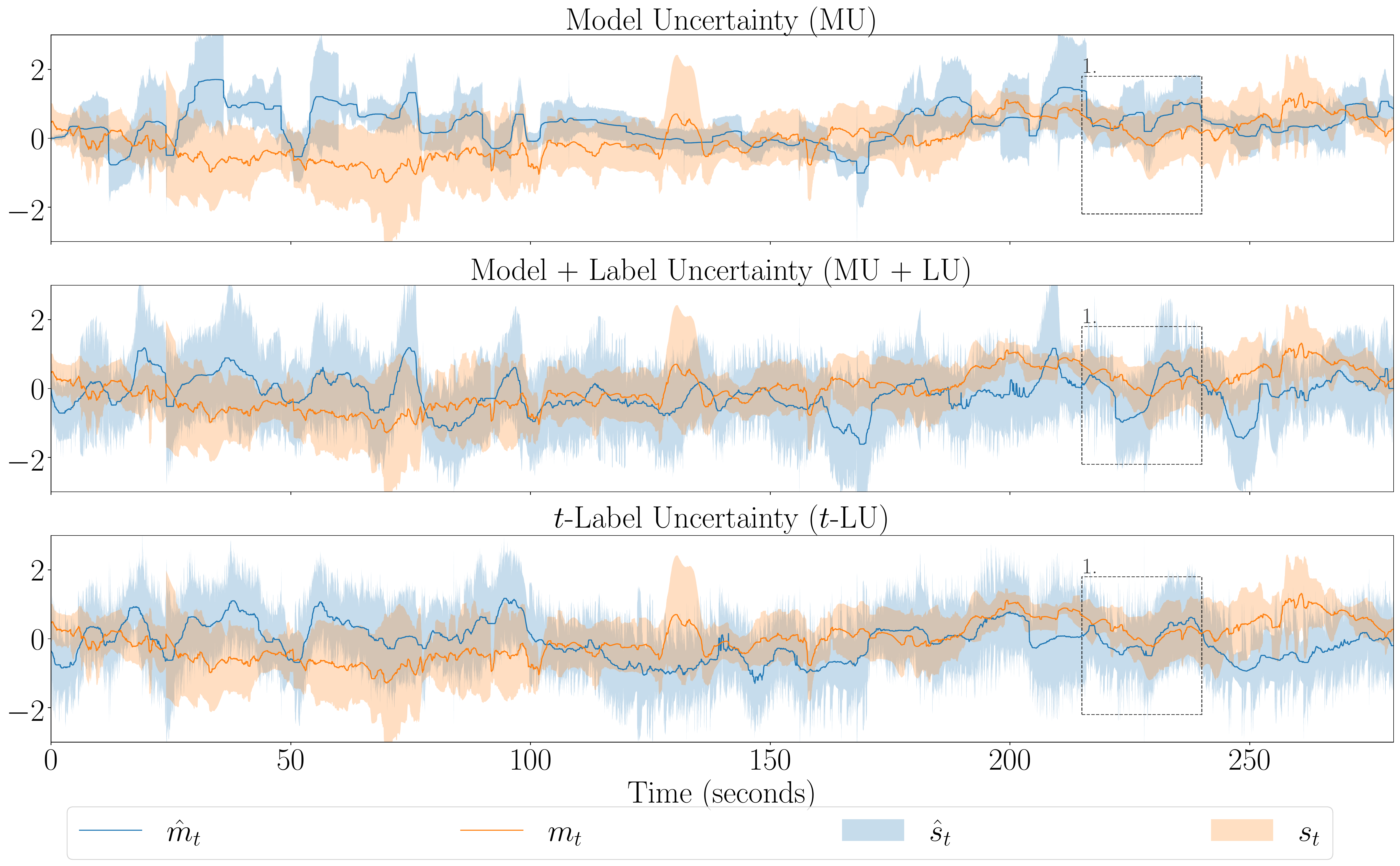}
        \caption{For valence, in \textbf{MSPConv} dataset.}
        \captionsetup{justification=centering}
        \label{Fig:msp-results-pred-valence}
     \end{subfigure}
\caption{Label distribution $\mathcal{Y}_t$ estimation  results for a test subject.}
\label{Fig:qual_results}
\end{figure*}


\subsection{Validation measures}
To validate the proposed method's \emph{mean} and \emph{standard deviation} estimates, we use $\mathcal{L}_\text{CCC}(m)$ and $\mathcal{L}_\text{CCC}(s)$ metrics, respectively, widely used in literature \cite{Tzirakis2018-speech, tzirakis2021-semspeech, han2020exploring}. However, $\mathcal{L}_\text{CCC}(m)$ and $\mathcal{L}_\text{CCC}(s)$ validate mean and standard deviation estimates \emph{separately}. To further \textit{jointly} validate mean and standard deviation estimates, as label distribution $\widehat{\mathcal{Y}}_t$, we use the $\mathcal{L}_\text{KL}$ measure. For a fair comparison, we validate all the models in comparison using $\mathcal{L}_\text{KL}$ based on their respective distribution assumptions on $\mathcal{Y}_t$, as the models are trained in a similar fashion. The proposed {$t$-LU} version is validated and trained on the $t$-distribution $\mathcal{L}_\text{KL}$ \eqref{loss:tdist-KL}, and the baselines on the Gaussian $\mathcal{L}_\text{KL}$ \eqref{loss:gauss-KL}. Nevertheless, from the experiments, we also noted that the proposed $t$-LU performs better in terms of both \eqref{loss:tdist-KL} and \eqref{loss:gauss-KL}. Finally, the statistical significance of results is estimated using a one-tailed $t$-test, asserting significance for \emph{p}-values $\leq0.05$.

\section{Results and Discussion}

\subsection{Quantitative analysis of estimates}\label{sec:quant-results}

Table \ref{tab:quant_results} shows the average performance of the baselines and the proposed models, in terms of their mean $m$, standard deviation $s$, and distribution $\widehat{\mathcal{Y}}_t$ estimations, $\mathcal{L}_{\text{CCC}}(m)$, $\mathcal{L}_{\text{CCC}}(s)$, and $\mathcal{L}_{KL}$, respectively. Results are presented with respect to two datasets, for AVEC'16 in Table \ref{tab:avec_qaunt_results}, and for MSPConv in Table \ref{tab:mspconv_qaunt_results}. From the analysis presented in Section \ref{sec:dataset}, we note that the MSPConv is a more complex dataset, in terms of modeling label uncertainty.




\begin{figure*}
     \centering
     \begin{subfigure}[b]{0.24\textwidth}
        \includegraphics[width=\textwidth]{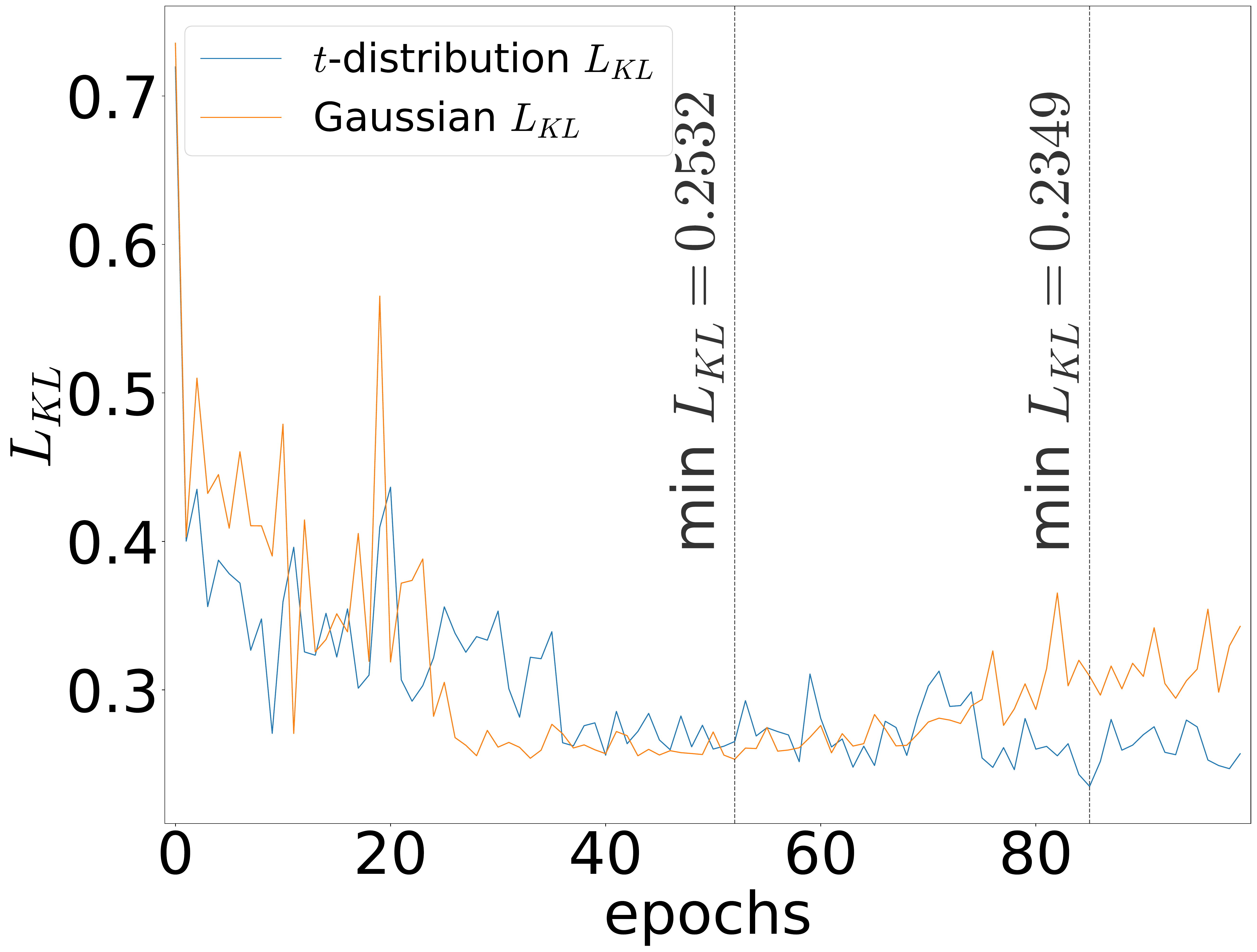}
        \caption{For arousal, in \textbf{AVEC'16}.}
        \captionsetup{justification=centering}
        \label{Fig:recola_loss_curves_arousal}
     \end{subfigure}
     \hfill
     \begin{subfigure}[b]{0.24\textwidth}
        \includegraphics[width=\textwidth]{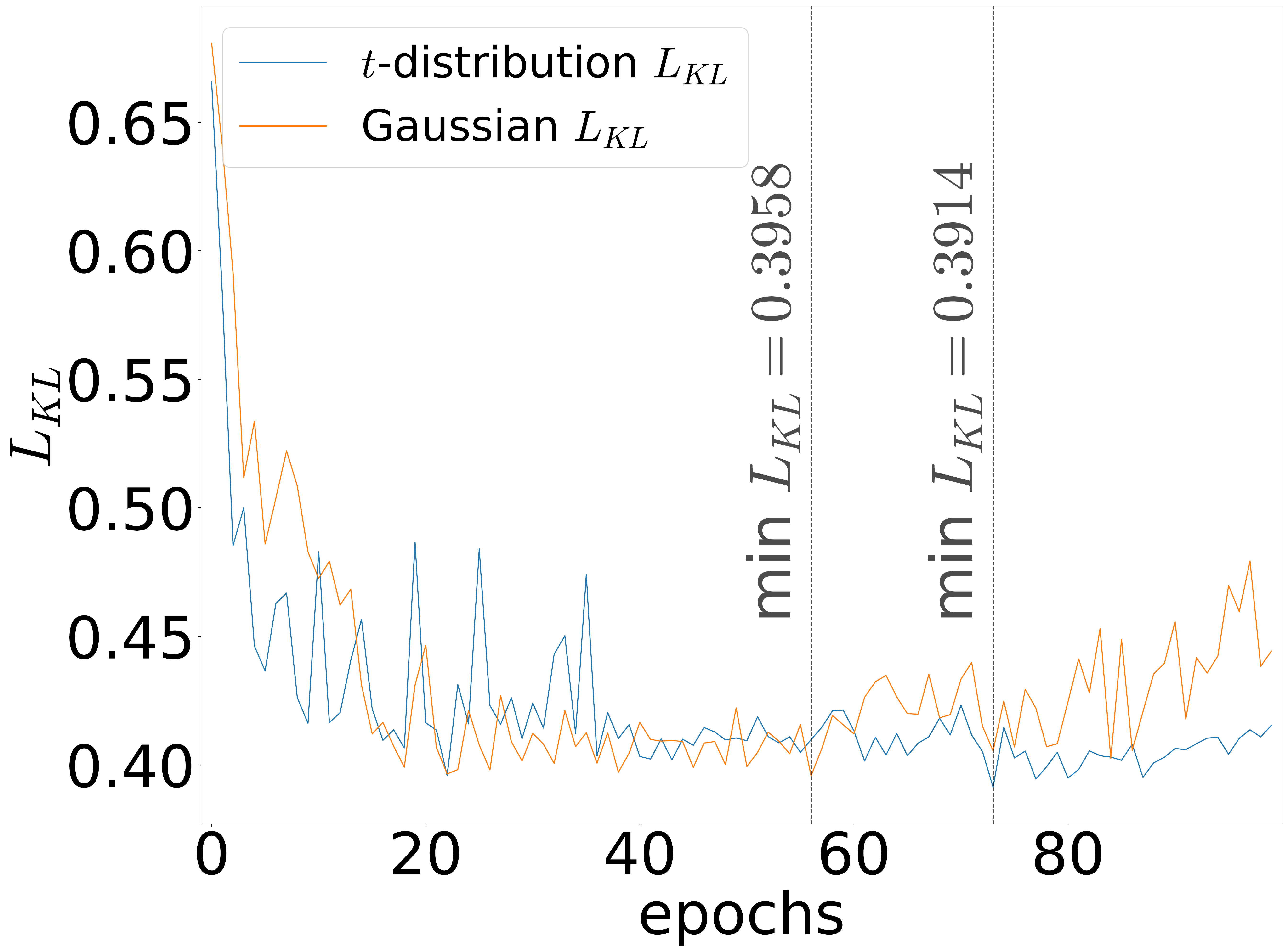}
        \caption{For valence, in \textbf{AVEC'16}.}
        \captionsetup{justification=centering}
        \label{Fig:recola_loss_curves_valence}
     \end{subfigure}
     \hfill
     \begin{subfigure}[b]{0.24\textwidth}
        \includegraphics[width=\textwidth]{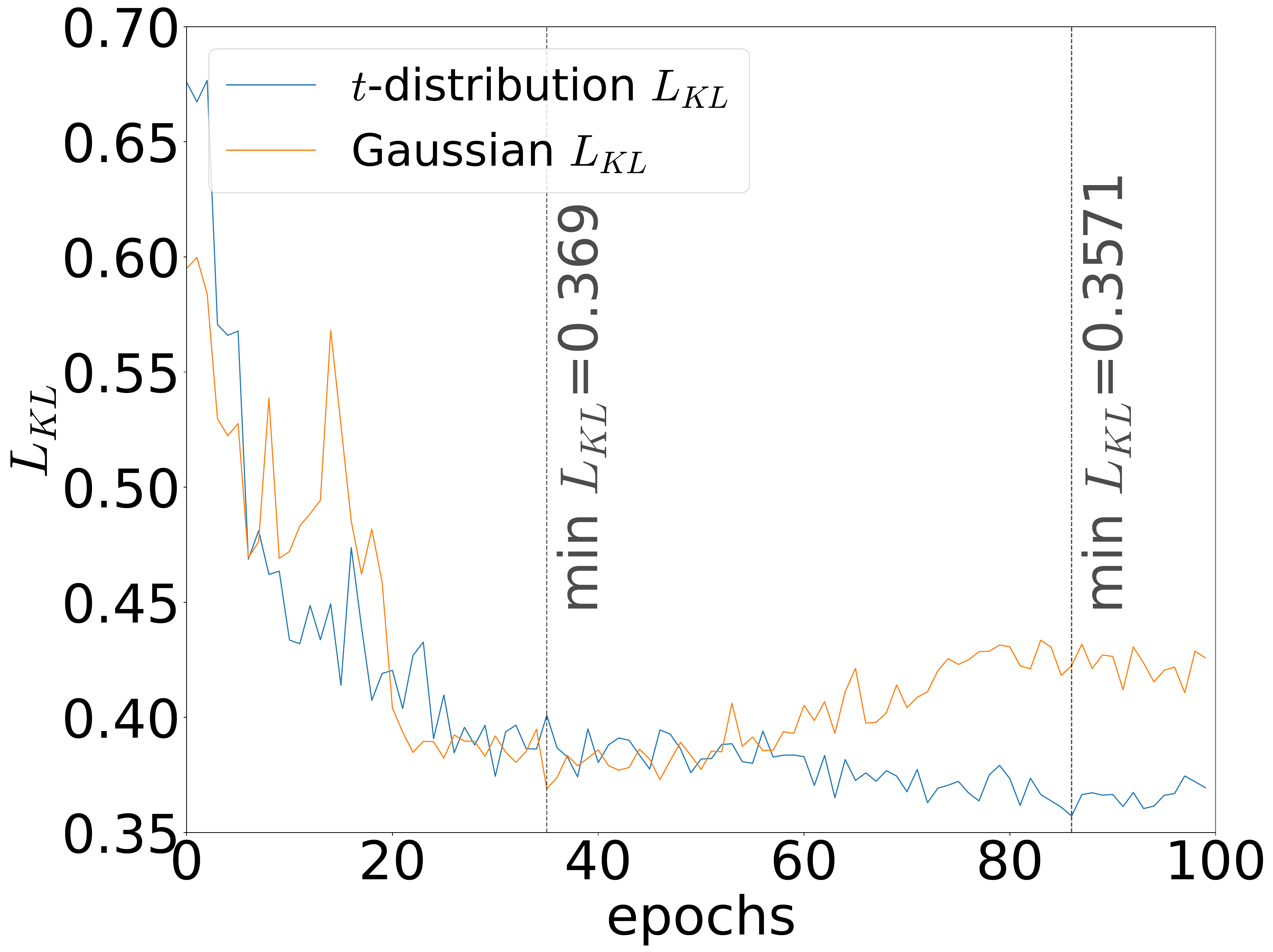}
        \caption{For arousal, in \textbf{MSPConv}.}
        \captionsetup{justification=centering}
        \label{Fig:mspconv_loss_curves_arousal}
     \end{subfigure}
     \hfill
     \begin{subfigure}[b]{0.24\textwidth}
        \includegraphics[width=\textwidth]{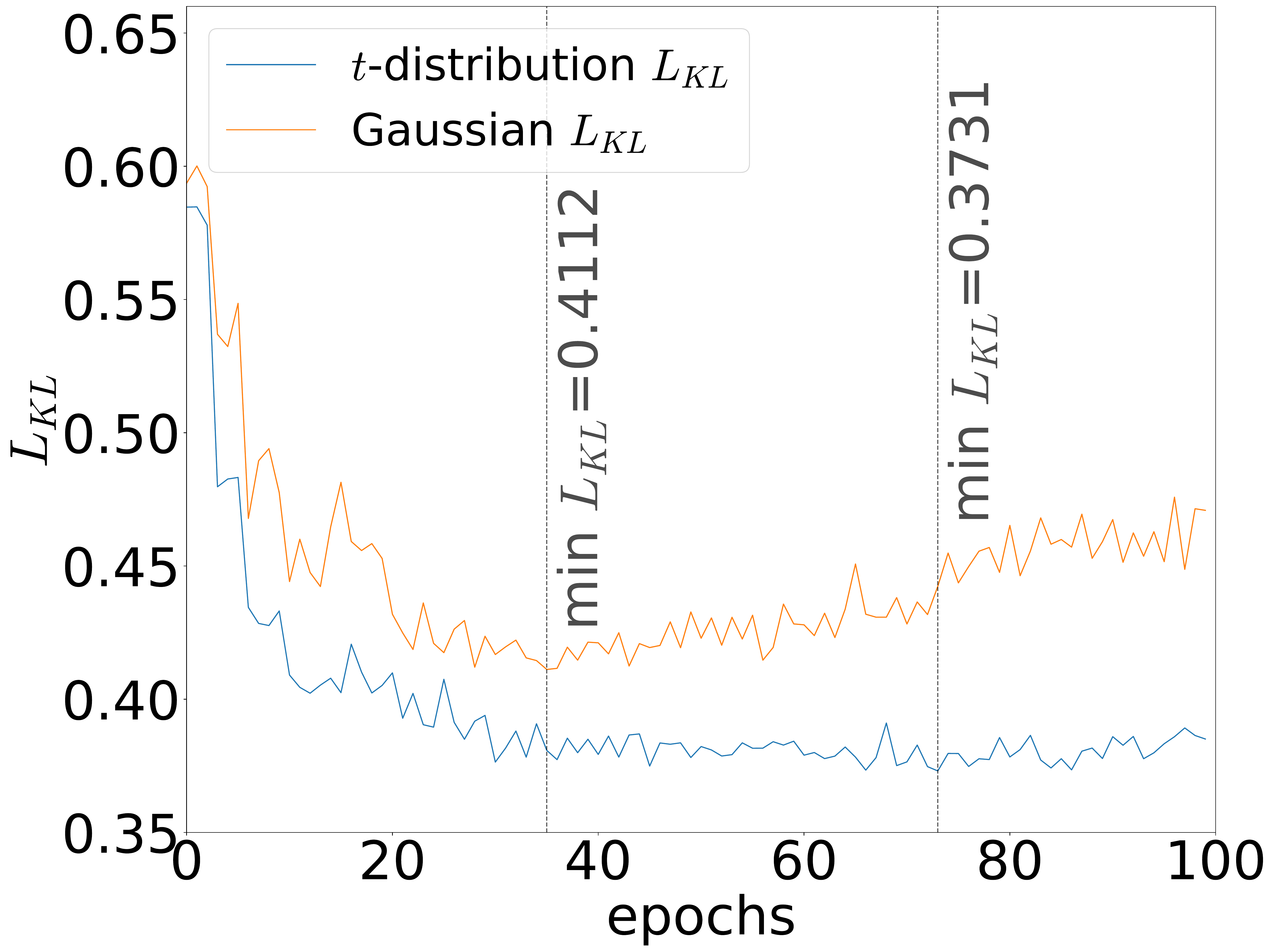}
        \caption{For valence, in \textbf{MSPConv}.}
        \captionsetup{justification=centering}
        \label{Fig:mspconv_loss_curves_valence}
     \end{subfigure}
\caption{Loss curve comparison between Gaussian $\mathcal{L}_{KL}$ \eqref{loss:gauss-KL} and proposed $t$-distribution $\mathcal{L}_{KL}$ \eqref{loss:tdist-KL}.}
\label{Fig:loss_curves}
\end{figure*}

\subsubsection{Comparison on mean estimates}\label{sec:quant-mean-est-discussion}

In terms of \emph{arousal}, Table \ref{tab:quant_results} shows that the proposed $t$-LU model performs the \textit{best} in comparison with the baselines, in both AVEC'16 (Table \ref{tab:avec_qaunt_results}) and MSPConv (Table \ref{tab:mspconv_qaunt_results}) datasets, with \textit{statistical significance}. Four key takeaways can be noted from the $\mathcal{L}_{\text{CCC}}(m)$ results for \emph{arousal}. \textit{Firstly}, the proposed BBB-LDL versions (MU, MU$+$LU, and $t$-LU) achieve better $\mathcal{L}_{\text{CCC}}(m)$ than the MTL baselines (STL, and MTL PU). In the more challenging MSPConv dataset, the performance improvement is even more evident, which highlights the robustness of the proposed approach. For example, while the $t$-LU improves over MTL PU by $0.042$ in AVEC'16, a larger improvement of $0.093$ $\mathcal{L}_{\text{CCC}}(m)$ can be noted in the MSPConv. \textit{Secondly}, between the BBB-LDL versions, the superiority of the proposed $t$-distribution $\mathcal{L}_{KL}$ \eqref{loss:tdist-KL} over the Gaussian $\mathcal{L}_{KL}$ \eqref{loss:gauss-KL} is noted, with $t$-LU outperforming MU$+$LU in both the datasets. \emph{Thirdly}, when incorporating uncertainty modeling in the E2E Baseline, a \textit{compromise} on $\mathcal{L}_{\text{CCC}}(m)$ is made with improving uncertainty estimates ($\mathcal{L}_{\text{CCC}}(s)$ and $\mathcal{L}_\text{KL}$). This can be noted when comparing the results of MU and MU$+$LU with that of the E2E Baseline. However, the proposed $t$-LU is free from this compromise, outperforming the E2E Baseline and other BBB-LDL versions. The $t$-LU achieves a $\mathcal{L}_{\text{CCC}}(m)$ of $0.782$ in AVEC'16 and $0.389$ in MSPConv, with E2E Baseline achieving $0.770$ and $0.373$, respectively. \emph{Finally}, the \emph{E2E Baseline w/o Temp} performs the worst in comparison. The improved performance of E2E Baseline and the proposed models over the \emph{E2E Baseline w/o Temp} emphasises the fact that temporal modeling exists in the proposed models and is achieved through the inclusion of the LSTM layers in their architecture.

In terms of \emph{valence}, in the AVEC'16, the MTL PU baseline performs significantly better than the proposed models. However, in the larger and more complex MSPConv dataset, the proposed $t$-LU performs the best with statistical significance. The MTL PU requires dataset dependent tuning of the loss using the average correlation between $m_t$ and $s_t$. For example, $\mathcal{L} = \mathcal{L}_\text{ccc}(m) - \mathcal{L}_\text{ccc}(s)$ for datasets with negative average correlation, and $\mathcal{L} = \mathcal{L}_\text{ccc}(m) + \mathcal{L}_\text{ccc}(s)$ for a positive one \cite{han2020exploring}. In AVEC'16 the average correlation between $m_t$ and $s_t$ is $+0.103$ (a positive correlation exists). In MSPConv the average correlation between $m_t$ and $s_t$ is $0.002$ where no correlation exists. With these statistics, we note that the MTL PU is robust only in datasets where a correlation between $m_t$ and $s_t$ exists, and not robust in cases of complex datasets like the MSPConv. Moreover, with the dataset dependent tuning of the loss, MTL PU is also not robust in cross-corpora evaluation (see Sec.~\ref{sec:cross-corpora}).

\begin{figure*}
     \centering
     \begin{subfigure}[b]{0.49\textwidth}
        \includegraphics[width=\textwidth]{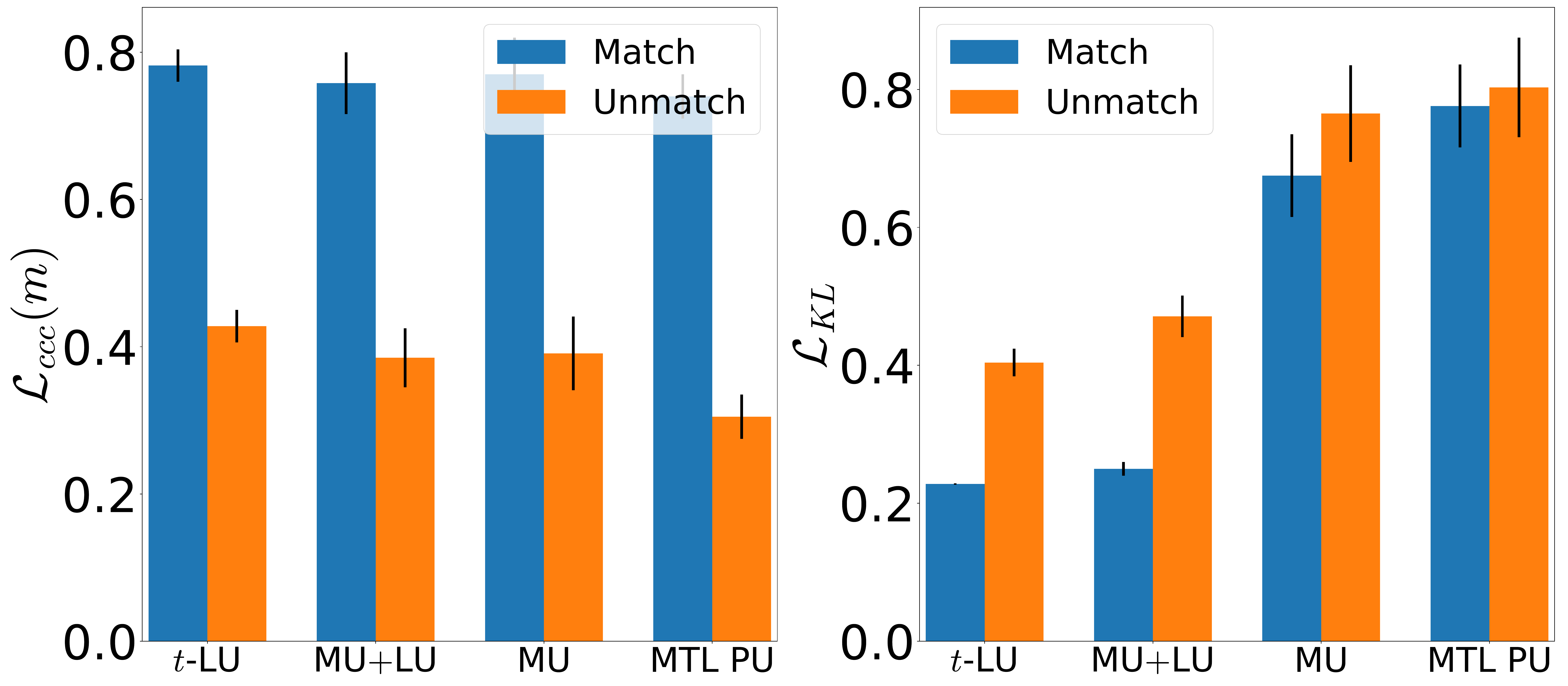}
        \caption{For arousal, in \textbf{AVEC'16} dataset.}
        \captionsetup{justification=centering}
        \label{Fig:crosscorpora_experiment_recola-arousal}
     \end{subfigure}
     \begin{subfigure}[b]{0.49\textwidth}
        \includegraphics[width=\textwidth]{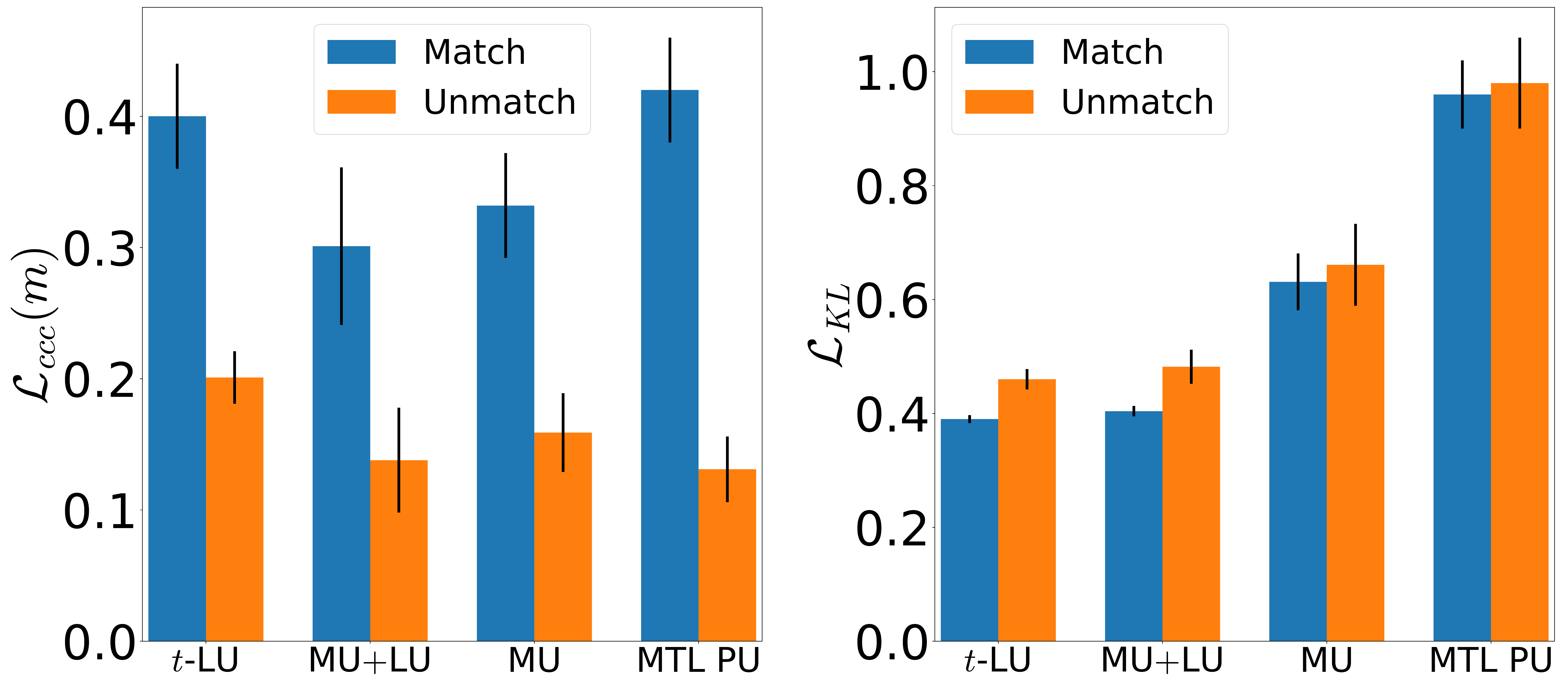}
        \caption{For valence, in \textbf{AVEC'16} dataset.}
        \captionsetup{justification=centering}
        \label{Fig:crosscorpora_experiment_recola-valence}
     \end{subfigure}
     \vfill
     \vspace*{0.2cm}
     \vfill
     \begin{subfigure}[b]{0.49\textwidth}
        \includegraphics[width=\textwidth]{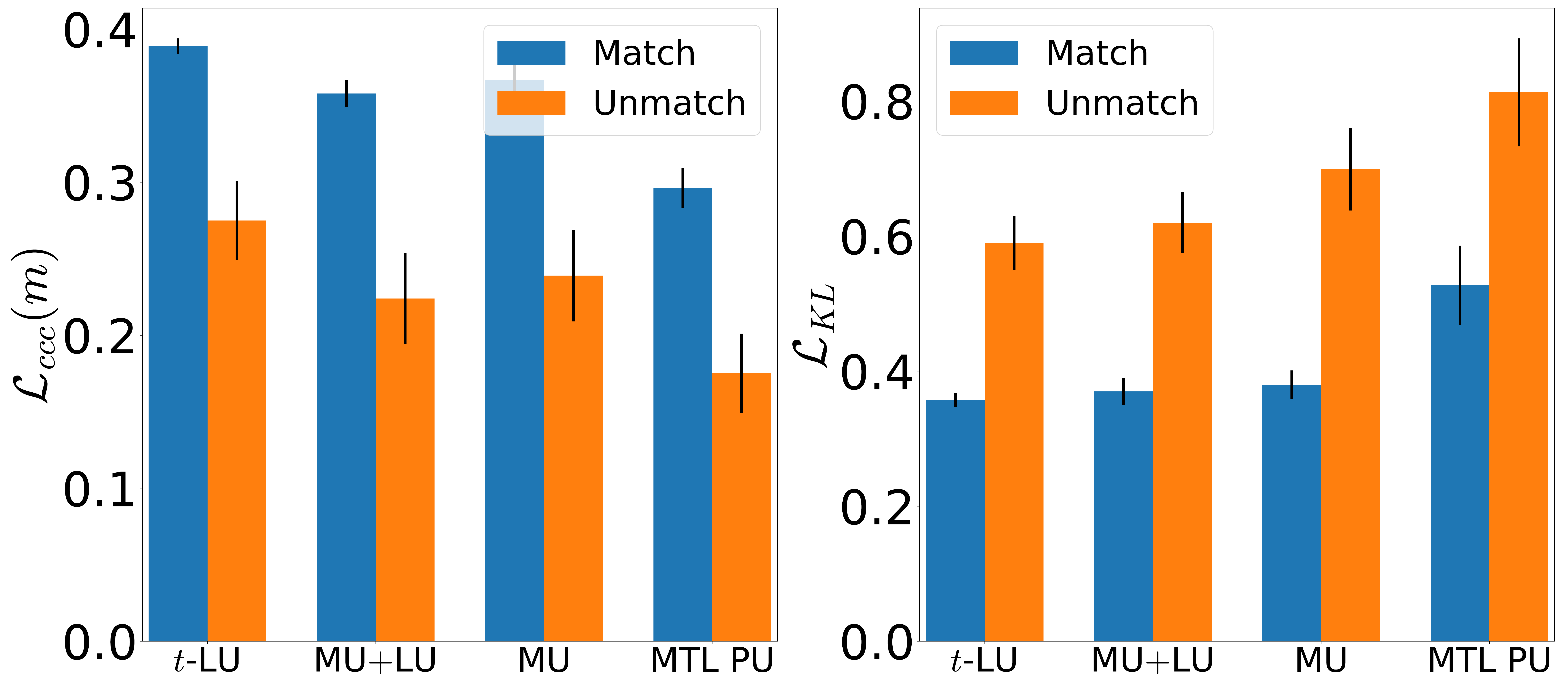}
        \caption{For arousal, in \textbf{MSPConv} dataset.}
        \captionsetup{justification=centering}
        \label{Fig:crosscorpora_experiment_mspconv-arousal}
     \end{subfigure}
     \begin{subfigure}[b]{0.49\textwidth}
        \includegraphics[width=\textwidth]{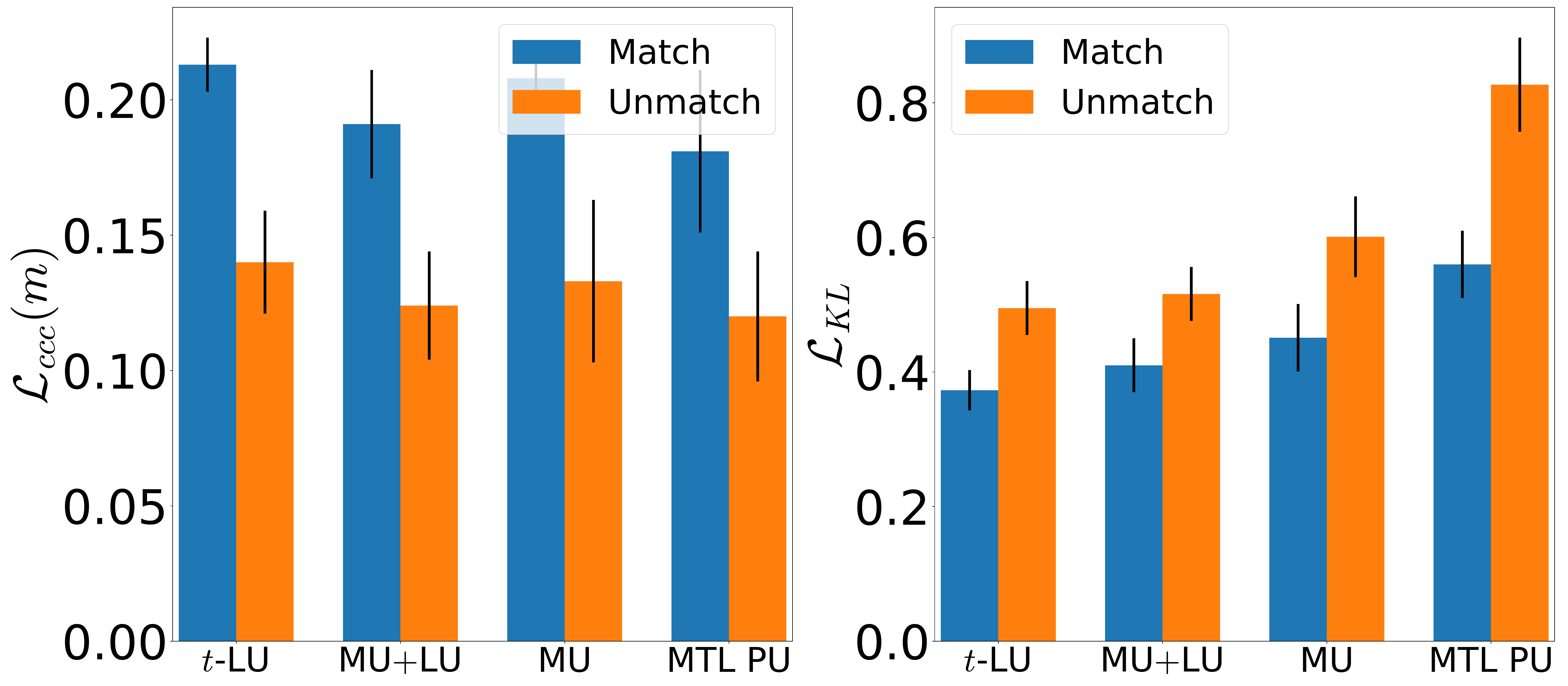}
        \caption{For valence, in \textbf{MSPConv} dataset.}
        \captionsetup{justification=centering}
        \label{Fig:crosscorpora_experiment_mspconv-valence}
     \end{subfigure}
\caption{Cross-corpora evaluations, for \emph{Match} and \emph{Unmatch} conditions in terms of $\mathcal{L}_{\text{ccc}(m)}$ and $\mathcal{L}_\text{KL}$.}
\label{Fig:crosscorpora_experiment}
\end{figure*}


\subsubsection{Comparison on uncertainty estimates}
Table \ref{tab:quant_results} shows that the proposed $t$-LU achieves the best uncertainty estimates across datasets, in terms of both $\mathcal{L}_{\text{CCC}}(s)$ and $\mathcal{L}_{KL}$. In AVEC'16, the improvements are \textit{statistically significant} over all baselines in comparison. In MSPConv, the improvements are statistically significant over all baselines only with respect to the $\mathcal{L}_{KL}$ measure. In terms of the $\mathcal{L}_{\text{CCC}}(s)$ measure, improvements are not statistically significant over the MU$+$LU baseline alone. For instance, in AVEC'16, $t$-LU achieves $0.381$ $\mathcal{L}_{\text{CCC}}(s)$ and $0.228$ $\mathcal{L}_{KL}$, improving with statistical significance. In MSPConv, $t$-LU achieves $0.118$ $\mathcal{L}_{\text{CCC}}(s)$ and $0.357$ $\mathcal{L}_{KL}$, where statistical significance over \textit{all} other baselines exists only for $\mathcal{L}_{KL}$. The reason for this trend is that, firstly, the MSPConv is more complex with larger levels of subjectivity (see Sec.~\ref{sec:dataset}). Secondly, the model is exclusively trained on $\mathcal{L}_{KL}$, so direct improvements over $\mathcal{L}_{KL}$ is expected rather than on $\mathcal{L}_{\text{CCC}}(s)$.





For \textit{valence} in AVEC'16, unlike the $\mathcal{L}_{\text{CCC}}(m)$ performances, Table \ref{tab:avec_qaunt_results} shows that the proposed $t$-LU achieves improved \emph{uncertainty estimates}, in terms of both the measures ($\mathcal{L}_{\text{CCC}}(s)$ and $\mathcal{L}_{KL}$). Moreover, the improvements are statistical significance over all other baselines in terms of the $\mathcal{L}_{KL}$ measure, but only over the MTL-based baselines in terms of the $\mathcal{L}_{\text{CCC}}(s)$ measure. Similar improvement trends can also be noted in the more complex MSPConv dataset (from Table \ref{tab:mspconv_qaunt_results}). This improved uncertainty estimates of the proposed $t$-LU across datasets emphasises the advantage of using the $t$-distribution based $\mathcal{L}_{KL}$ loss \eqref{loss:tdist-KL} for label uncertainty modeling. The $t$-distribution, as seen in Figure \ref{fig:kl-analysis}, promotes the model to fit on a more relaxed $s_t$, thereby more robust in capturing the whole label distribution. The fitting on a relaxed $s_t$ leads to increased robustness towards outliers, as noted in \cite{bishop2006pattern}.

\subsection{Qualitative analysis of estimates}

For qualitative analyses, we plot the mean $\widehat{m}_t$ and standard deviation $\widehat{s}_t$ estimates of $\widehat{\mathcal{Y}}_t$ against the ${m}_t$ and standard deviation ${s}_t$ of ground-truth distribution ${\mathcal{Y}}_t$. Plots for a test subject from AVEC'16, in terms of arousal and valence, can be seen in figures \ref{Fig:results-pred-arousal} and \ref{Fig:results-pred-valence}, respectively, and, for MSPConv, in figures \ref{Fig:msp-results-pred-arousal} and \ref{Fig:msp-results-pred-valence}, respectively. Parts of the plots are boxed and numbered to note clear performance differences.

For \textit{arousal}, in figures \ref{Fig:results-pred-arousal} and \ref{Fig:msp-results-pred-arousal}, further backing the results in Table \ref{tab:quant_results}, the proposed $t$-LU model best captures $m_t$ and $s_t$ of the annotation distribution ${\mathcal{Y}}_t$, in comparison with MU and MU$+$LU. For example, in AVEC'16 (see Fig. \ref{Fig:results-pred-arousal}), in boxes \emph{2} and \emph{3}, $t$-LU best captures the whole distribution $\mathcal{Y}_t$, where $\widehat{s}_t$ best resembles $s_t$. This further highlights the robustness of training on a relaxed $s_t$ through a $t$-distribution. Backing the quantitative results in Table \ref{tab:quant_results}, improvements are more evident in MSPConv, noted from boxes \emph{2} and \emph{3} in Fig. \ref{Fig:msp-results-pred-arousal}). Crucially, along with the $\widehat{s}_t$ improvements by $t$-LU, notable improvements are also seen on mean estimates $\widehat{m}_t$. 

For \textit{valence}, figures \ref{Fig:results-pred-valence} and \ref{Fig:msp-results-pred-valence} show that the proposed $t$-LU evidently improves on mean estimates $\widehat{m}_t$ on both datasets, with only small improvements on standard deviation estimates $\widehat{s}_t$. This can be seen for instance in box \emph{1} of Fig. \ref{Fig:results-pred-valence}. Hence, capturing $s_t$ in valence by only relying on audio is a challenging task, and more complex in datasets such as the MSPConv where some frames have a very high subjectivity (see log histograms in Fig.\ref{Fig:s_dist_mspconv}). It is a common trend in the literature that the audio modality insufficiently explains ground-truth valence $m_t$ \cite{tzirakis2021-mm, kusha_valence}, and this trend is even more challenging for modeling $s_t$ in valence. 



\subsection{Analysis on training loss curve}

To further study the advantages of the proposed $t$-distribution $\mathcal{L}_{KL}$ \eqref{loss:tdist-KL} during the training phase, we compare the testing loss curve of \eqref{loss:tdist-KL} with the Gaussian $\mathcal{L}_{KL}$ in MU$+$LU \eqref{loss:gauss-KL}. The comparisons can be seen in Figure \ref{Fig:loss_curves}.


Figure \ref{Fig:loss_curves} illustrates two crucial advantages of the proposed $t$-distribution $\mathcal{L}_{KL}$ loss term \eqref{loss:tdist-KL} during training in both datasets. Firstly, we see that in the initial epochs, before epoch $20$, the proposed loss converges quicker than the Gaussian $\mathcal{L}_{KL}$ \eqref{loss:gauss-KL}. This is the result of the proposed $\mathcal{L}_{KL}$ \eqref{loss:tdist-KL} loss term which penalizes more for lower $s_t$ values, in comparison to the Gaussian $\mathcal{L}_{KL}$ \eqref{loss:gauss-KL} (see Sec. \ref{Section:kl-analysis}), thereby achieving faster convergence. Secondly, during the later epochs, after epoch $70$, the \textit{Gaussian} $\mathcal{L}_{KL}$ \eqref{loss:gauss-KL} shows signs of overfitting, which is more evident in the MSPConv dataset. However, at the same time, the proposed $t$-distribution $\mathcal{L}_{KL}$ \eqref{loss:tdist-KL} converges to the best minima during the later epochs. For instance, in MSPConv, the proposed \eqref{loss:tdist-KL} achieves minima $\mathcal{L}_{KL}$ at epoch $86$, with $\mathcal{L}_{KL}$ of $0.357$ for arousal and $0.373$ for valence, while the Gaussian achieves a minima well before the later epochs, at epoch $35$, with $\mathcal{L}_{KL}$ of $0.369$ for arousal and $0.411$ for valence. The proposed $\mathcal{L}_{KL}$ \eqref{loss:tdist-KL} is free from overfitting in the later stages of training and also learns the optima at this stage, noticed across two datasets. This behaviour can be attributed to the nature of the proposed $\mathcal{L}_{KL}$ \eqref{loss:tdist-KL} which promotes the model to learn a more relaxed $s_t$, thereby introducing more regularization to the model, preventing overfitting and converging on an improved $s_t$.

\begin{figure*}
     \centering
     \begin{subfigure}[b]{0.49\textwidth}
        \includegraphics[width=\textwidth]{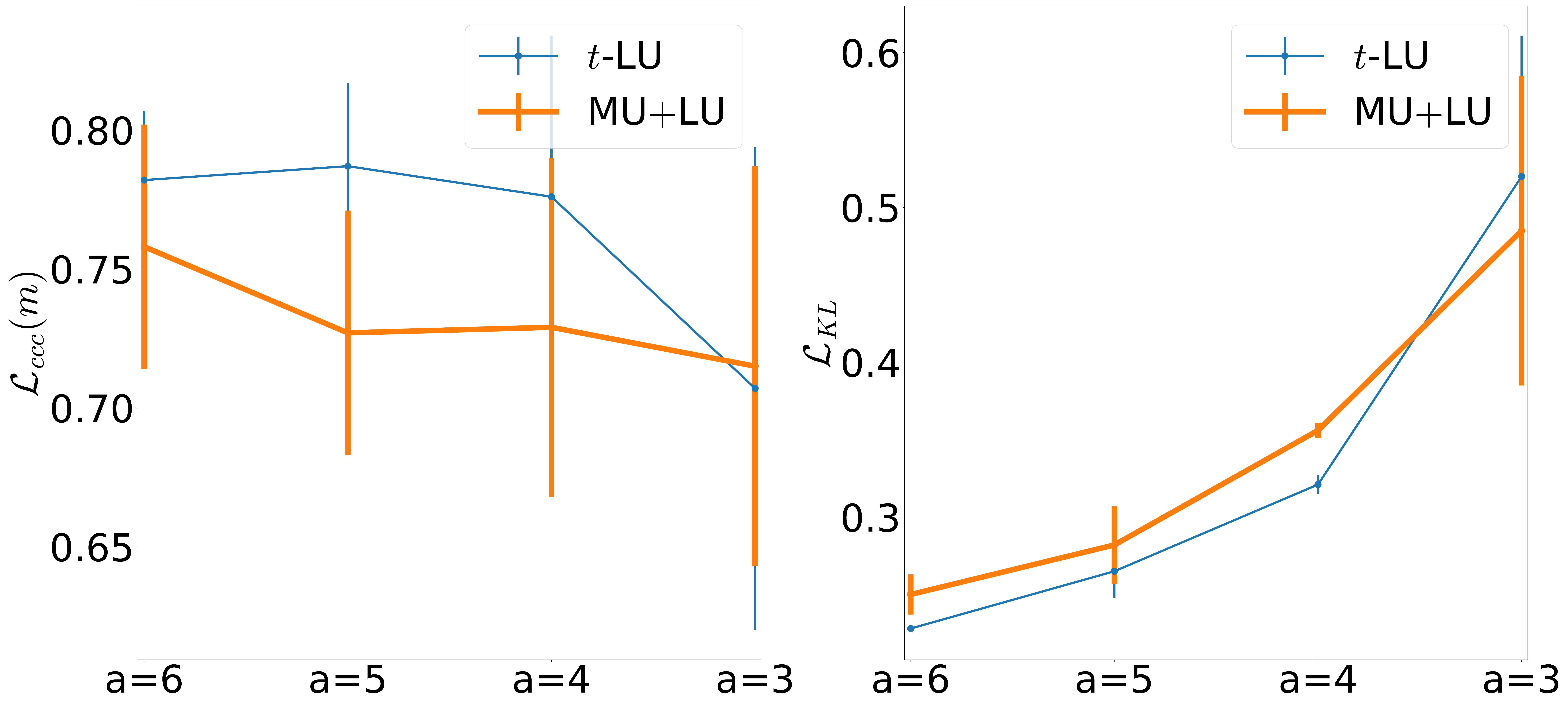}
        \caption{For arousal, in \textbf{AVEC'16} dataset.}
        \captionsetup{justification=centering}
        \label{Fig:nannot_experiment_recola-arousal}
     \end{subfigure}
     \begin{subfigure}[b]{0.49\textwidth}
        \includegraphics[width=\textwidth]{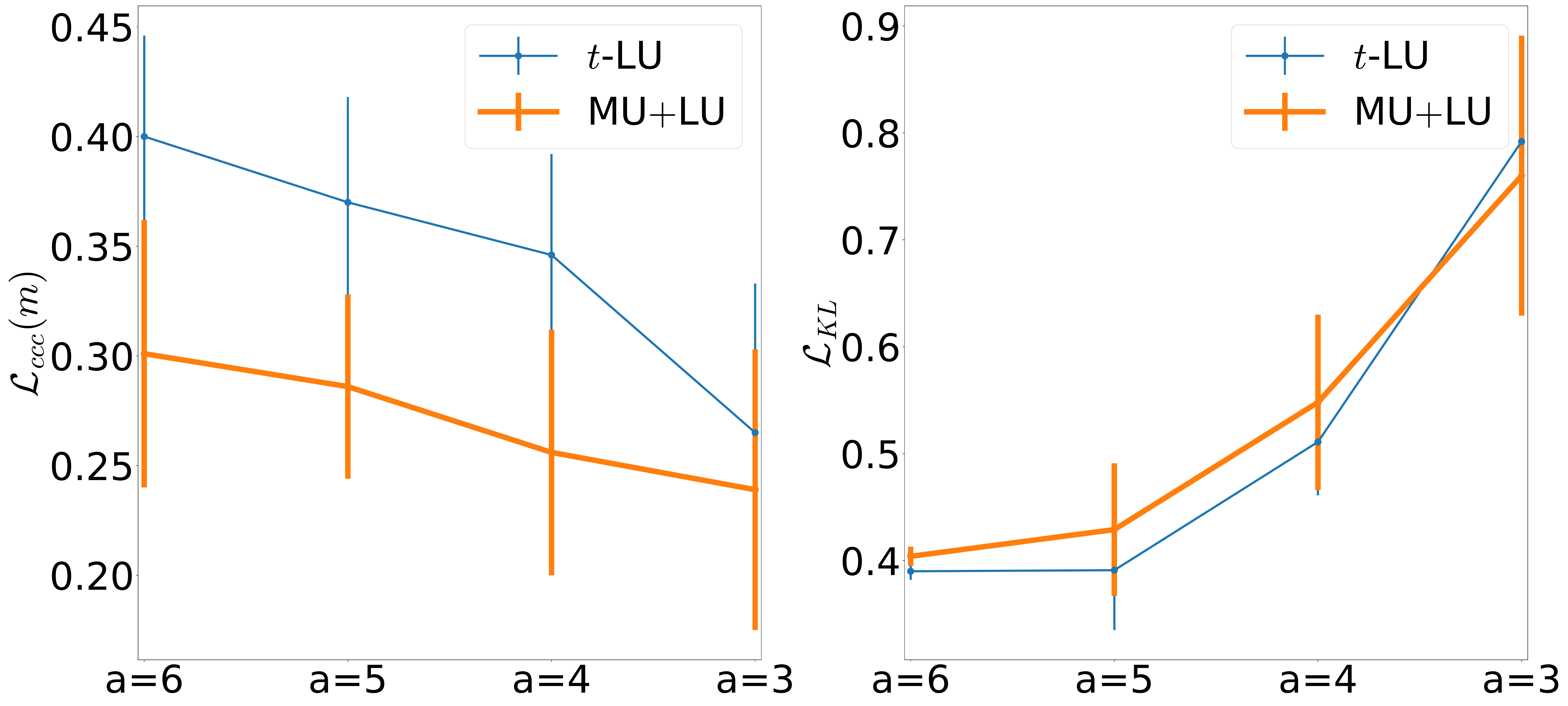}
        \caption{For valence, in \textbf{AVEC'16} dataset.}
        \captionsetup{justification=centering}
        \label{Fig:nannot_experiment_recola-valence}
     \end{subfigure}
     \vfill
     \vspace*{0.2cm}
     \vfill
     \centering
     \begin{subfigure}[b]{0.49\textwidth}
        \includegraphics[width=\textwidth]{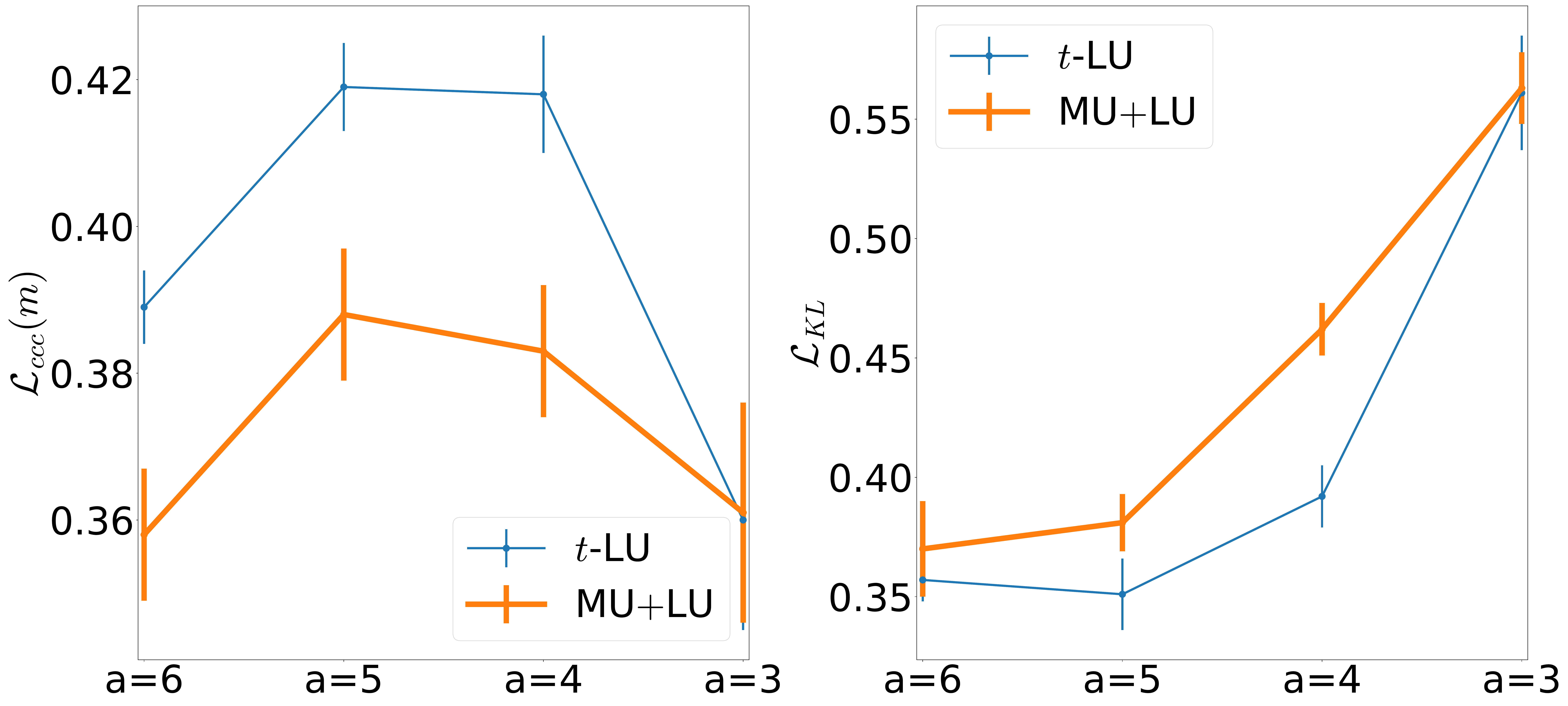}
        \caption{For arousal, in \textbf{MSPConv} dataset.}
        \captionsetup{justification=centering}
        \label{Fig:nannot_experiment_mspconv-arousal}
     \end{subfigure}
     \begin{subfigure}[b]{0.49\textwidth}
        \includegraphics[width=\textwidth]{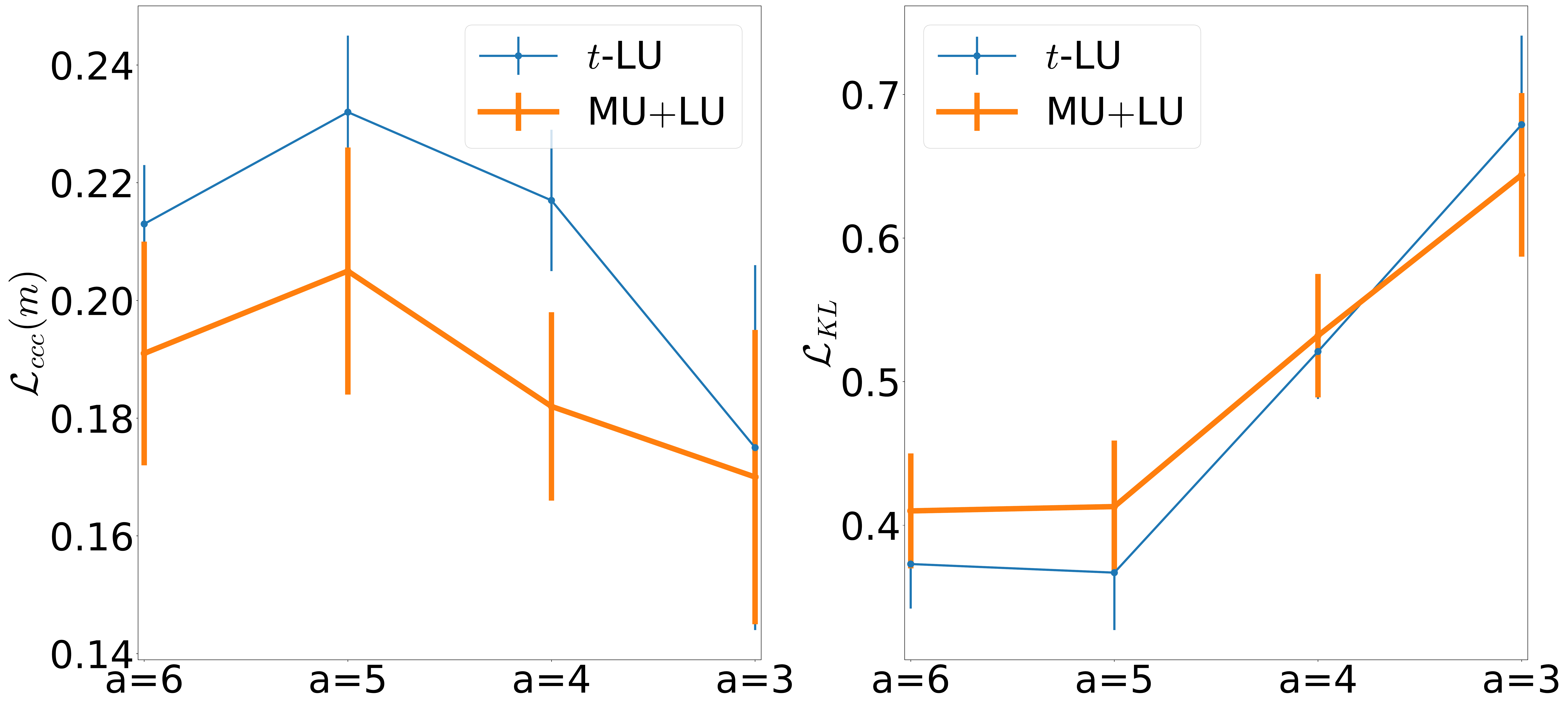}
        \caption{For valence, in \textbf{MSPConv} dataset.}
        \captionsetup{justification=centering}
        \label{Fig:nannot_experiment_mspconv-valence}
     \end{subfigure}
\caption{Impact of number of annotations available $a=6,5,4,3$ on $\mathcal{L}_{\text{ccc}(m)}$ and $\mathcal{L}_\text{KL}$.}
\label{Fig:nannot_experiment}
\end{figure*}

\subsection{Cross-corpora evaluation} \label{sec:cross-corpora}

To validate the robustness and generalisation capabilities of the models, we performed cross-corpora evaluations. In Figure \ref{Fig:crosscorpora_experiment}, results are presented in terms of $\mathcal{L}_{\text{ccc}}(m)$ and $\mathcal{L}_{KL}$, under two conditions. The \emph{Match} condition where the train and the test partitions come from the \emph{same} dataset, and the \emph{Unmatch} condition where the \textit{train} partition is from a \textit{different} dataset. Apart from the dataset size, other dataset-specific factors, such as population demographics and social context, severely challenge the cross-corpora performances because human behaviour varies across group-sizes \cite{ramanrajprabhu2023, gatica-perez_automatic_2009} and social contexts \cite{MspPod}. Crucial differences exist between the AVEC'16 and MSPConv datasets. While the social context of AVEC'16 is a \textit{dyadic} interaction in a \textit{virtual} setting, MSPConv comprises of \textit{larger groups} in a \textit{face-to-face} setting. Moreover, AVEC'16 was collected from \textit{French}-speaking persons, while MSPConv from \textit{English}-speaking persons. 



Figure \ref{Fig:crosscorpora_experiment} illustrates that the proposed $t$-LU achieves the best cross-corpora performances on both datasets, and MU with the second best performances. Under the \textit{Unmatch} condition, for \textit{arousal} in AVEC'16 (see Fig. \ref{Fig:crosscorpora_experiment_recola-arousal}), $t$-LU achieves $0.421$ $\mathcal{L}_{\text{ccc}}(m)$ and $0.409$ $\mathcal{L}_{KL}$, while MU achieves $0.342$ and $0.490$, respectively. Similarly, in MSPConv (see Fig. \ref{Fig:crosscorpora_experiment_mspconv-arousal}), $t$-LU achieves $0.260$ $\mathcal{L}_{\text{ccc}}(m)$ and $0.600$ $\mathcal{L}_{KL}$, while MU achieves $0.216$ and $0.655$, respectively. 

All models degrade in performance from the \textit{Match} to \textit{Unmatch} conditions. For both arousal and valence, across datasets and metrics, $t$-LU achieves the \textit{least degrade percentage} while the MTL PU results in the \textit{highest} degrade. For instance, in AVEC'16, in terms of \textit{arousal} mean-estimates $\mathcal{L}_{\text{ccc}}(m)$ (see Fig. \ref{Fig:crosscorpora_experiment_recola-arousal}), $t$-LU achieves the least degradation of $46\%$ and MTL PU degrades the most with $61\%$. Similarly, for \textit{valence} (see Fig. \ref{Fig:crosscorpora_experiment_recola-valence}), $t$-LU degrades least with $53\%$, and MTL PU degrades the most with $62\%$. This further emphasises on the robustness of the proposed $t$-LU model and clearly highlights the lack of robustness of the MTL PU baseline. The MTL PU which achieves the best $\mathcal{L}_{\text{ccc}}(m)$ for valence on the AVEC'16 (see Table \ref{tab:avec_qaunt_results}), degrades the most on cross-corpora evaluations. This drawback of the MTL PU baseline stems from the dataset-dependent tuning of loss function that it relies on. The proposed $t$-LU is free from such dataset-dependent tuning and hence more robust. The degrade \textit{percentage} in $\mathcal{L}_{KL}$ is not comparable as the scale of the measure is not linear (depicted in Fig. \ref{fig:kl-analysis}). Also notable is that, for all models, the degrade percentage is larger for valence than for arousal.


%


\begin{table*}[h!]
\caption{Ablation study results of the \emph{t-LU} model, on the AVEC'16 \cite{avec16} and MSP-Conversation \cite{MspConv} datasets. Modules included in the ablation study are the Uncertainty Layer (BBB), the end-to-end Feature Extractor (E2E), and the Label Distribution Learning loss (KL). $\checkmark$ denotes the \emph{inclusion} of the respective module, and $\times$ its \emph{omission}. \textbf{Bold} results denote the \emph{best two} results for a particular metric, and \underline{underline} denotes the \emph{least two}. $*$ indicates statistically significant better results over non-bold results. Absence of $*$ indicates that the improvements are not statistically significant.}
\centering
\begin{tabular}{c|ccc|ccc|ccc}
\toprule
                         & \multicolumn{3}{c|}{\textbf{Modules}} & \multicolumn{3}{c}{\textbf{Arousal}}  & \multicolumn{3}{c}{\textbf{Valence}}\\
                         & E2E & BBB & KL & $\mathcal{L}_{\text{ccc}}(m) \uparrow$ & $\mathcal{L}_{\text{ccc}}(s) \uparrow$ & $\mathcal{L}_\text{KL} \downarrow$ & $\mathcal{L}_{\text{ccc}}(m) \uparrow$ & $\mathcal{L}_{\text{ccc}}(s) \uparrow$ & $\mathcal{L}_\text{KL} \downarrow$    \\ 
                         \midrule
\multirow{ 8}{*}{AVEC'16}&

        $\checkmark$  & $\checkmark$     & $\checkmark$      & \textbf{0.782*}    & \textbf{0.381*}   & \textbf{0.228*}   & {0.400} & 0.050 & \textbf{0.390*}  \\
        & $\checkmark$  & $\checkmark$     & $\times$ & {0.743} & 0.356          & 0.412   & \underline{0.329} & \textbf{0.054} & 0.594  \\
        & $\checkmark$  & $\times$         & $\checkmark$ & \underline{0.704}           & \underline{0.315}          & 0.299   & 0.373 & 0.039 & 0.426  \\
        & $\checkmark$  & $\times$         & $\times$          & 0.721           & \textbf{0.392*}          & \underline{0.512}    & \textbf{0.401} & \textbf{0.064} & \underline{0.863}  \\
        & $\times$      & $\checkmark$     & $\checkmark$      & \textbf{0.772*}  & {0.330}   & \textbf{0.276*}   
                                                                                & 0.366 & \underline{0.033} & \textbf{0.411}  \\
        & $\times$      & $\checkmark$     & $\times$          & 0.758          & 0.329          & 0.446  
                                                                                & \underline{0.330} & 0.050 & 0.601  \\
        & $\times$      & $\times$         & $\checkmark$      & \underline{0.716}           & 0.329          & 0.318   
                                                                                & {0.381} & {0.039} & 0.446  \\
        & $\times$      & $\times$         & $\times$    &      0.740  & \underline{0.310}     & \underline{0.776}       & \textbf{0.420*}  & \underline{0.032}  & \underline{0.960}  \\ 
                         \midrule
                         
\multirow{ 8}{*}{MSPConv}& $\checkmark$  & $\checkmark$     & $\checkmark$      & \textbf{0.389*}  & \textbf{0.118*}   & \textbf{0.357*} & \textbf{0.213*} & \textbf{0.032}   & \textbf{0.373*}   \\
                         & $\checkmark$  & $\checkmark$     & $\times$          & 0.286    & 0.097 & 0.412   & 0.180 & 0.029    & 0.495    \\
                         & $\checkmark$  & $\times$         & $\checkmark$      & \underline{0.163}    & {0.051}  & 0.515   & \underline{0.122}  & \underline{0.009} & 0.537  \\
                         & $\checkmark$  & $\times$         & $\times$          & 0.271    & 0.100  & 0.489   & 0.174   &  0.012  & \underline{0.593}          \\
                         & $\times$      & $\checkmark$     & $\checkmark$      & \textbf{0.401*}    & 0.056   & \textbf{0.392*}   & \textbf{0.230*}  & 0.017 & \textbf{0.391*}   \\
                         & $\times$      & $\checkmark$     & $\times$          & 0.308 & 0.078     & \underline{0.551}   & 0.181   & 0.026 & 0.416          \\
                         & $\times$      & $\times$         & $\checkmark$      & \underline{0.247}  & \underline{0.040}  & 0.490   & \underline{0.140}   & \underline{0.005}   & {0.549} \\
                         & $\times$      & $\times$         & $\times$          & 0.296 & \textbf{0.107} & \underline{0.527}   & 0.181  & \textbf{0.030}  & \underline{0.560}  
                         
\end{tabular}
\label{tab:ablation_study}
\end{table*}

\subsection{Impact of number of annotations $a$ available}
In Sec. \ref{sec:quant-results}, we noted the benefits of modeling $\mathcal{Y}_t$ as a $t$-distribution, with \textit{six} available annotations. To capitalise on the benefits of the $t$-distribution $t$-LU over the Gaussian MU$+$LU, especially when \textit{fewer annotations} are available, we performed experiments by varying $a$ and thereby the degrees of freedom $\nu$. The results are presented in Figure \ref{Fig:nannot_experiment}, under $4$ settings, $a=3$, $a=4$, $a=5$, and, $a=6$. Annotations were ignored to achieve conditions of $a\leq5$. The order of annotation to be ignored was handled based on Pearson's correlations measure. For instance, under setting $a=4$, annotations from two annotators, with the least inter-annotator correlation, for the whole audio file were ignored to model ground-truth annotation distribution $\mathcal{Y}_t$.

Figure \ref{Fig:nannot_experiment} shows that, \textit{especially when} $a\geq4$, the $t$-distribution based $t$-LU shows superior performance over the Gaussian MU$+$LU on both datasets. Crucially, the improvements are larger and more evident when $a=4$ and $a=5$ than when $a=6$. In the case of $\mathcal{L}_\textrm{CCC}(m)$, a non-monotonic behavior with the available number of annotations is notable; $\mathcal{L}_\textrm{CCC}(m)$ initially increases from $a=6$ to $a=5$ and subsequently decreases with reducing annotations $a\leq4$. The initial increase is noticed as annotations are ignored in the order of reducing Pearson's correlation, hence we can expect better consensus in $m_t$ for $a=5$ than $a=6$. The subsequent decrease can be associated with the reduced number of annotations to model a stable distribution $\mathcal{Y}_t$. This emphasises the advantage of $t$-distribution over the Gaussian with increasing inter-annotator correlation and reducing number of available annotations.
In the case of $a=3$, the performance of $t$-LU drops below that of the Gaussian MU$+$LU, as $t$-LU becomes highly uncertain with only $3$ annotations because of the large relaxation on $s_t$ introduced by the scaling in Equation \ref{eq:scale-sigma} (see Supplementary Sec.~2 for theoretical analysis). This behaviour is similar to the $t$-test calculation, where models become more uncertain with reducing $\nu$.  For modeling emotion annotations as a distribution and uncertainty modeling, we therefore recommend the $t$-distribution over the Gaussian when more than $3$ annotations are available. Noting that both $t$-distribution and Gaussian drop in performances with only $3$ annotations, we suggest collecting at least $4$ annotations to obtain a reliable annotation distribution and its ground-truth consensus.






\subsection{Ablation study}

The proposed end-to-end label uncertainty model has three essential modules, namely 1) feature extractor, 2) uncertainty layer, and, 3) label uncertainty loss. To understand the modules' specific contributions, we perform an ablation study and present its results in Table \ref{tab:ablation_study}. In case of the feature extractor, \emph{E2E}, $\checkmark$ denotes usage of an end-to-end feature extractor and $\times$ the hand-crafted features \cite{eyben2015geneva, t21_interspeech}. In case of the uncertainty layer, \emph{BBB}, $\checkmark$ denotes the usage of the BBB-based uncertainty layer and $\times$ the MTL-based $s_t$ estimator. For label uncertainty loss, \emph{KL}, $\checkmark$ denotes using $\mathcal{L}_{KL}$ loss and $\times$ denotes usage of $\mathcal{L}_{\text{ccc}}(s)$ loss. 


Table \ref{tab:ablation_study} firstly shows that end-to-end models achieve better uncertainty estimates than hand-crafted feature models. For instance, in AVEC'16, \textit{E2E} based BBB-KL model achieves $0.381$ $\mathcal{L}_{\text{ccc}}(s)$ and $0.228$ $\mathcal{L}_{KL}$, improving over hand-crafted features based BBB-KL model which achieves $0.330$ and $0.276$, respectively. Similarly, in the larger and more complex MSPConv, the E2E based BBB-KL model achieves the best uncertainty estimate performances, against all other models in comparison, with $0.118$ $\mathcal{L}_{\text{ccc}}(s)$ and $0.357$ $\mathcal{L}_{KL}$. This trend is inline with literature that suggests end-to-end learning, that learns emotion representations in a data-driven manner, for uncertainty modeling \cite{alisamir2021evolution}. Secondly, the combination of BBB-based uncertainty layer and KL-based loss term (BBB + KL) results in improved performances in both mean and uncertainty estimates, recommending the combination of BBB-layer and KL-loss for label uncertainty modeling in SER. The performance of BBB-layer with a $\mathcal{L}_{\text{ccc}}(s)$ loss term degrades performance across metrics. An intuition behind this is that KL-based \textit{distribution} loss is apt for optimizing the weight \textit{distributions} $P(w|\mathcal{D})$, rather than a loss with only optimizes for $s_t$ of label distribution. Thirdly, across datasets, for both arousal and valence, the KL-based loss term contributes to the improvement of both uncertainty and mean estimates, as the KL loss jointly optimizes for $m_t$ and $s_t$. For instance, in terms of arousal, the inclusion of KL loss to the E2E+BBB architecture results in a $5\%$ improvement on mean estimates $\mathcal{L}_{\text{ccc}}(m)$ in AVEC'16 and $26\%$ in MSPConv. At the same time, improvements on uncertainty estimates are also noted, $7\%$ improvement of $\mathcal{L}_{\text{ccc}}(s)$ in AVEC'16 and $18\%$ in MSPConv.

Finally, MTL-based $s_t$ estimating model achieves the best $\mathcal{L}_{\text{ccc}}(m)$ performance for valence, but only in AVEC'16 (see last row in Table \ref{tab:ablation_study}). However, in MPSConv, the proposed BBB+KL based models achieve better results. This improvement, noted only for valence in the AVEC'16, again stems from the dataset-dependent tuning of the loss that is required by MTL-based $s_t$ estimating models (see Sec.~\ref{sec:quant-mean-est-discussion}). However, this tuning also results in MTL-based $s_t$ estimating models losing their robustness and generalisation capabilities, as shown in cross-corpora evaluations (see Sec.~\ref{sec:cross-corpora}). Moreover, the MTL-based uncertainty models collapse when not trained on $\mathcal{L}_{\text{ccc}}(s)$ loss, and are not capable of distribution learning using the $\mathcal{L}_{KL}$ loss. Overall, these trends suggest that BBB-based $\mathcal{Y}_t$ learning models are to be preferred over MTL-based $s_t$ estimating models for label uncertainty modeling in SER.

\section{Conclusion}

We introduced an end-to-end BNN capable of modeling emotion annotations as a label distribution, thereby accounting for the inherent subjectivity-based label uncertainty. In the literature, emotion annotations are commonly modeled using a Gaussian distribution or a histogram representation, however with assumptions based on only limited annotations. In contrast, in this work, we modeled ground-truth emotion annotations as a Student's $t$-distribution, which also accounts for the number of annotations available. Specifically, we derived a $t$-distribution based KL divergence loss that, for limited and sparse annotations, produces robust mean estimates and standard deviation estimates that well capture the outliers. We showed that the proposed $t$-distribution loss term leads to training on a relaxed standard deviation, which is adaptable with respect to the number of annotations available. We validated our approach on two publicly available in-the-wild datasets. Quantitative analysis of the results showed that our proposed approach achieved state-of-the-art results for mean and uncertainty estimations, in terms of both CCC and KL divergence measures, which were also consistent for cross-corpora evaluations. By analysing the loss curves, we showed that the proposed loss term yields faster and improved convergence, and is less prone to overfitting than the Gaussian loss term. Our results further revealed that the advantage of $t$-distribution over the Gaussian grows with increasing inter-annotator correlation and decreasing numbers of available annotations. Finally, our ablation study suggests that, for modeling label uncertainty in SER, BBB-based label distribution learning models are to be preferred over estimating standard deviation as an auxiliary task.

\subsection{Limitations and Future Avenues}
In our work, we modeled the emotion annotations as a label distribution using a BNN. However, the BNN introduced here, both MU+LU and $t$-LU, \textit{jointly} captures the two types of uncertainty--- model and label uncertainty. In future work, it would be interesting to focus on disentangling the two types of uncertainty for reliable label uncertainty aware SER systems. One possible way to achieve this concerns \textit{Prior Networks} (PNs) \cite{malinin2018predictive}, a variant of BNNs, which could be employed to exclusively capture the label uncertainty. PNs do this by parameterizing a prior distribution over predictive label distributions.

 
This work specifically focused on modeling emotion annotations in a time- and value-continuous manner. In future work, the proposed methodology can be directly extended to model emotion annotations at the utterance-level, as opposed to time-continuous annotations, by simply adding a pooling layer to the feature extractor. However, the method cannot be directly extended to modeling discrete emotion annotations (e.g., classification tasks). Note that the model architecture introduced here (Fig. \ref{Fig:speechEmoBnn}) can be modified to classify discrete emotion labels, but the introduced label uncertainty loss \eqref{loss:tdist-KL} operates only on value-continuous annotations samples. To further extend the introduced loss function for classification tasks, future work may focus on the \textit{discrete} variant of $t$-distributions. In that case, similar to the loss function derivation in Sec.~\ref{sec:kl-loss-derivation}, KL divergence loss for \textit{discrete} $t$-distributions would need to be derived.

While the proposed model achieved significantly better state-of-the-art performances in terms of the arousal dimension of emotion across datasets, in one of the datasets (AVEC'16) it did not achieve state-of-the-art performance in terms of \textit{valence}. Note however that state-of-the-art valence performance was achieved in the more complicated MSPConv dataset. It is well documented in the literature that the audio modality insufficiently explains the valence dimension of emotions \cite{tzirakis2021-semspeech}. This is likely also the reason why the best performing $t$-LU model, in terms of valence in tables \ref{tab:avec_qaunt_results}, \ref{tab:mspconv_qaunt_results}, and \ref{tab:ablation_study}, did not achieve statistical significance in some of the metrics despite its improved performance. To overcome this drawback, future work could also include the video and semantic modalities in the feature extractor module, thereby achieving multimodality.

\bibliographystyle{IEEEtran}
\bibliography{references}

\vspace{-1cm}

\begin{IEEEbiography}[{\includegraphics[width=1in,height=1.25in,clip,keepaspectratio]{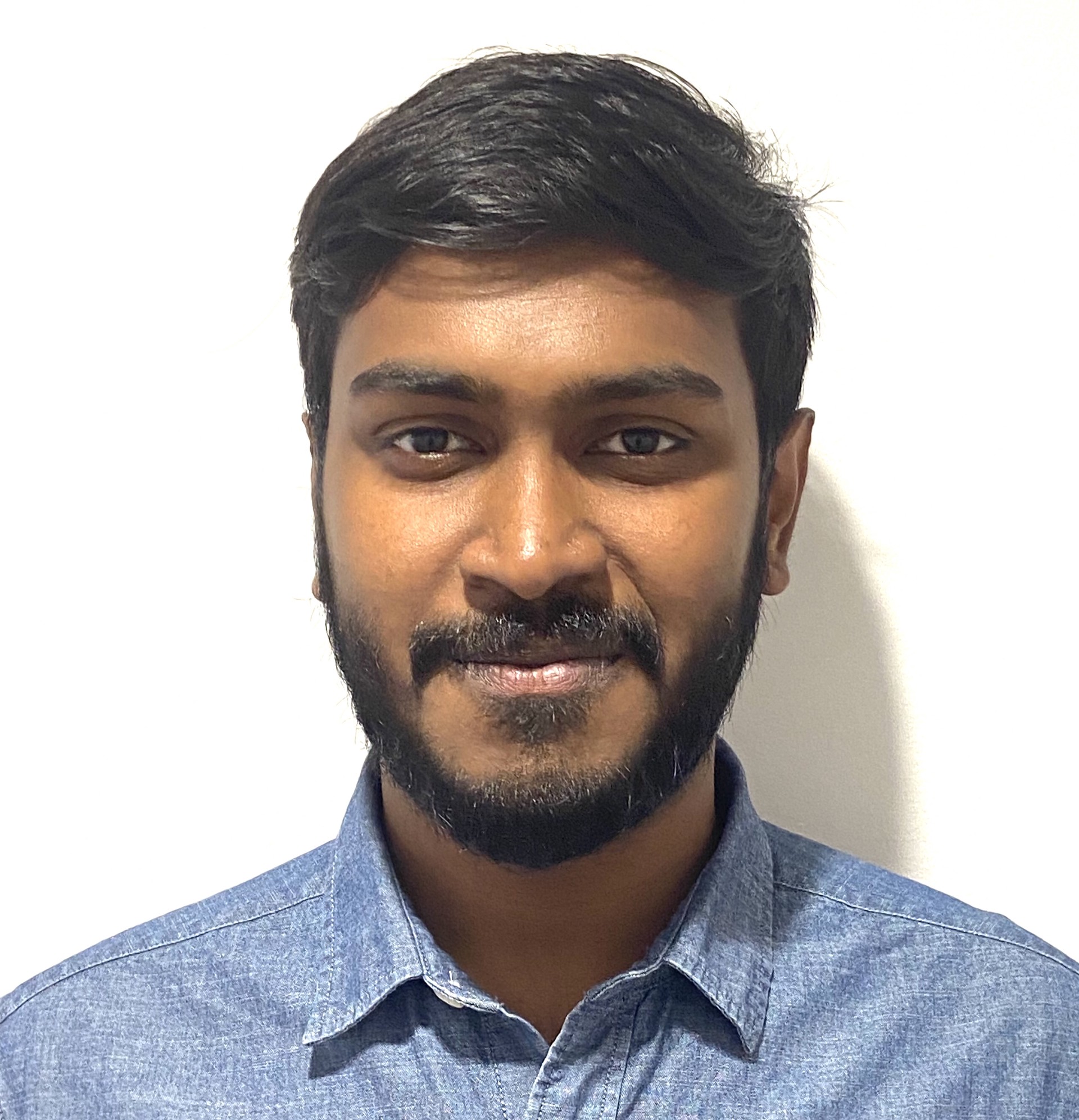}}]{Navin Raj Prabhu}
received a B.Tech degree in Computer Science from SRM University, India, in 2015, and the MS degree in Computer Science from the Intelligent Systems Department at the Delft University of Technology, Delft, The Netherlands, in 2020. Currently, he is a PhD student at the Signal Processing Lab and Organisation Psychology Lab, University of Hamburg, Hamburg, Germany. His research interests include affective computing, social signal processing, deep learning, uncertainty modelling, speech signal processing, and group affect. 
\end{IEEEbiography}

\vspace{-1cm}

\begin{IEEEbiography}[{\includegraphics[width=1in,height=1.25in,clip,keepaspectratio]{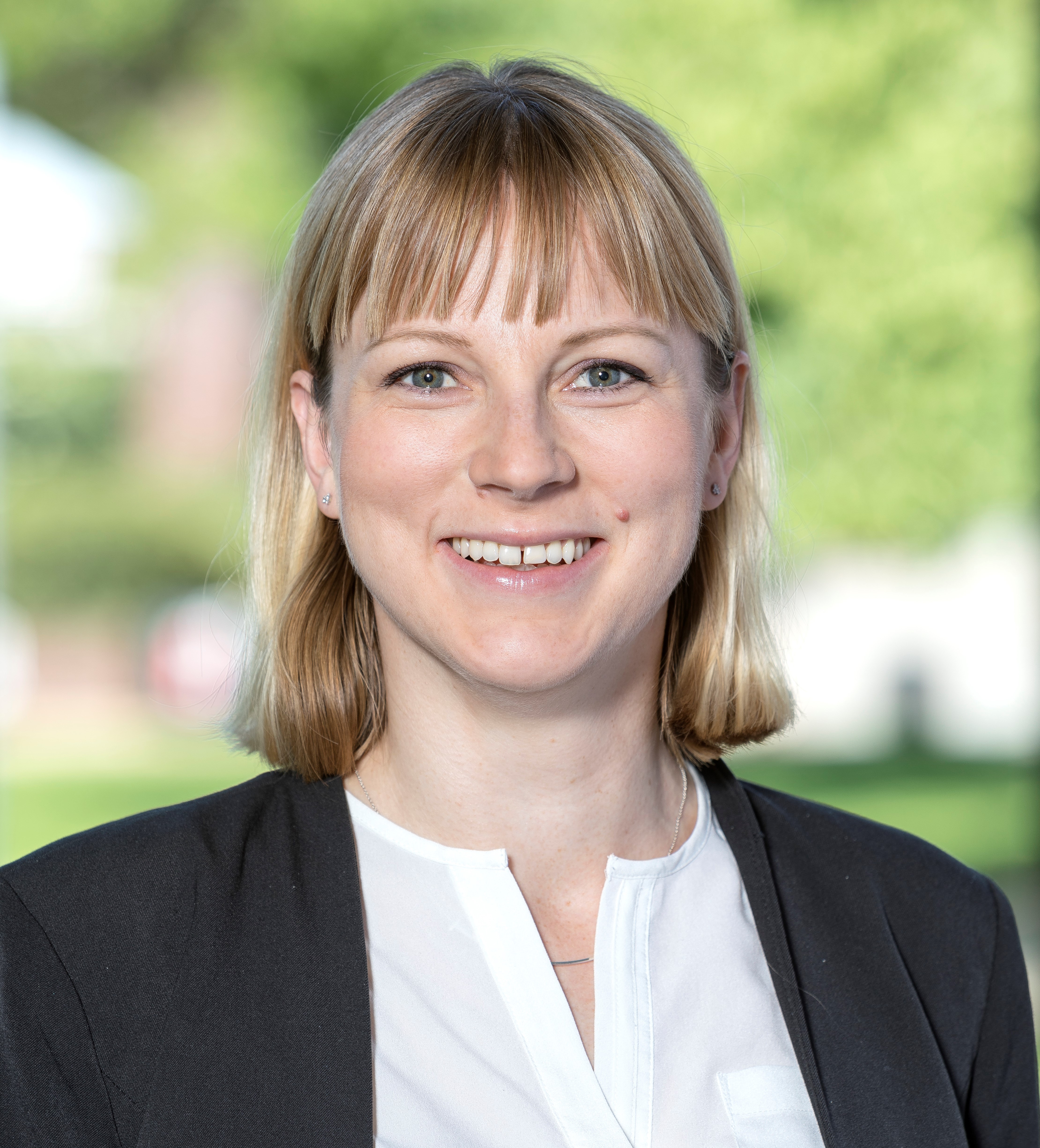}}]{Nale Lehmann-Willenbrock}
studied Psychology at the University of Goettingen and University of California, Irvine. She holds a PhD in Psychology from Technische Universität Braunschweig (2012). After several years working as an assistant professor at Vrije Universiteit Amsterdam and Associate Professor at the University of Amsterdam, she joined Universität Hamburg as a full professor and chair of Industrial/Organizational Psychology in 2018, where she also leads the Center for Better Work. She studies emergent behavioral patterns in organizational teams, social dynamics among leaders and followers, and meetings at the core of organizations. Her research program blends organizational psychology, management, communication, and social signal processing. She serves as associate editor for the Journal of Business and Psychology as well as for Small Group Research.
\end{IEEEbiography}


\vspace{-1cm}

\begin{IEEEbiography}[{\includegraphics[width=1in,height=1.25in,clip,keepaspectratio]{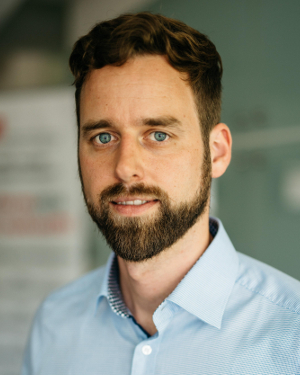}}]{Timo Gerkmann}
(S’08–M’10–SM’15) studied Electrical Engineering and Information Sciences at the Universität Bremen and the Ruhr-Universität Bochum in Germany. He received his Dipl.-Ing. degree in 2004 and his Dr.-Ing. degree in 2010 both in Electrical Engineering and Information Sciences from the Ruhr-Universität Bochum, Bochum, Germany. In 2005, he spent six months with Siemens Corporate Research in Princeton, NJ, USA. During 2010 to 2011 Dr. Gerkmann was a postdoctoral researcher at the Sound and Image Processing Lab at the Royal Institute of Technology (KTH), Stockholm, Sweden. From 2011 to 2015 he was a professor for Speech Signal Processing at the Universität Oldenburg, Oldenburg, Germany. During 2015 to 2016 he was a Principal Scientist for Audio \& Acoustics at Technicolor Research \& Innovation in Hanover, Germany. Since 2016 he is a professor for Signal Processing at the Universität Hamburg, Germany. His main research interests are on statistical signal processing and machine learning for speech and audio applied to communication devices, hearing instruments, audio-visual media, and human-machine interfaces. Timo Gerkmann serves as an elected member of the IEEE Signal Processing Society Technical Committee on Audio and Acoustic Signal Processing and as an Associate Editor of the IEEE/ACM Transactions on Audio, Speech and Language Processing. He received the VDE ITG award 2022.
\end{IEEEbiography}

\includepdf[pages=-]{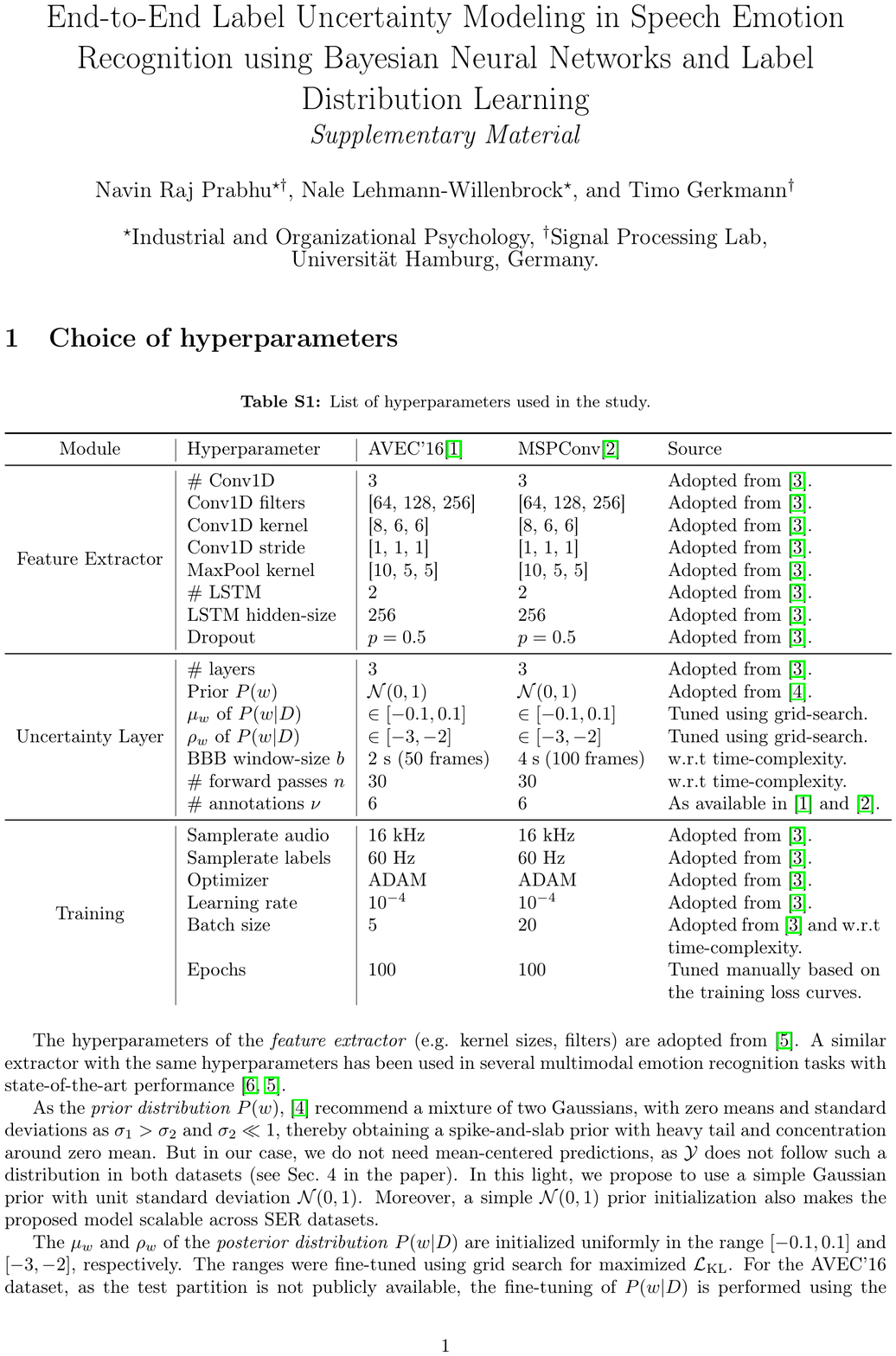}




\end{document}


\maketitle

\section{Choice of hyperparameters}
\begin{table*}[h!]
\caption{List of hyperparameters used in the study.}
\centering
\begin{tabular}{c|l|p{2.4cm}p{2.4cm}p{4cm}}
\toprule
    Module & Hyperparameter  & AVEC'16\cite{avec16} & MSPConv\cite{MspConv} & Source \\\midrule
\multirow{ 8}{*}{Feature Extractor}& 

          $\#$ Conv1D  & 3    & 3   & Adopted from \cite{Tzirakis2018-speech}.  \\
        & Conv1D filters  & [64, 128, 256]    & [64, 128, 256]   & Adopted from \cite{Tzirakis2018-speech}.  \\  
        & Conv1D kernel  & [8, 6, 6]    & [8, 6, 6]   & Adopted from \cite{Tzirakis2018-speech}.  \\
        & Conv1D stride  & [1, 1, 1]    & [1, 1, 1]   & Adopted from \cite{Tzirakis2018-speech}.  \\
        & MaxPool kernel & [10, 5, 5]   & [10, 5, 5]  & Adopted from \cite{Tzirakis2018-speech}.  \\
        
        & $\#$ LSTM   & 2     & 2   & Adopted from \cite{Tzirakis2018-speech}.  \\
        & LSTM hidden-size  & 256     & 256   & Adopted from \cite{Tzirakis2018-speech}.  \\

        & Dropout  & $p=0.5$    & $p=0.5$   & Adopted from \cite{Tzirakis2018-speech}.  \\
        
                         \midrule
                         
\multirow{ 7}{*}{Uncertainty Layer}& 

          $\#$ layers & 3  & 3   & Adopted from \cite{Tzirakis2018-speech}. \\
        & Prior $P(w)$  & $\mathcal{N}(0, 1)$    & $\mathcal{N}(0, 1)$   & Adopted from \cite{blundell2015weight}.  \\
        & $\mu_w$ of $P(w|D)$ & $\in[-0.1, 0.1]$ & $\in[-0.1, 0.1]$ & Tuned using grid-search.  \\
        & $\rho_w$ of $P(w|D)$ & $\in[-3, -2]$    & $\in[-3, -2]$   & Tuned using grid-search.  \\
        & BBB window-size $b$  & 2~s ($50$ frames)    & 4~s ($100$ frames)   &  w.r.t time-complexity.  \\
        & $\#$ forward passes $n$  & 30    & 30   &  w.r.t time-complexity. \\ 
        & $\#$ annotations $\nu$  & 6    & 6   & As available in \cite{avec16} and \cite{MspConv}.  \\ 
                         
                          \midrule
                         
\multirow{ 8}{*}{Training}& 
          Samplerate audio & 16 kHz    & 16 kHz   & Adopted from \cite{Tzirakis2018-speech}.  \\
        & Samplerate labels & 60 Hz    & 60 Hz   & Adopted from \cite{Tzirakis2018-speech}.  \\
        & Optimizer  & ADAM    & ADAM   & Adopted from \cite{Tzirakis2018-speech}.  \\
        & Learning rate  & $10^{-4}$    & $10^{-4}$   & Adopted from \cite{Tzirakis2018-speech}.  \\
        & Batch size  & 5  & 20   & Adopted from \cite{Tzirakis2018-speech} and w.r.t time-complexity.  \\
        & Epochs  & 100    & 100   &  Tuned manually based on the training loss curves.  \\ 
                         
\end{tabular}
\label{tab:hyperparameter-list}
\end{table*}

The hyperparameters of the \emph{feature extractor} (e.g. kernel sizes, filters) are adopted from \cite{tzirakis2021-semspeech}. A similar extractor with the same hyperparameters has been used in several multimodal emotion recognition tasks with state-of-the-art performance \cite{tzirakis2021-mm, tzirakis2021-semspeech}. 

As the \emph{prior distribution} $P(w)$, \cite{blundell2015weight} recommend a mixture of two Gaussians, with zero means and standard deviations as $\sigma_1 > \sigma_2$ and $\sigma_2 \ll 1$, thereby obtaining a spike-and-slab prior with heavy tail and concentration around zero mean. But in our case, we do not need mean-centered predictions, as $\mathcal{Y}$ does not follow such a distribution in both datasets (see Sec.~4 in the paper). In this light, we propose to use a simple Gaussian prior with unit standard deviation $\mathcal{N}(0, 1)$. Moreover, a simple $\mathcal{N}(0, 1)$ prior initialization also makes the proposed model scalable across SER datasets.

The $\mu_w$ and $\rho_w$ of the \emph{posterior distribution} $P(w|D)$ are initialized uniformly in the range $[-0.1, 0.1]$ and $[-3, -2]$, respectively. The ranges were fine-tuned using grid search for maximized $\mathcal{L}_\text{KL}$. For the AVEC'16 dataset, as the test partition is not publicly available, the fine-tuning of $P(w|D)$ is performed using the train partition. For the MSPConv dataset, the development partition is used. Also, note that the \emph{posterior distribution} $P(w|D)$ and time-shift for post-processing are the only parameters tuned using the partitions.


It is computationally expensive to sample new weights at every time-step ($40$~ms) and also the level of uncertainties varies rather slowly. In this light, for the AVEC'16 dataset, we set the \emph{BBB window-size} $b = 2$~s ($50$ frames). As the MSPConv dataset is comparatively larger, a compromise was made for computational simplicity and $b = 4$~s ($100$ frames) is used. For median filtering, a window-size of $2$~s is used. In this work, we assume a Gaussian on $\widehat{\mathcal{Y}}_t$, and noted previously that $n \ge 30$ is required for the assumption to hold. In this light, and keeping the time-complexity in mind, we fixed $n = 30$. 

For training, we use the Adam optimizer with a learning rate $10^{-4}$. The batch size used was 5 and 20, for AVEC'16 and MSPConv, respectively, with a sequence length of 300 frames, $40$~ms each. All models were trained for a fixed 100 epochs. The complete list of hyperparameters used by this work is listed in Table \ref{tab:hyperparameter-list}.

\section{Post-processing}


For all the baselines and models proposed in this work, two post-processing techniques are applied, namely, median filtering \cite{Tzirakis2018-speech} with window-size same as the BBB window-size $b$, and time-shifting \cite{mariooryad2014correcting} (with shifts between 0.04s and 10s). To find the best time-shift, a grid-search was performed between 0.04s and 10s using the training partition in AVEC'16 and the development partition in MSPConv. Specifically, the grid-search is performed to maximize $\mathcal{L}_{\text{ccc}}(m)$ metric in the respective partition, and subsequently the best time-shift is used to also recalculate the $\mathcal{L}_{\text{ccc}}(s)$ and $\mathcal{L}_\textrm{KL}$ measures.

The following trends were noticed during the post-processing of accounting for the annotator lag. On one hand, in the \emph{MSPConv} dataset, correction for annotator lag was not required for most of the models and baselines. This is because, as detailed in Sec.~4.1.2, we performed median and low-pass filtering on the continuous annotations, for a uniform sampling rate and to remove periodic distortions noticed in the dataset. These filtering techniques which inherently use sliding-windows might have already filtered out the annotator lags. On average across the baselines and models, correction for a lag of 0.24 was sufficient to achieve the best results. On the other hand, in the \emph{AVEC'16} dataset, on average across the baselines and models, a rather large correction of 1.36s was required to achieve the best results.

\section{Mean- and Mode- seeking KL divergence}

\begin{figure}[H]
     \centering
     \begin{subfigure}[b]{0.4\textwidth}
        \includegraphics[width=\textwidth]{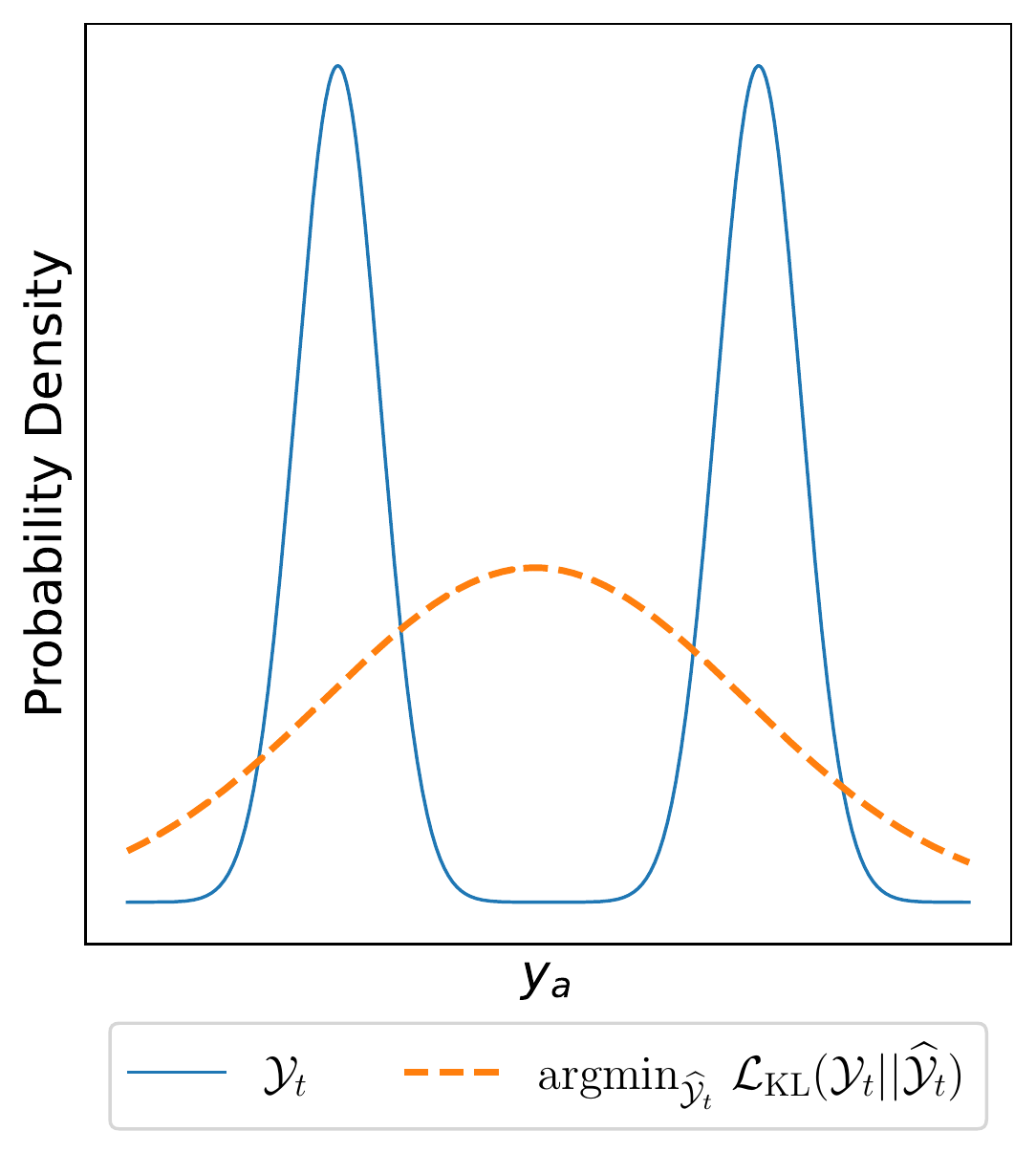}
        \caption{\textit{Mean}-seeking approximation: $\mathcal{L}_\mathrm{KL}(\mathcal{Y}_t || \widehat{\mathcal{Y}}_t)$}
        \label{Fig:mean-seek-approx}
     \end{subfigure}
     \hspace{2cm}
     \begin{subfigure}[b]{0.4\textwidth}
        \includegraphics[width=\textwidth]{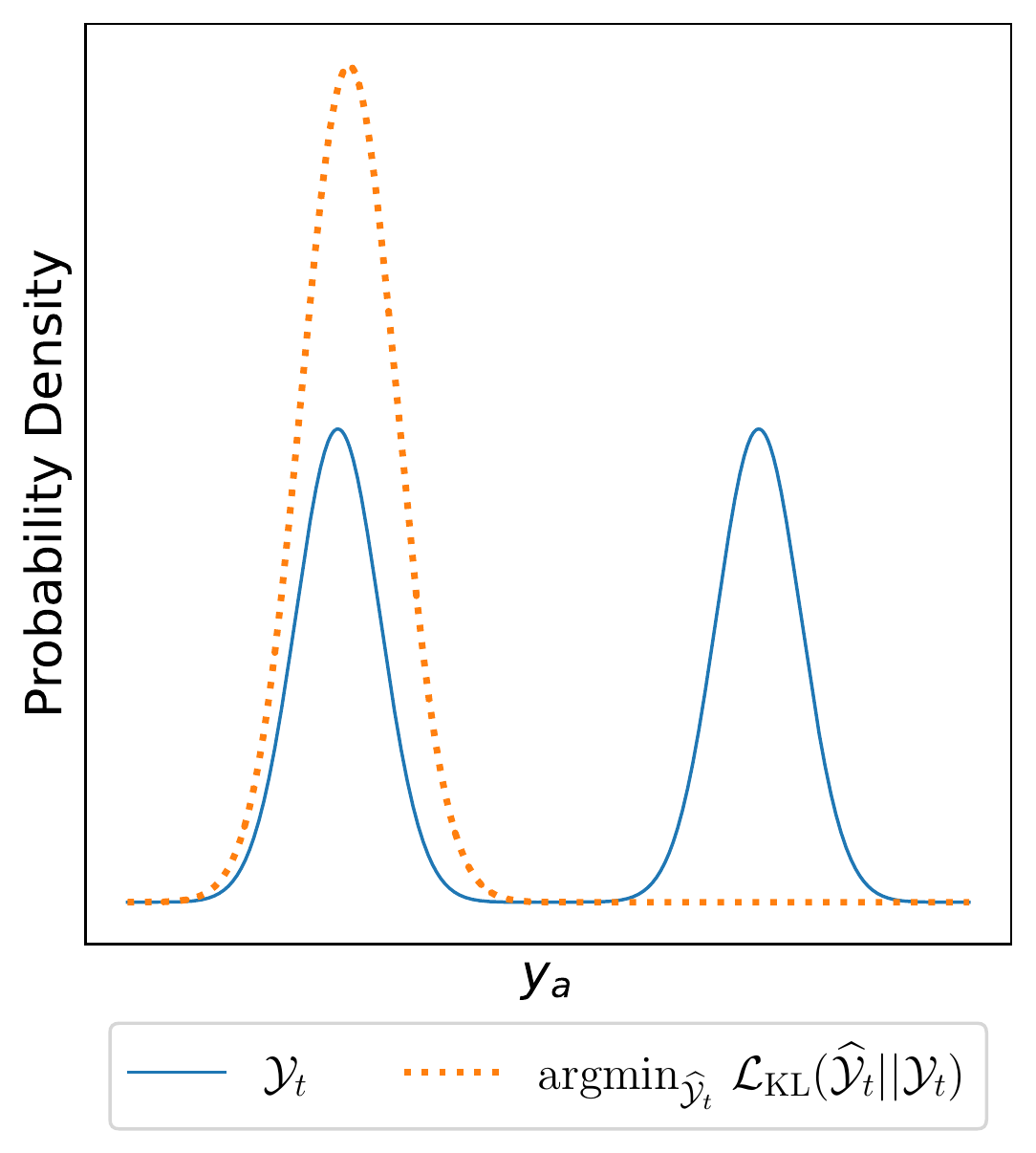}
        \caption{\textit{Mode}-seeking approximation: $\mathcal{L}_\mathrm{KL}(\widehat{\mathcal{Y}}_t || \mathcal{Y}_t)$}
        \label{Fig:mode-seek-approx}
     \end{subfigure}
\caption{Comparison between the mean- and mode- seeking approximations of KL divergence $\mathcal{L}_\textrm{KL}$.}
\label{Fig:kl_approx}
\end{figure}

The KL divergence $\mathcal{L}_\textrm{KL}$ is asymmetric. We have the choice of minimizing either $\mathcal{L}_\mathrm{KL}(\mathcal{Y}_t || \widehat{\mathcal{Y}}_t)$ or $\mathcal{L}_\mathrm{KL}(\widehat{\mathcal{Y}}_t || \mathcal{Y}_t)$. In Figure \ref{Fig:kl_approx}, we illustrate the difference between the two choices of approximations: the \textit{mean}-seeking approximation, where the ground-truth distribution $\mathcal{Y}_t$ is followed by its estimate distribution $\widehat{\mathcal{Y}}_t$, and the mode-seeking approximation, where the order is reversed and the estimate $\widehat{\mathcal{Y}}_t$ is followed by the ground-truth $\mathcal{Y}_t$. In case of the \textit{mean}-seeking approximation (Figure \ref{Fig:mean-seek-approx}), when $\mathcal{Y}_t$ has multiple modes, the estimate $\widehat{\mathcal{Y}}_t$ blurs the modes together by estimating high probability mass on all of them, thereby capturing the whole distribution \cite{Goodfellow-et-al-2016}. But in case of the \textit{mode}-seeking approximation (Figure \ref{Fig:mode-seek-approx}), when $\mathcal{Y}_t$ has multiple modes, $\mathcal{L}_\textrm{KL}$ is minimized by 
fitting on a \textit{single mode}, thereby not capturing the whole distribution \cite{Goodfellow-et-al-2016}. However we argue that, in our case of modeling emotion annotations $y_a$ as a distribution $\mathcal{Y}_t$, we require the estimate distribution $\widehat{\mathcal{Y}}_t$ to capture the whole distribution without fitting on a single mode. Intuitively, when $\widehat{\mathcal{Y}}_t$ is fit on a single mode it fails to produce reliable mean and standard-deviation estimates, a crucial goal in uncertainty modeling for emotion recognition research. Moreover, our preliminary experiments comparing the mean- and mode-seeking approximations also indicated that the mean-seeking approximation tends to achieve better distribution modeling results than the mode-seeking one.

\section{Modeling distributions with only 3 samples: Theoretical analysis}
\begin{figure}[h!]
     \centering
     \begin{subfigure}[b]{0.4\textwidth}
        \includegraphics[width=\textwidth]{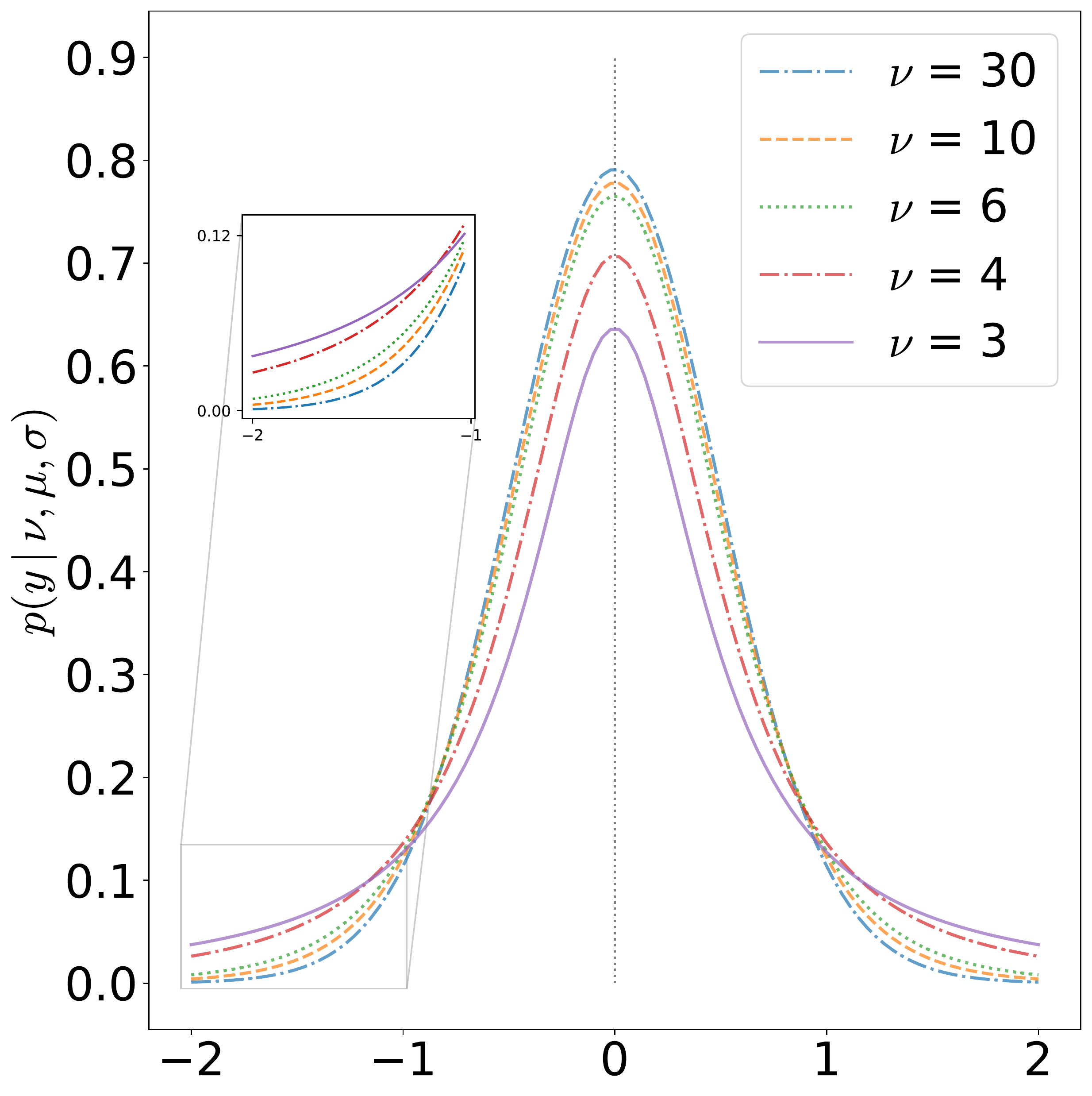}
        \caption{Highly-uncertain distribution as $\nu$ decreases.}
        \label{Fig:dist_uncertain_nu}
     \end{subfigure}
     \hspace{1cm}
     \begin{subfigure}[b]{0.49\textwidth}
        \includegraphics[width=\textwidth]{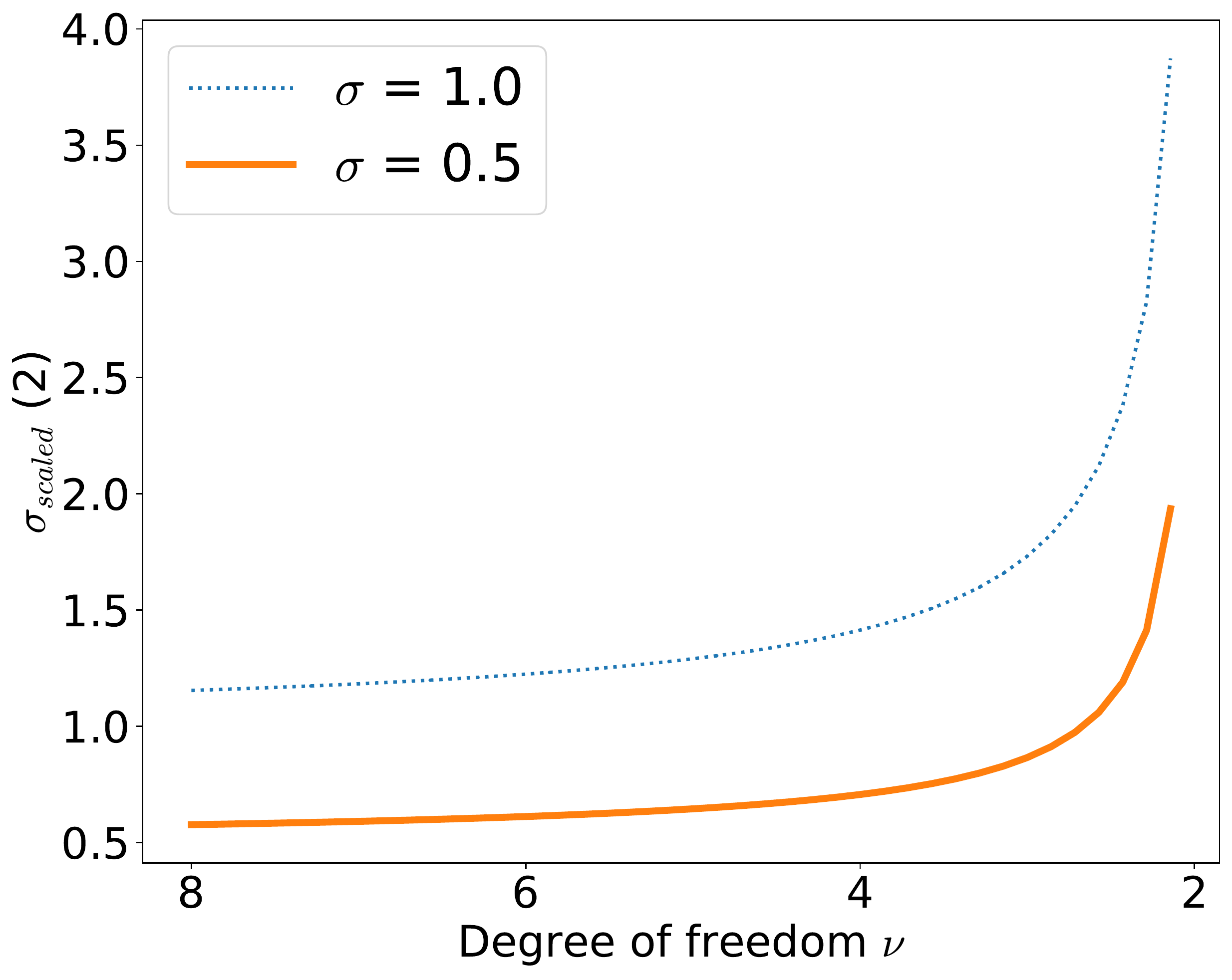}
        \caption{Increasing standard deviation $\sigma_{scaled}$ as $\nu$ decreases}
        \label{Fig:std_trend_nu}
     \end{subfigure}
\caption{Effect of degree of freedom $\nu$ on the scaling of $\sigma$ and highly-uncertain distribution.}
\label{Fig:anal_sigma_nu}
\end{figure}

Unlike the Gaussian distribution, the $t$-distribution also accounts for the number of samples used to model the distribution through the degree of freedom $\nu$ included in its probability density function,
\begin{equation}\label{pdf:tstudent}
    p(y\mid \nu ,{ {\mu }},{ {\sigma }})=\frac{1}{\mathrm{B} (\frac{1}{2}, \frac{\nu}{2})} \frac{1} {\sqrt{\nu {\sigma }^2}} \left(1+\frac{(y- {\mu })^2}{\nu {\sigma }^2}\right)^{-{\frac {\nu +1}{2}}},
\end{equation}
where B(., .) is the Beta function, for Gamma function $\Gamma$, formulated as $B(i, j) = \dfrac{\Gamma(i)\,\Gamma(j)}{\Gamma(i+j)}$. Furthermore, the standard deviation ${\sigma}$ of the $t$-distribution, in \eqref{pdf:tstudent}, takes the scaled form, where $\sigma$ is scaled using the normality parameter $\nu$:
\begin{equation}\label{eq:scale-sigma}
     { {\sigma } \, \sqrt{\frac {\nu }{\nu -2}} {\text{  for }}}\nu >2.
\end{equation}
Let us denote this scaled form of the standard deviation as $\sigma_{scaled}$. Through this scaling \eqref{eq:scale-sigma}, the standard deviation of the $t$-distribution $\sigma_{scaled}$ \textit{increases} with decreasing number of annotation samples available to model the distribution \cite{villa2018objective}. Bishop \cite{bishop2006pattern} associates this scaled standard deviation $\sigma_{scaled}$ towards the increased robustness of the $t$-distribution towards outliers and sparse distributions. This is also noticed from the results of the experiments presented where the $t$-distribution is superior in modeling the annotation distribution over the Gaussian. However, a caveat of this $\sigma_{scaled}$ is that when the number of annotations samples is \emph{less than 4}, the $t$-distribution associates this as a highly-uncertain distribution with a highly scaled $\sigma$. Figure \ref{Fig:anal_sigma_nu} further illustrates the impact of $\sigma_{scaled}$ on the probability density function of the $t$-distribution \eqref{pdf:tstudent}. Figure \ref{Fig:dist_uncertain_nu} depicts the increasing uncertainty in the $t$-distribution as the degrees of freedom $\nu$ decreases. The zoomed region further highlights the case of $\nu=3$ (only 3 annotations available) where a relatively high likelihood is associated along the tails, thereby the distribution becomes highly-uncertain. Figure \ref{Fig:std_trend_nu} illustrates the increasing scaled standard deviation $\sigma_{scaled}$ with reducing $\nu$.  It is noted here that, for $\nu\leq3$, the rate of increase in standard deviation further enlarges, thereby explaining the reason why label distribution modeling fails when only 3 annotation samples are available. Note that this highly-uncertain distribution and scaled standard deviation also affects the Kullback–Leibler divergence loss thereby affecting the training process.



\section{Effect of $\alpha$: regularization with label uncertainty loss term $\mathcal{L}_\textrm{KL}$}

\begin{figure}[h!]
     \centering
     \begin{subfigure}[b]{0.24\textwidth}
        \includegraphics[width=\textwidth]{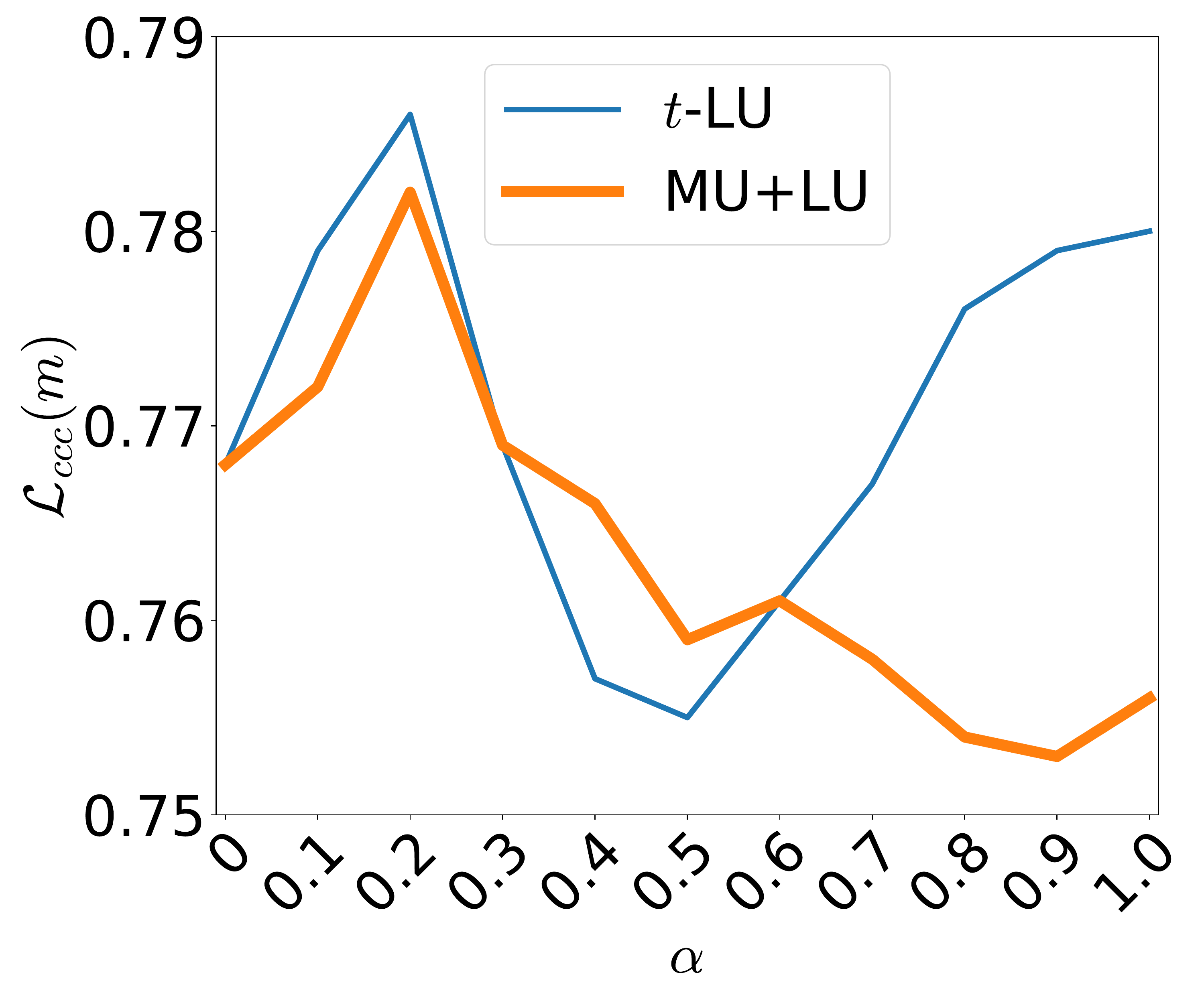}
        \caption{$\mathcal{L}_{\text{ccc}}(m)$ in \textit{arousal}}
        \label{Fig:rec_arousal_ccc_m}
     \end{subfigure}
    \hfill
     \begin{subfigure}[b]{0.24\textwidth}
        \includegraphics[width=\textwidth]{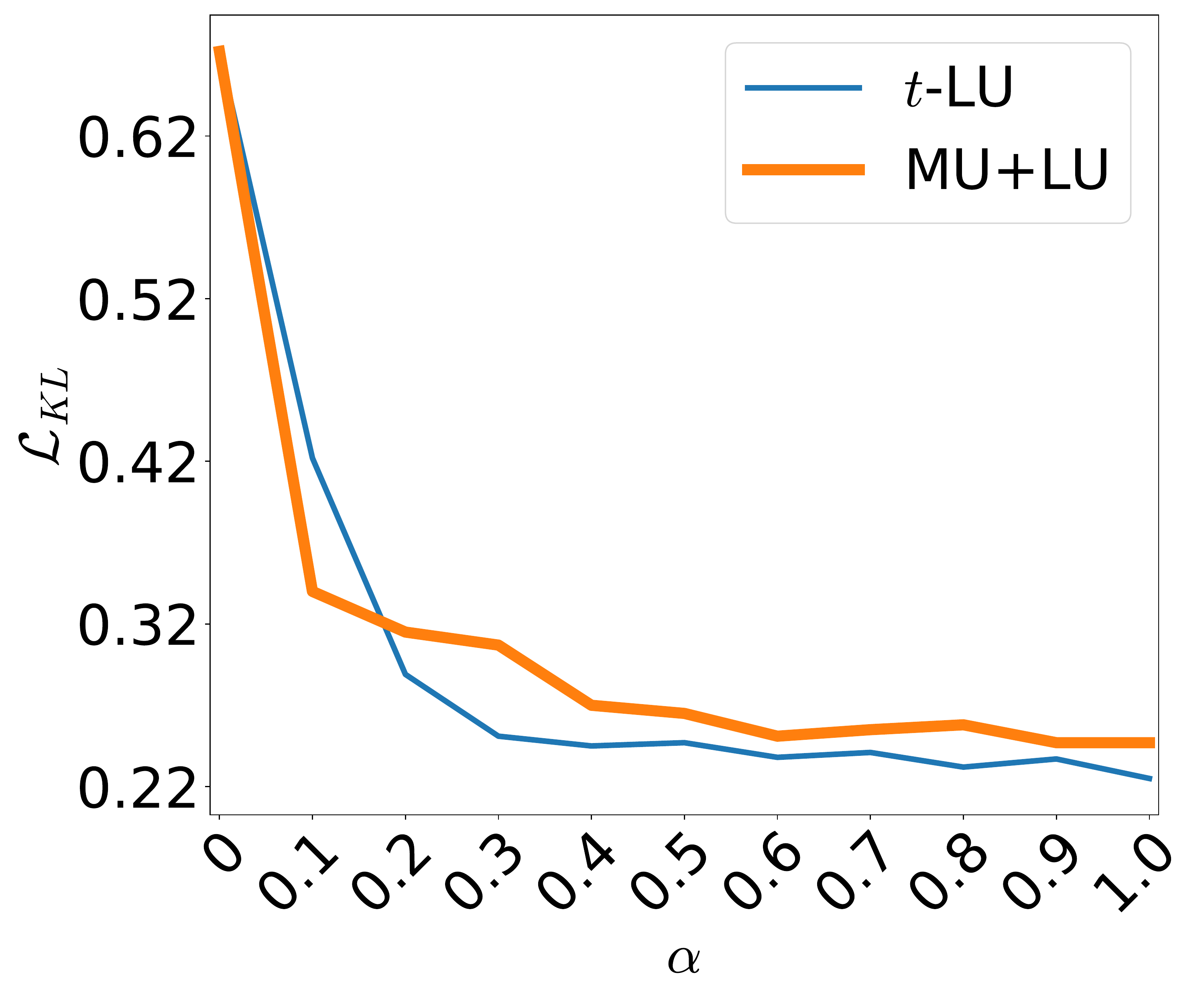}
        \caption{$\mathcal{L}_{\text{KL}}$ in \textit{arousal}}
        \label{Fig:rec_arousal_kl}
     \end{subfigure}
    \hfill
     \begin{subfigure}[b]{0.24\textwidth}
        \includegraphics[width=\textwidth]{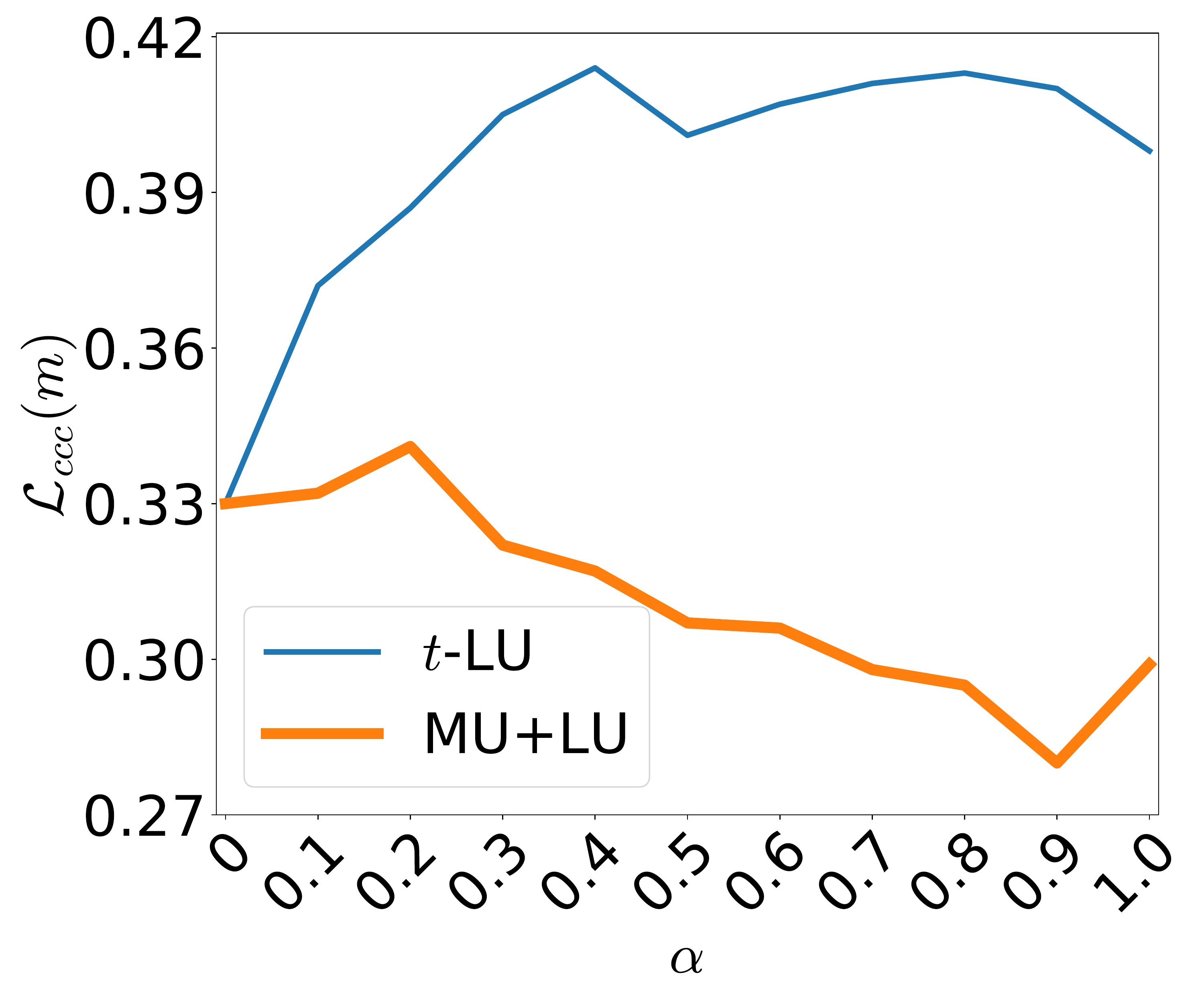}
        \caption{$\mathcal{L}_{\text{ccc}}(m)$ in \textit{valence}}
        \label{Fig:rec_valence_ccc_m}
     \end{subfigure}
    \hfill
     \begin{subfigure}[b]{0.24\textwidth}
        \includegraphics[width=\textwidth]{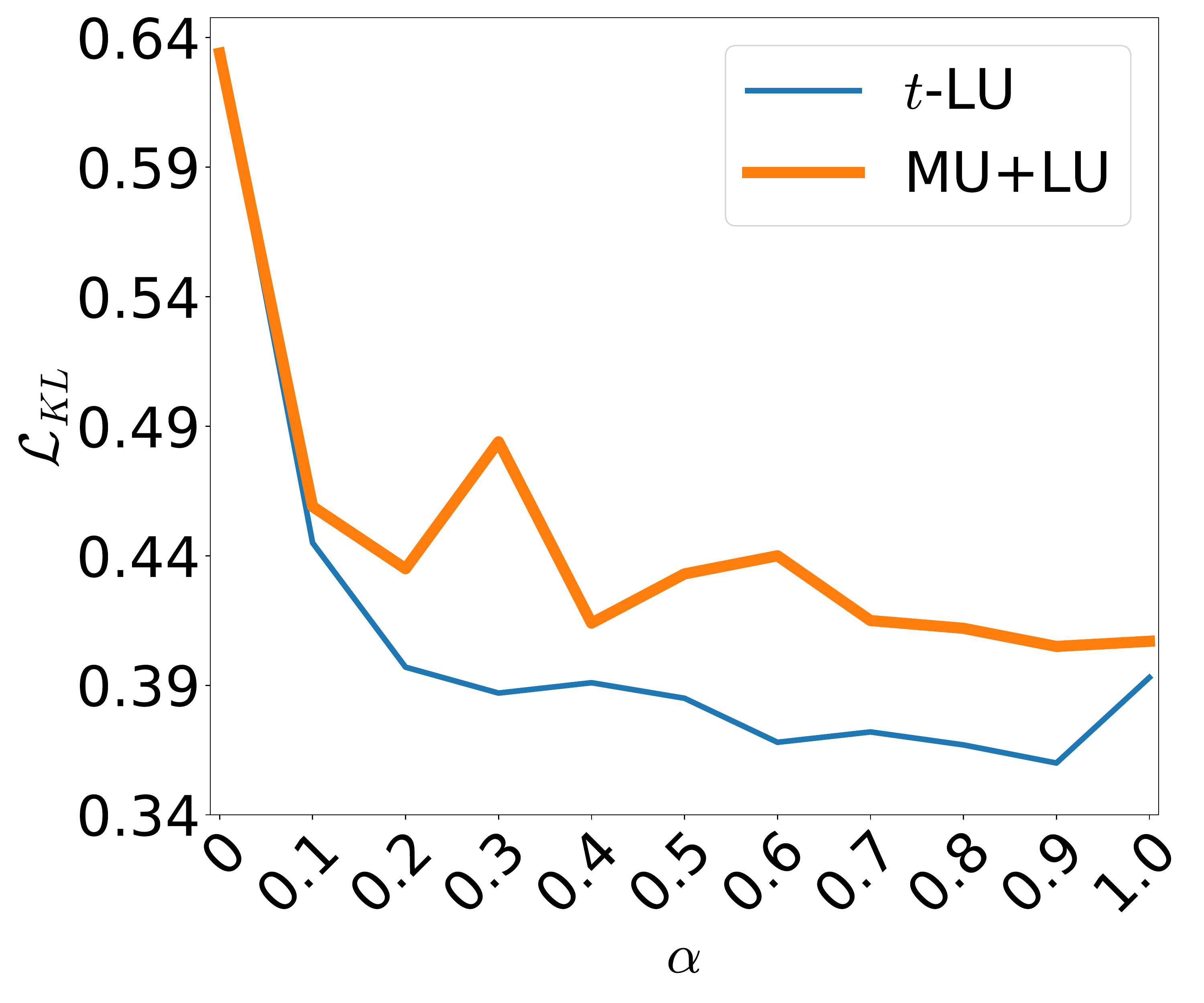}
        \caption{$\mathcal{L}_{\text{KL}}$ in \textit{valence}}
        \label{Fig:rec_valence_kl}
     \end{subfigure}
\caption{Effect of $\alpha$: regularization with label uncertainty loss term $\mathcal{L}_\textrm{KL}$, in the \textbf{AVEC'16} dataset.}
\label{Fig:alpha_test_recola}
\end{figure}

\begin{figure}[h!]
     \centering
     \begin{subfigure}[b]{0.24\textwidth}
        \includegraphics[width=\textwidth]{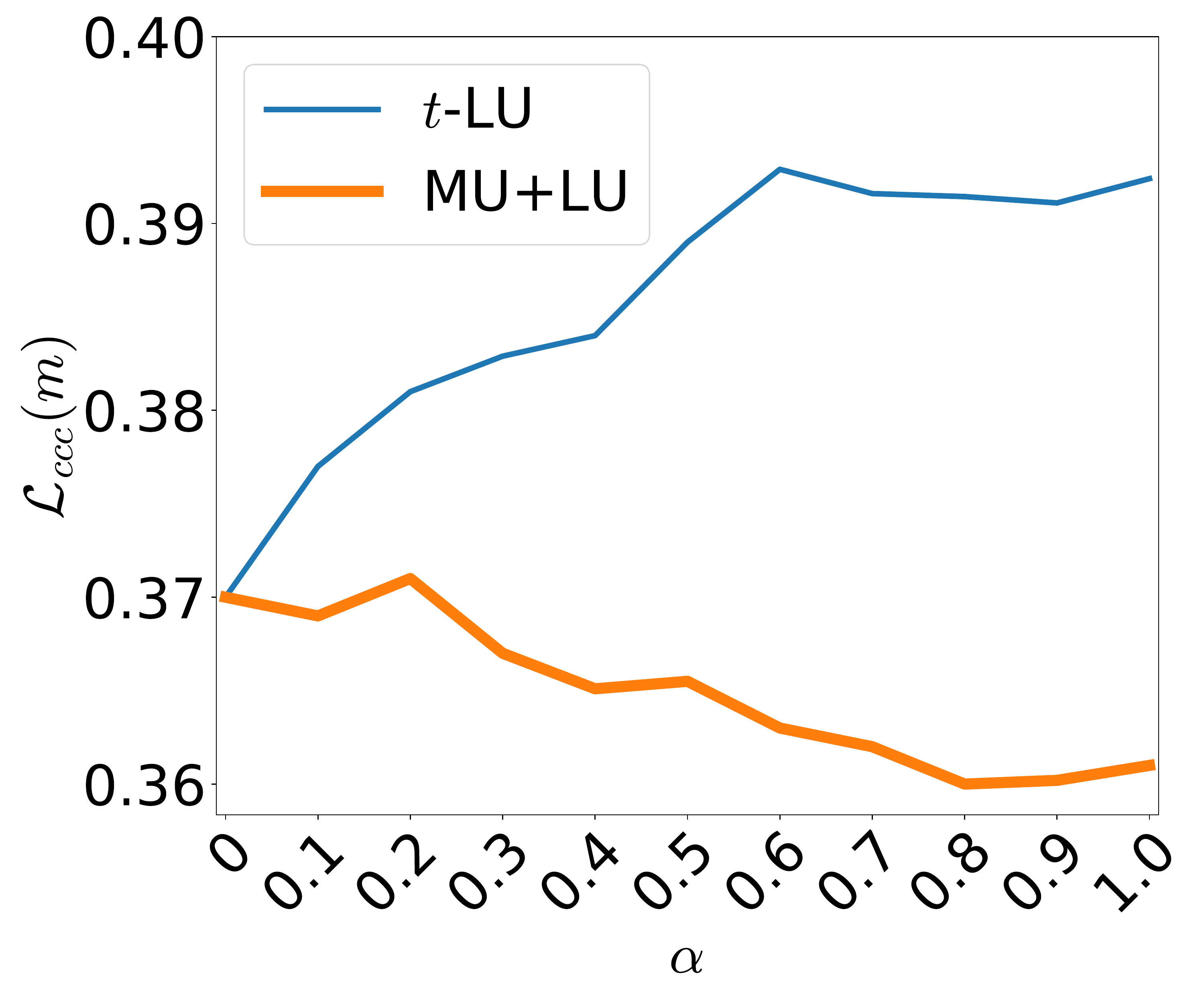}
        \caption{$\mathcal{L}_{\text{ccc}}(m)$ in \textit{arousal}}
        \label{Fig:msp_arousal_ccc_m}
     \end{subfigure}
    \hfill
     \begin{subfigure}[b]{0.24\textwidth}
        \includegraphics[width=\textwidth]{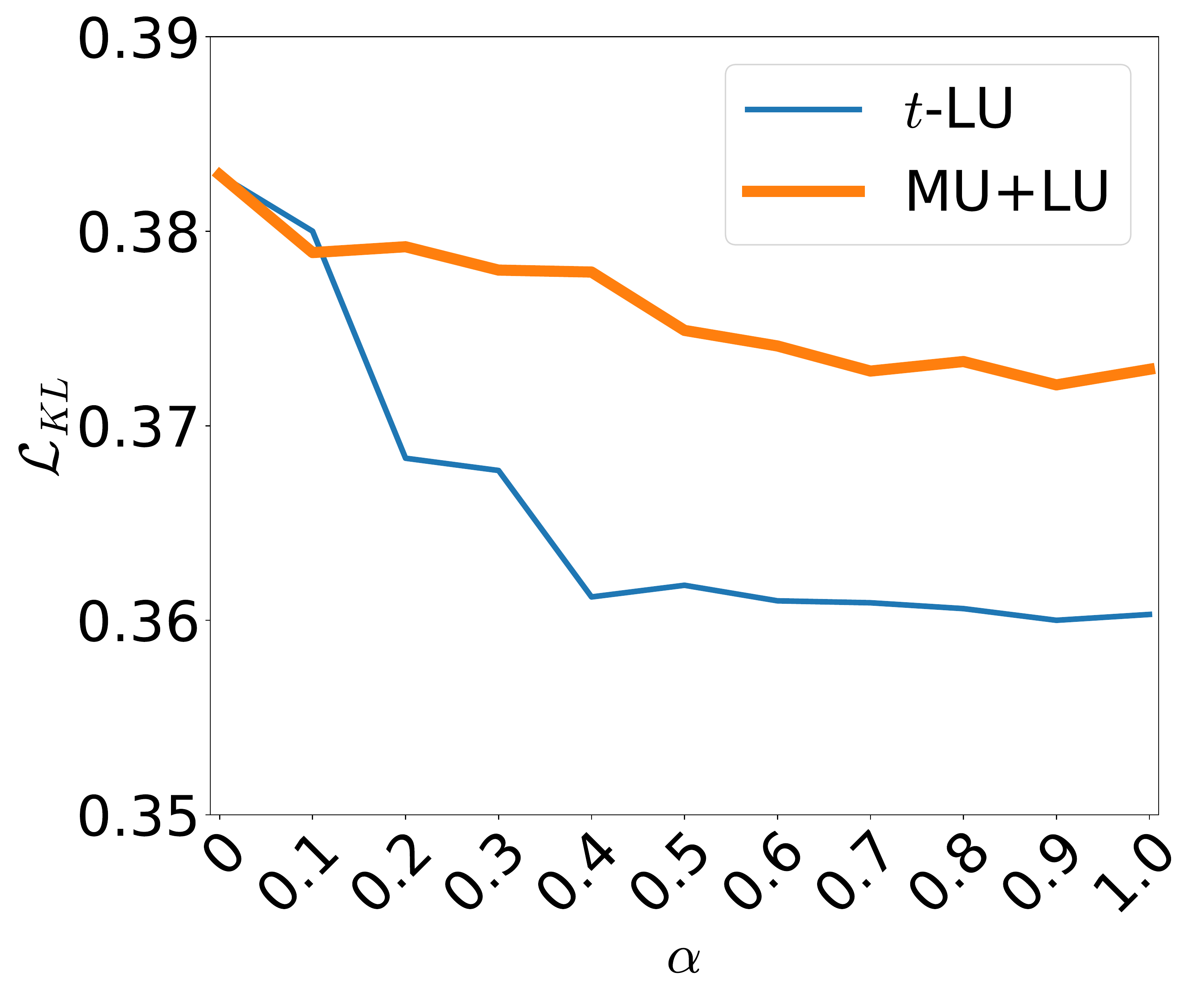}
        \caption{$\mathcal{L}_{\text{KL}}$ in \textit{arousal}}
        \label{Fig:msp_arousal_kl}
     \end{subfigure}
    \hfill
     \begin{subfigure}[b]{0.24\textwidth}
        \includegraphics[width=\textwidth]{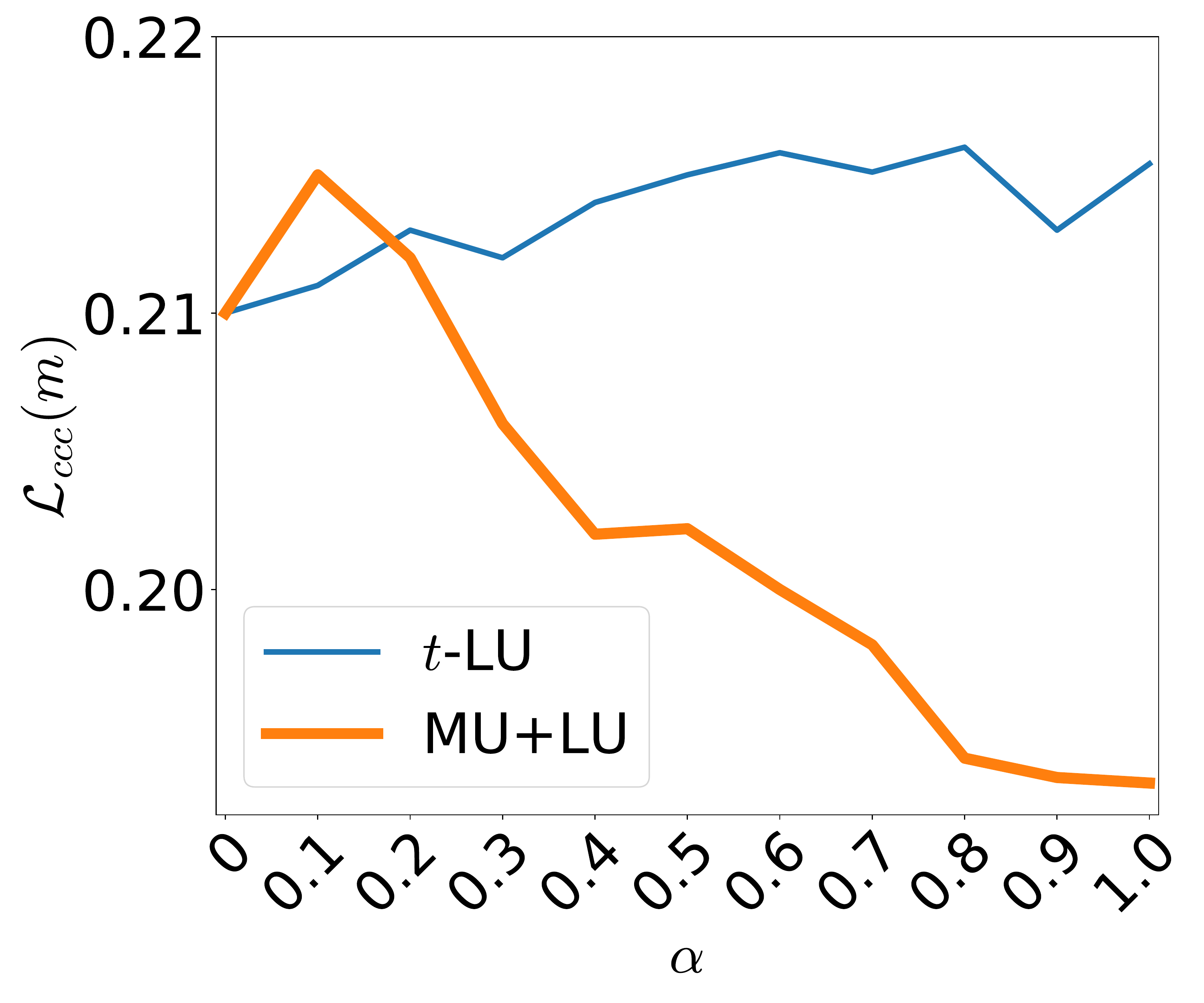}
        \caption{$\mathcal{L}_{\text{ccc}}(m)$ in \textit{valence}}
        \label{Fig:msp_valence_ccc_m}
     \end{subfigure}
    \hfill
     \begin{subfigure}[b]{0.24\textwidth}
        \includegraphics[width=\textwidth]{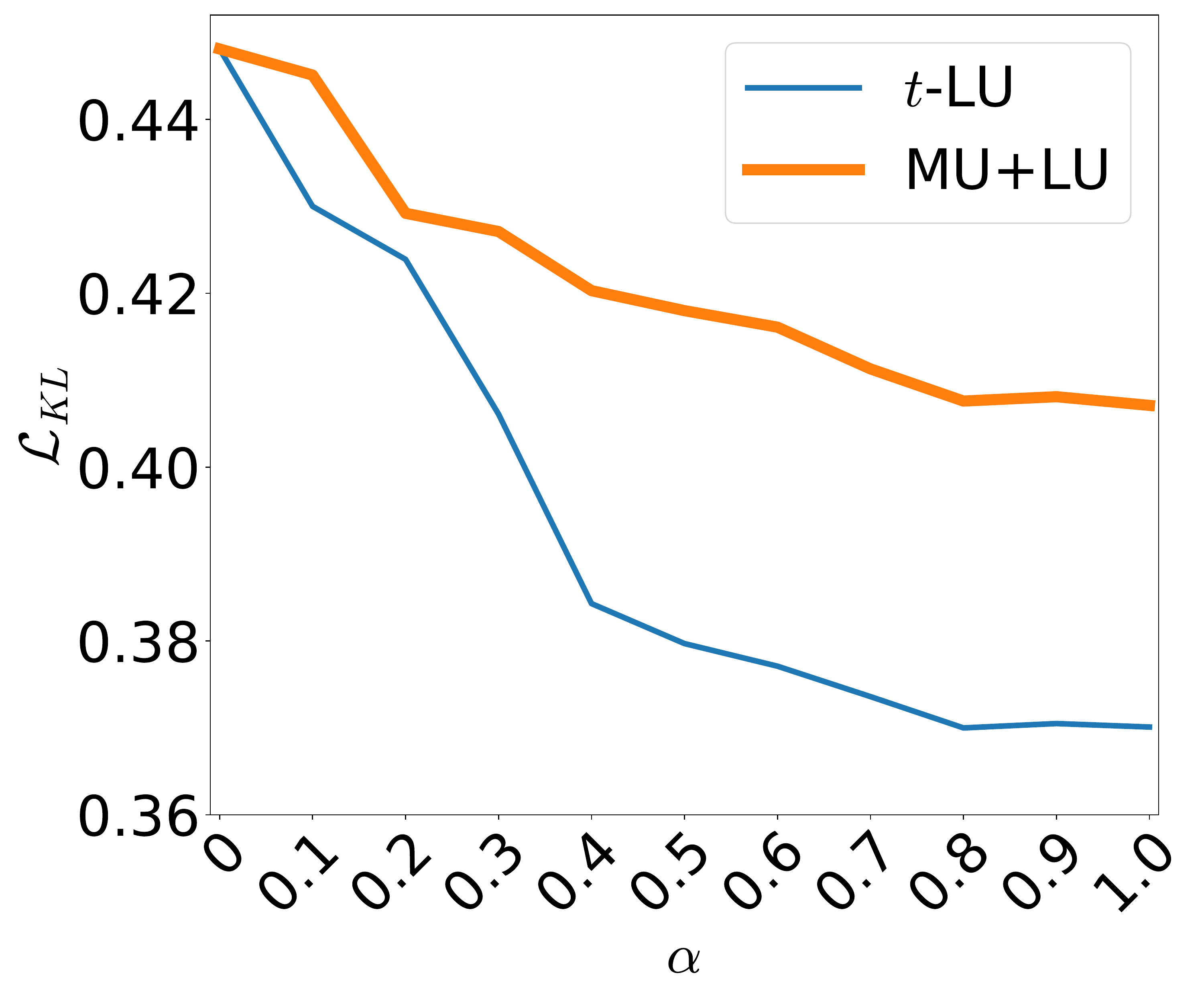}
        \caption{$\mathcal{L}_{\text{KL}}$ in \textit{valence}}
        \label{Fig:msp_valence_kl}
     \end{subfigure}
\caption{Effect of $\alpha$: regularization with label uncertainty loss term $\mathcal{L}_\textrm{KL}$, in the \textbf{MSPConv} dataset.}
\label{Fig:alpha_test_msp}
\end{figure}

The proposed end-to-end uncertainty loss is,
\begin{equation}\label{eq:end-to-end_loss}
   \mathcal{L} = (1 - \mathcal{L}_{\text{CCC}}(m)) + \mathcal{L}_{\text{BBB}} + \alpha \mathcal{L}_{\text{KL}}.
\end{equation}
Intuitively, $\mathcal{L}_{\text{CCC}}(m)$ optimizes for mean predictions $m$, $\mathcal{L}_{\text{BBB}}$ optimizes for BBB weight distributions, and $\mathcal{L}_{\text{KL}}$ optimizes for the label distribution $\mathcal{Y}_t$. The variable $\alpha$ controls the degree to which the model is regularized on the label uncertainty loss term $\mathcal{L}_\textrm{Kl}$. For $\alpha=0$, the model only captures model uncertainty (\emph{MU}). For $\alpha=1$, the model also captures \emph{label uncertainty} (\emph{MU$+$LU} or \emph{$t$-LU}). To further understand the effect of the regularization weighting factor $\alpha$, we performed experiments with varying $\alpha$ from 0 to 1 and with a hop of 0.1. The results of the experiments, in-terms of the $\mathcal{L}_{\text{CCC}}(m))$ and $\mathcal{L}_{\text{KL}}$ metrics, for the AVEC'16 \cite{avec16} and MSPConv \cite{MspConv} datasets can be seen in Figures \ref{Fig:alpha_test_recola} and \ref{Fig:alpha_test_msp}, respectively.

From figures \ref{Fig:alpha_test_recola} and \ref{Fig:alpha_test_msp}, we observe the following trends with respect to the regularization factor $\alpha$. Firstly, with increasing $\alpha$, as expected, the $\mathcal{L}_\textrm{KL}$ also decreases, with a sharp decrease until $\alpha=0.3$ and gradually decreasing for $\alpha\geq0.3$ until it starts plateauing from $\alpha=0.7$ (seen from figures \ref{Fig:rec_arousal_kl}, \ref{Fig:rec_valence_kl}, \ref{Fig:msp_arousal_kl}, \ref{Fig:msp_valence_kl}). This indicates that both the $t$-LU and MU+LU models reach their maximum capacity in-terms of modeling the label distribution $\mathcal{Y}_t$ with an $\alpha$ greater than 0.7. Furthermore, in-terms of the mean-estimates $\mathcal{L}_{\text{ccc}}(m)$, crucially we note that, with increasing regularization on the label uncertainty loss $\mathcal{L}_\textrm{KL}$, while the $t$-LU model performance increases with increasing $\alpha$, the MU+LU model performance drops gradually with increasing $\alpha$ (seen from figures \ref{Fig:rec_arousal_ccc_m}, \ref{Fig:rec_valence_ccc_m}, \ref{Fig:msp_arousal_ccc_m}, \ref{Fig:msp_valence_ccc_m}). This behaviour further emphasises that $t$-LU is free from the compromises MU+LU make on mean-estimates $\mathcal{L}_{\text{ccc}}(m)$ while modeling label uncertainty (detailed in Sec.~5.1.1 in the paper). Similarly to the plateauing of $\mathcal{L}_\textrm{KL}$ for $\alpha$ greater than 0.7, the $\mathcal{L}_{\text{ccc}}(m)$ also starts plateauing from 0.8. Overall, from figures \ref{Fig:alpha_test_recola} and \ref{Fig:alpha_test_msp}, with respect to both mean-estimate modeling $\mathcal{L}_{\text{ccc}}(m)$ and label distribution modeling $\mathcal{L}_\textrm{KL}$, we recommend using a regularization factor of $\alpha\in[0.8, 1.0]$.

\newpage
\bibliographystyle{IEEEtran}
\bibliography{references}
\printbibliography